\documentclass{aa} 

\usepackage{natbib}
\bibpunct{(}{)}{;}{a}{}{,}
\usepackage{graphicx}
\usepackage{txfonts}
\DeclareRobustCommand{\uppartial}{\text{\rotatebox[origin=t]{20}{\scalebox{1}[1]{$\partial$}}}\hspace{-1pt}} 
\usepackage{geometry}
\usepackage{amsmath,amssymb}

\usepackage{bm}
\DeclareMathAlphabet{\mathbfsf}{\encodingdefault}{\sfdefault}{bx}{n}

\usepackage{subcaption}
\usepackage{minibox}
\usepackage{color}
\usepackage{upgreek}
\usepackage{pifont}
\usepackage{empheq}
\usepackage{physics}

\usepackage{hyperref}

\numberwithin{equation}{section}

\usepackage{pgf, tikz}
\usepackage{tikz-3dplot}
\usepackage{pgfplots}
\usetikzlibrary{calc}
\usetikzlibrary{shapes.misc}
\usetikzlibrary{shapes,snakes}
\usetikzlibrary{arrows}
\usetikzlibrary{automata,positioning}
\tikzset{
  load/.style   = {ultra thick,-latex},
  stress/.style = {-latex},
  dim/.style    = {latex-latex},
  axis/.style   = {-latex},
}
\tikzset{dimetric2/.style={
  x={(0.935cm,-0.118cm)},
  y={(0.354cm, 0.312cm)},
  z={(0.000cm, 0.943cm)},
}}
\usepackage{color}
\definecolor{auburn}{rgb}{0.43, 0.21, 0.1}
\definecolor{oxfordblue}{rgb}{0.0, 0.13, 0.28}
\definecolor{coolblack}{rgb}{0.0, 0.18, 0.39}
\definecolor{darkcerulean}{rgb}{0.03, 0.27, 0.49}
\definecolor{denim}{rgb}{0.08, 0.38, 0.74}
\definecolor{purpletaupe}{rgb}{0.31, 0.25, 0.3}
\definecolor{purpletaupemodif}{rgb}{0, 0, 0}
\definecolor{applegreen}{rgb}{0.55, 0.71, 0.0}
\definecolor{chaptercolor}{rgb}{0.2, 0.2, 0.2}
\definecolor{blue(ryb)}{rgb}{0.01, 0.28, 1.0}
\definecolor{ao}{rgb}{0.0, 0.0, 1.0}
\definecolor{electricultramarine}{rgb}{0.25, 0.0, 0.95}
\definecolor{electricpurple}{rgb}{0.75, 0.0, 1.0}
\definecolor{palatinatepurple}{rgb}{0.41, 0.16, 0.38}
\definecolor{aureolin}{rgb}{0.99, 0.93, 0.0}
\definecolor{fluorescentorange}{rgb}{1.0, 0.85, 0.0} 

\begin{document}

   \title{Layered semi-convection and tides in giant planet interiors}

   \subtitle{I. Propagation of internal waves}

   \author{Q. Andr\'e
          \inst{1}\fnmsep\inst{2}\fnmsep\inst{3}
          \and
          A. J. Barker\inst{2}\fnmsep\inst{4}
          \and S. Mathis\inst{1}\fnmsep\inst{5}
          }
   \institute{Laboratoire AIM Paris-Saclay, CEA/DRF, CNRS, Univ. Paris-Diderot, IRFU/SAp Centre de Saclay, 91191 Gif-sur-Yvette, France\\
              \email{quentin.andre@cea.fr}
         \and
         	Department of Applied Mathematics and Theoretical Physics, University of Cambridge, Centre for Mathematical Sciences, Wilberforce Road, Cambridge CB3 0WA, UK
         \and
         	D\'epartement de Physique, ENS Paris-Saclay, Universit\'e Paris-Saclay, 61 Avenue du Pr\'esident Wilson, 94230 Cachan, France
         \and
             Department of Applied Mathematics, School of Mathematics, University of Leeds, Leeds, LS2 9JT, UK\\
             \email{a.j.barker@leeds.ac.uk}
         \and
         	LESIA, Observatoire de Paris, PSL Research University, CNRS, Sorbonne Universit\'es, UPMC Univ. Paris 06, Univ. Paris-Diderot, Sorbonne Paris Cit\'e, 5 place Jules Janssen, 92195 Meudon, France\\
         	\email{stephane.mathis@cea.fr}
             }

   \date{Received XXXX; accepted YYYY}

 
  \abstract
   {Layered semi-convection is a {possible} candidate to explain Saturn's luminosity excess and the abnormally large radius of some hot Jupiters. In giant planet interiors, {it could lead to the creation of density staircases}, which are convective layers separated by thin stably stratified interfaces. These are also observed on Earth in some lakes and in the Arctic Ocean.}
   {We study the propagation of internal waves in a region of layered semi-convection, with the aim to predict energy transport by internal waves incident upon a density staircase. {The goal is then to understand the resulting tidal dissipation when these waves are excited by other bodies} {such as moons in giant planets systems.}}
   {We use a local Cartesian analytical model, taking into account the complete Coriolis acceleration at any latitude, thus generalizing previous works. We use a model in which stably stratified interfaces are infinitesimally thin, before relaxing this assumption with a second model that assumes a piecewise linear stratification.}
   {We find transmission of incident internal waves to be strongly affected by the presence of a density staircase, even if these waves are initially pure inertial waves {(which are restored by the Coriolis acceleration)}. In particular, low-frequency waves of all wavelengths are perfectly transmitted near the critical latitude, {defined by ${\theta_{\text{c}} = \sin^{-1}({\omega}/{2\Omega})}$, where ${\omega}$ is the wave's frequency and ${\Omega}$ is the rotation rate of the planet}. {Otherwise,} short-wavelength waves are only efficiently transmitted if they are resonant with a free mode (interfacial gravity wave or short-wavelength inertial mode) of the staircase.  {In all other cases, waves are primarily reflected} unless their wavelengths are longer than the vertical extent of the {entire} staircase (not just a single step).}
   {We expect incident internal waves to be strongly affected by the presence of a density staircase in a frequency-, latitude- and wavelength-dependent manner. {First, this could lead to new criteria to probe the interior of giant planets by seismology; and second, this may have important consequences for tidal dissipation and our understanding of the evolution of giant planet systems.}}


   \maketitle
%
\section{Introduction}\label{SEC:intro}
%

Since the first discovery of a planet orbiting a star outside our solar system \citep{MayorQueloz1995}, astronomy has experienced an epoch of remarkable expansion: more than 3000 extrasolar planets are now confirmed, including more than 500 planets in multi-planetary systems {(see \url{http://exoplanet.eu})}. Planet formation is thus a universal physical process. Planetary systems subsequently evolve dynamically by gravitational and magnetic interactions over astronomical time-scales {\citep[e.g.][]{LaskarEtal2012, BolmontMathis2016, Strugarek2016}}. {The induced evolution of the orbital, rotational, thermal and compositional properties of the planets due to these interactions depends strongly on the internal structure of the planets involved} {\citep[e.g.][]{OgilvieLin2004, EfroimskyLainey2007, APM2014}}. {For example}, the convective instability, which is expected to operate in giant planet gaseous envelopes (which are the focus of attention here), {is efficient at transporting heat, homogenises mean density profiles and mixes chemical elements}. {This may strongly impact tidal friction in these planets \citep[][]{Zahn1966, Zahn1989, OgilvieLin2004, OgilvieLesur2012, MathisEtal2016}.} However, whether the gaseous envelope is fully convective remains an open question.

Planetary interiors are poorly constrained. The \textsc{juno} spacecraft, orbiting Jupiter since July, 4, 2016, should provide high precision measurements of Jupiter's gravitational potential, aiming to constrain its interior {\citep{MilitzerEtal2016}}. Giant planet seismology, on the other hand, is very difficult because radial velocities associated with modes that can potentially be observed in Jupiter or Saturn are of very small amplitude {\citep[][]{GaulmeEtal2011}}. Based on Saturn's ring seismology, \citet{Fuller2014} inferred that there could be a region of stable stratification in the deep interior of the giant planet, departing from the standard model of planetary interiors, which considers a large H/He convective gaseous envelope sitting on a rocky/icy core {\citep[that could be either fluid or solid, see e.g.][]{MazevetEtal2015}} expected from planet formation by core accretion \citep[see e.g.][]{PollackEtal1996}.

Moreover, it has been showed that a stabilizing compositional gradient could exist in certain regions of giant planet interiors as a natural outcome of planet formation and thermal evolution, thus competing with the destabilizing entropy gradient that drives the convective instability. Namely, this is expected to occur in two different regions:
\begin{itemize}
\item just outside the core, {the erosion of part of the core of giant planets was shown to be energetically plausible in \cite{GuillotEtal2004}. This is because} in the conditions of temperature and pressure that reign in {their central regions}, some of the heavy elements composing the core (e.g. silicates) are thermodynamically unstable \citep[e.g.][]{WilsonMilitzer2012a, WilsonMilitzer2012b, WahlEtal2013, Gonzalez-CataldoEtal2014, MazevetEtal2015}. Thus, erosion and redistribution of core materials in the envelope must be taken into account, and could provide a stable compositional gradient. {We also note that \cite{Stevenson1985} suggested that impacts of planetesimals with a giant planet (this could happen before the proto-planetary disc clears) could lead to the formation of a stabilizing gradient of heavy elements just outside the core.}
\item in the transition region between molecular and metallic H/He ices: the helium rain region \citep[see][]{Salpeter1973, Stevenson1975}, the phase separation between H and He could provide a stabilizing compositional gradient \citep{StevensonSalpeter1977} which could in turn trigger double diffusive convection \citep{NettelmannEtal2015}.
\end{itemize}

Thanks to laboratory experiments, it is well known that the presence of a compositional gradient (i.e. a gradient {of} the mean molecular weight) can change the mean density profile that develops in a stratified fluid. This is due to the fact that, like temperature, the mean molecular weight influences the buoyancy of the fluid \citep{Ledoux1947}.
But because of diffusive processes, even a density stratification {stable} with respect to the convective instability can be {unstable}. This so-called {double-diffusive instability}, first theorized by \cite{Stern1960}, can arise if the diffusivity of one of the quantities (in general heat) is significantly greater than the other (heavy elements). In the case of oscillatory double-diffusive convection \citep[also referred to as {semi-convection}, see][for a review]{Garaud2013}, the entropy gradient is destabilizing while the compositional gradient is stabilizing (the gradient of heavy elements is directed towards the planet center). Note that without this stable chemical gradient, the envelope would be convectively unstable {in the usual sense}.

For particular parameter values, as the instability grows, the system can quickly develop a layered structure of well mixed, convective layers separated by thin stably stratified interfaces where both temperature and density undergo a sudden jump \citep{Radko2003}. This so-called {layered semi-convection} is thus associated with a {density staircase}-like profile. This is confirmed by local three-dimensional non linear numerical simulations \citep[e.g.][]{RosemblumEtal2011, StellmachEtal2011, MirouhEtal2012, WoodEtal2013} and by direct observation on Earth, e.g. in the Canada basin in the Arctic Ocean \citep{GhaemsaidiEtal2016}. Note that in {local} numerical experiments, these layers are observed to merge {\citep[see e.g.][]{Garaud2013}, but} the long-term evolution of such a configuration is not currently understood.

Could this kind of layered structure exist in giant planet interiors and affect the dissipative processes at play? Several decades ago, \citet{Stevenson1985} pointed out that such layered interior profiles could be relevant for solar system giant planets. Following that suggestion, \cite{LeconteChabrier2012} proposed a giant planet interior model involving layered semi-convection, with the aim to verify whether such a model would be consistent with observational constraints. The new picture {obtained} departs from the standard picture of giant planets that assumes a three-layer structure and a {rocky/icy} core surrounded by a metallic H/He layer with a molecular H$_2$/He envelope on top. Indeed, because of its ability to hamper large-scale convection, we expect the presence of layered semi-convection in giant planet interiors to deeply modify the long-term interior evolution of planets. Saturn's infrared luminosity shows an excess compared to what is expected from the inherent gravitational contraction and cooling of the body, which cannot be explained invoking standard models of giant planet interiors \citep{PollackEtal1977, FortneyEtal2011}. \cite{LeconteChabrier2013} proposed that this could be explained invoking layered semi-convection. Before that, \citet{ChabrierBaraffe2007} showed that layered semi-convection {could} also play a role in explaining the abnormally large radius of some hot Jupiters {(first noticed by \citealt{BodenheimerEtal2001} and \citealt{GuillotShowman2002}, see also \citealt{BaraffeEtal2005}), though its efficiency in practice has been questioned by \cite{KurokawaInutsuka2015}}. 
{In any case, those findings seem to} suggest that layered semi-convection could be a crucial ingredient in realistic models of giant planet internal structures.

If layered semi-convection and its associated density staircases are present in giant planet interiors, it is very important to determine ultimately how this complex structure affects the rates of tidal dissipation, which has not been studied before. In particular, it is crucial to determine whether layered semi-convection in giant planet interiors could account for the higher tidal dissipation than previously thought in Jupiter and Saturn found by \cite{LaineyEtal2009, LaineyEtal2012, LaineyEtal2017} based on astrometric measurments spanning more than a century, including {some from the Cassini spacecraft}. In a series of papers, we will thus be driven by the following question:
\textit{how would the presence of density staircases in giant planet interiors affect the propagation of internal waves and modify the rates of tidal dissipation?} We also note that semi-convection is thought to be able to produce a layered state in massive stars of mass $M_*\gtrsim 15M_{\odot}$, ($M_{\odot}$ denoting the mass of the Sun) outside their convective core, which contracts over time, potentially leaving a stabilizing $\mathrm{He}$ gradient in the hydrogen envelope of the star \citep[see][]{SchwarzscildHarm1958, SakashitaHayashi1959}. As a consequence, the results of these papers, which are focused on giant planets, may {also} be relevant for those massive stars.

Before evaluating the rates of tidal dissipation in a layered profile (which will be the focus of attention of a second paper), we need to understand how density staircases affect the propagation and transmission of gravito-inertial waves that are potentially excited by tidal forcing. This paper aims to determine how density staircases modify the linear propagation of internal waves in a rotating planet. This is done by extending and generalizing two previous studies: \cite*{BelyaevQuataertFuller2015} (BQF15 hereafter), who derived the dispersion relation for the free modes of a staircase and considered the effects of rotation at the pole and equator; and \cite{Sutherland2016} (S16 hereafter), who studied the transmission of an incident internal wave upon a density staircase embedded in a stably stratified medium under the traditional approximation. {The traditional approximation consists in neglecting the horizontal component of the rotation vector in the Coriolis acceleration, which is mostly valid in strongly stratified fluids \citep[see e.g.][]{Friedlander1987}}.
Based on a similar model of a plane-parallel density staircase, we study in detail the effects of rotation (by including the complete Coriolis force) at any latitude, to determine its effects on the free modes and transmission of incident waves. On Fig. \ref{fig:modelOverview}, we give an overview of our reference physical model (and its regions of applicability discussed above), Convective layers of size $d$, in which density is uniform, are separated by infinitesimally thin stably stratified interfaces across which the density undergoes a discontinuous jump by a value $\Delta\rho$.

\pgfdeclareradialshading{pizza}{\pgfpointorigin}{%
  color(0cm)=(red!80);
  color(1.1cm)=(red!80);
  color(1.2cm)=(red!75);
  color(2.5cm)=(fluorescentorange!60);
  color(4.1cm)=(fluorescentorange!55);
  color(4.5cm)=(blue!30);
  color(4.9cm)=(fluorescentorange!55);
  color(6.0cm)=(fluorescentorange!20)
}
\begin{figure}
\centering
\begin{tikzpicture}
  \begin{scope}
    \clip (0,0) -- +(60:6) arc (60:120:6) --cycle;
    \pgfuseshading{pizza}
  \end{scope}
  \draw[thick, black!70] (-0.1,1.6) -- (0.1,1.6) -- (0.1,1.8) -- (-0.1,1.8) -- cycle;
  \draw[thick, black!60] (-0.1,4.4) -- (0.1,4.4) -- (0.1,4.6) -- (-0.1,4.6) -- cycle;
  \draw[thick, black!80] (2,0) -- (4,0) -- (4,2.5) -- (2,2.5) -- cycle;
  \draw[axis, black!80] (2,0) -- (2,2.85) node[right] {$r$};
  \draw[axis, black!80] (2,2.5) -- (4.3,2.5) node[below] {$\rho$};
  \draw[densely dotted, black!70] (0.1,1.6) -- (2,0);
  \draw[densely dotted, black!70] (0.1,1.8) -- (2,2.5);
  \draw[densely dotted, black!60] (0.1,4.4) -- (2,0);
  \draw[densely dotted, black!60] (0.1,4.6) -- (2,2.5);
  \pgfmathsetmacro{\d}{2.5/6}
  \draw[thick, color=orange] (2.52,2.5) -- ++ (0,-\d) -- ++ (0.2,0) -- ++ (0,-\d) -- ++ (0.2,0) -- ++ (0,-\d) -- ++ (0.2,0) -- ++ (0,-\d) -- ++ (0.2,0) -- ++ (0,-\d) -- ++ (0.2,0) -- ++ (0,-\d);
  \node at (-1,0.5) {\small \textsf{Z (core)}};
  \node at (-1.65,1.45) {\small \textsf{Z + H/He}};
  \node at (-2.3,2.85) {\small \textsf{H/He}};
    \node at (-2.3,2.55) {\small \textsf{(metallic)}};
  \node at (-3.15,3.85) {\small \textsf{H + He rain}};
  \node at (-3.5,4.75) {\small \textsf{H/He}};
  	\node at (-3.5,4.45) {\small \textsf{(molecular)}};
\end{tikzpicture}
\caption[Internal structure of a giant planet]{Overview of our model. In the helium rain region or in the region just outside the core (where heavy elements, symbolized here by Z, could be released into the gaseous envelope), layered semi-convection could operates. The resulting density profile is staircase-like.}
\label{fig:modelOverview}
\end{figure}
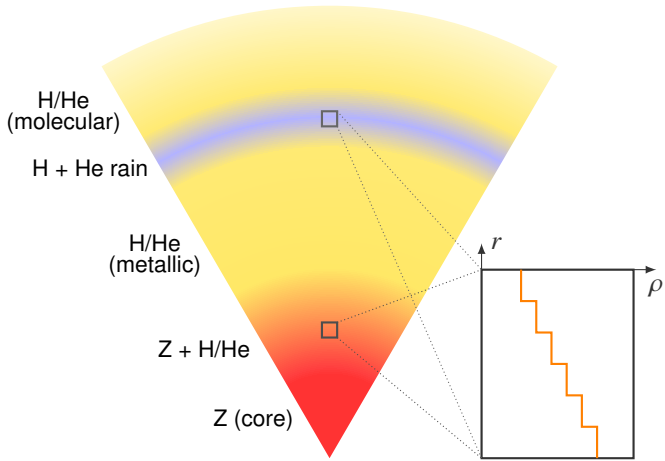

\defcitealias{BelyaevQuataertFuller2015}{BQF15}
\defcitealias{Sutherland2016}{S16}
The outline of this paper is as follows. In Section \ref{SEC:IW}, we present {the important} mathematical and physical aspects of the linear propagation of internal waves, using a formalism first introduced by \cite{GerkemaShrira2005}. Some energetical aspects are discussed in Section \ref{subsec:GIW:energy}, to yield a general expression of the transmission coefficient. Section \ref{SEC:layeredCase} then presents our study of the layered case. In particular, we generalize the dispersion relation {obtained by} \citetalias{BelyaevQuataertFuller2015} in Section \ref{sec:FREE:disprel}, and derive a series of analytical expression for the transmission coefficient of an internal wave incident upon a density staircase in Section \ref{sec:FREE:transmission}, with the aim to predict which waves will be able to penetrate into deeper regions of giant planets. A link is made between the bands of perfect transmission that arise and the free modes of the staircase given by the dispersion relation. In addition, we extend in Section \ref{sec:FREE:t_FiniteSizeInterface} the physical model to a more realistic one for which stably stratified interfaces have a finite size. Finally, we summarise our main results and discuss their astrophysical implications{, particularly for giant planets seismology,} in Section \ref{SEC:conclusion}.

\section{Internal waves in giant planet interiors}\label{SEC:IW}
\subsection{Main assumptions}\label{sec:MainAssumptions}
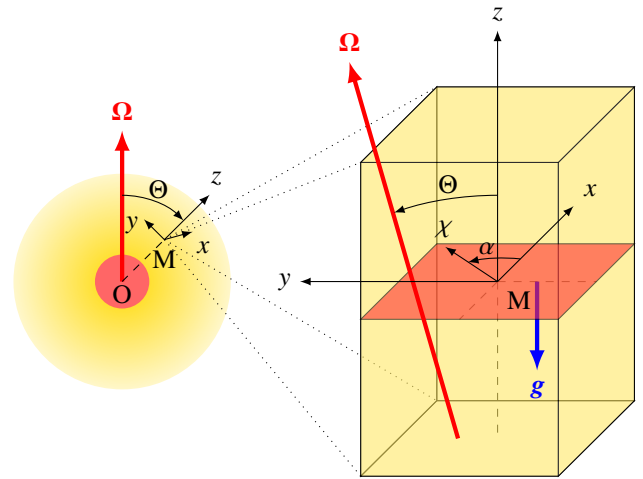
\begin{figure}
\centering
\begin{tikzpicture}[scale=0.52]
\pgfmathsetmacro{\cubex}{5}
\pgfmathsetmacro{\cubey}{8}
\pgfmathsetmacro{\cubez}{5}
\pgfmathsetmacro{\centerx}{-2.4*\cubex}
\pgfmathsetmacro{\centery}{-\cubey/2}
\pgfmathsetmacro{\centerz}{-\cubez/2}
\pgfmathsetmacro{\planetradius}{1.1*\cubex/2}
%
\shadedraw[shading=radial,outer color=fluorescentorange!20,inner color=fluorescentorange,draw=none] (\centerx,\centery,\centerz) circle (\planetradius);
\fill[color=red!60] (\centerx,\centery,\centerz) circle (\planetradius/4);
\node at (\centerx,1.07*\centery-0.05,\centerz) {O};
\draw[load,color=red] (\centerx,\centery,\centerz) -- ++ (0,1.4*\planetradius,0) node[above]{$\bm{\Omega}$};
\draw[dotted] (\centerx+0.55*\planetradius/1.4142,\centery+0.55*\planetradius/1.4142,\centerz) -- (-\cubex,0,-\cubez);
\draw[dotted] (\centerx+0.55*\planetradius/1.4142,\centery+0.55*\planetradius/1.4142,\centerz) -- (-\cubex,0,0);
\draw[dotted] (\centerx+0.55*\planetradius/1.4142,\centery+0.55*\planetradius/1.4142,\centerz) -- (-\cubex,-\cubey,-\cubez);
\draw[dotted] (\centerx+0.55*\planetradius/1.4142,\centery+0.55*\planetradius/1.4142,\centerz) -- (-\cubex,-\cubey,0);
\draw[dashed] (\centerx,\centery,\centerz) -- ++ (0.55*\planetradius/1.4142,0.55*\planetradius/1.4142,0);
\draw[axis] (\centerx+0.55*\planetradius/1.4142,\centery+0.55*\planetradius/1.4142,\centerz) -- ++ (\planetradius/2.4,\planetradius/2.4,0) node[above]{$~~z$};
\draw[axis] (\centerx+0.55*\planetradius/1.4142,\centery+0.55*\planetradius/1.4142,\centerz) -- ++ (-\planetradius/5.3,\planetradius/5.3,0);
\node at (\centerx+0.25,\centery+1.45,\centerz) {$y$};
\draw[axis] (\centerx+0.55*\planetradius/1.4142,\centery+0.55*\planetradius/1.4142,\centerz) -- ++ (\planetradius/5.3,0,-\planetradius/5.3) node[below]{$~~~x$};
\node at (\centerx+0.55*\planetradius/1.4142,\centery+0.55*\planetradius/1.4142-0.5,\centerz) {M};
\draw[axis,color=black] (\centerx,\centery+0.8*\planetradius,\centerz) arc (90:45:0.8*\planetradius);
\node at (\centerx+\planetradius/3,\centery+0.855*\planetradius,\centerz) {$\Theta$};
%
\draw[dashed] (-\cubex/2,-\cubey/2,-\cubez/2) -- ++ (\cubex/2,0,0);
\draw[dashed] (-\cubex/2,-\cubey/2,-\cubez/2) -- ++ (0,0,\cubez/2);
\draw[dashed] (-\cubex/2,-\cubey/2,-\cubez/2) -- ++ (0,-\cubey/2,0);
\draw[black] (0,0,-\cubez) -- ++(-\cubex,0,0) -- ++(0,-\cubey,0) -- ++(\cubex,0,0) -- cycle;
\draw[black] (-\cubex,0,0) -- ++(0,0,-\cubez) -- ++(0,-\cubey,0) -- ++(0,0,\cubez) -- cycle;
\draw[black] (0,-\cubey,0) -- ++(-\cubex,0,0) -- ++(0,0,-\cubez) -- ++(\cubex,0,0) -- cycle;
\draw[black,fill=fluorescentorange!60,opacity=0.6] (0,0,0) -- ++(-\cubex,0,0) -- ++(0,-\cubey,0) -- ++(\cubex,0,0) -- cycle;
\draw[black,fill=fluorescentorange!60,opacity=0.6] (0,0,0) -- ++(0,0,-\cubez) -- ++(0,-\cubey,0) -- ++(0,0,\cubez) -- cycle;
\draw[black,fill=fluorescentorange!60,opacity=0.6] (0,0,0) -- ++(-\cubex,0,0) -- ++(0,0,-\cubez) -- ++(\cubex,0,0) -- cycle;
\draw[load,color=blue] (-0.9*\cubex/3,-\cubey/2,-\cubez/2) -- ++ (0,-0.85*\cubey/3,0) node[below]{$\bm{g}$};
\draw[black,fill=red!80,opacity=0.6] (0,-\cubey/2,0) -- ++(-\cubex,0,0) -- ++(0,0,-\cubez) -- ++(\cubex,0,0) -- cycle;
\draw[load,color=red] (-3.5*\cubex/5,-\cubey,-\cubez/2) -- ++ (-0.55*\cubex,1.2*\cubey,0) node[above]{$\bm{\Omega}$};
\draw[axis] (-\cubex/2,-\cubey/2,-\cubez/2) -- ++ (-1.25*0.7071*\cubex/2,0,-1.3*0.7071*\cubez/2) node[above]{$\chi$};
\draw[axis,color=black] (-\cubex/2,-\cubey/2,-4*\cubez/5) arc (84:103.18:4);
\node at (-2.12*\cubex/3,-0.98*\cubey/2,-4.5*\cubez/5) {$\alpha$};
\draw[axis] (-\cubex/2,-\cubey/2,-\cubez/2) -- ++ (-\cubex,0,0) node[left] {$y$};
\draw[axis] (-\cubex/2,-\cubey/2,-\cubez/2) -- ++ (0,0,-\cubez) node[above right] {$x$};
\draw[axis] (-\cubex/2,-\cubey/2,-\cubez/2) node[below right]{M} -- ++ (0,0.8*\cubey,0) node[above] {$z$};
\draw[axis,color=black] (-\cubex/2,-0.9*\cubey/4,-\cubez/2) arc (90:109.15:\cubey);
\node at (-3*\cubex/4,-0.95*\cubey/5,-\cubez/2) {$\Theta$};
\end{tikzpicture}
\caption[Local Cartesian box]{{Left:} global view of a giant planet: the gaseous envelope (in yellow, the shading denoting density), lies on top of {the core} (in red). {Right:} magnified picture of the local Cartesian box, centred on a point M of a giant planet envelope, corresponding to a colatitude $\Theta$. The local box is tilted with respect to the spin axis, and its vertical axis $z$, corresponding to the local radial direction, is thus anti-aligned with gravity. The $x$ and $y$ axes correspond to the local azimuthal and latitudinal directions, respectively, while the $\chi$ axis makes an angle $\alpha$ with respect to the $x-$axis.}
\label{fig:cartesianBox}
\end{figure}

We wish to study the propagation of short-wavelength internal waves. Therefore we adopt a local Cartesian model {\citep{GerkemaShrira2005,MathisNeinerTranminh2014}} that represents a small-patch of a giant planet {\citep[see the appendix of][for the cases of pure inertial waves and gravito-inertial waves, respectively]{OgilvieLin2004,AuclairDesrotourEtal2015}}, simplifying the global spherical geometry. We centre our box on a point M of the gaseous envelope (see Fig. \ref{fig:cartesianBox}).

The local system of coordinates $(x,y,z)$ corresponds to the local azimuthal, latitudinal and radial directions, respectively. The rotation vector $\bm{\Omega}$ makes an angle $\Theta$ with respect to the gravity vector $\bm{g}$, aligned with the vertical direction. Thus, in our local system of coordinates, the latitudinal and vertical components of the rotation vector are, respectively, 
\begin{align}
\tilde{f} &= 2\Omega\sin\Theta,\label{eq:ftilde}\\
f &= 2\Omega\cos\Theta,\label{eq:f}
\end{align}
so that $2\bm{\Omega} = (0, \tilde{f}, f).$ 

Both the rotation rate of the planet $\Omega$ and the local gravity $g$ are assumed to be uniform and constant. {We assume the rotation rate to be far below the breakup angular velocity ($\Omega_{\text{K}} = \sqrt{\mathcal{G}M_{\text{p}}/R_{\text{p}}^3}$, where $\mathcal{G}$ is the gravitational constant, $M_{\text{p}}$ and $R_{\text{p}}$ are the mass and radius of the planet), and accordingly we can ignore the centrifugal acceleration.} We also introduce a reduced horizontal coordinate, $\chi$, that makes an angle $\alpha$ with respect to the $x-$axis: $\chi =x\cos\alpha + y\sin\alpha$.

One must keep in mind that the local approach is valid only for a box size of negligible extent compared to the characteristic length scale of the planet,
\begin{equation}
L_x, L_y, L_z \ll R,
\label{cond:cartesianBoxSize1}
\end{equation}
otherwise curvature effects due to the spherical geometry should be taken into account. Here, $L_x$, $L_y$ and $L_z$ are the lengths of the box in the $x$, $y$ and $z$ directions, respectively, and $R$ is the radius of the planet. In addition, considering a constant gravity vector restricts us to consider dynamical phenomena with length scales $\lambda$ far below the pressure scale height $H_\text{p}$,
\begin{equation}
\lambda \ll H_{\text{p}}.
\label{cond:cartesianBoxSize2}
\end{equation}
In giant planet deep interiors, we have 
$H_{\text{p}} \sim R$ \citep{LeconteChabrier2012},
so that if condition (\ref{cond:cartesianBoxSize1}) is fulfilled, condition (\ref{cond:cartesianBoxSize2}) is as well, since $\lambda < L_x, L_y, L_z$. Finally, such an approach is suitable because it is expected that tidally excited waves have a small-scale structure \citep{OgilvieLin2004}.
\\

Our other main assumptions are the followings:
\begin{itemize}
\item We adopt the Boussinesq approximation. The fluid is assumed to be quasi-incompressible {with a reference density value $\rho_0$}, and accordingly we restrict our study to low Mach numbers, i.e. $|\bm{u}| \ll c_{\text{s}}$, where {$|\bm{u}|$ is the velocity and} $c_{\text{s}}$ is the sound speed. In addition, the vertical extent occupied by the fluid is far below the pressure scale height: $L_z \ll H_{\text{p}}$, a condition that is fulfilled because we use a local approach.
\item Dissipative processes (viscosity and thermal diffusion) are not taken into account, but they will be in our second paper, which will be focused on tidal dissipation.
\item The spatial dependence of the background quantities is assumed {to be fixed, as resulting from double-diffusive instabilities as described by e.g. \cite{LeconteChabrier2012}}. Thus, the back-reaction of internal waves on the layered structure is not taken into account, nor is the possible excitation of internal waves by double-diffusive convection \citep[see e.g.][]{MollGaraudStellmach2016}.
\item Non-linear effects are entirely neglected.
\end{itemize}

\subsection{Equations of motion}
\label{sec:linearSetOfEquations}
{Before studying the propagation (and transmission) of internal waves in density staircases associated with layered semi-convection, we need to introduce the formalism that allows us to treat gravito-inertial waves (GIWs) with the complete Coriolis acceleration, as well as {compute the} corresponding energetic quantities. In Section \ref{SEC:layeredCase}, which is the heart of this paper, we will {use these} results.}

We study the linear propagation of GIWs in the local Cartesian model. First, let us introduce the velocity field, 
\begin{equation}
\bm{u}(\bm{r},t) = \left(
\begin{array}{c}
u(\bm{r},t)\\
v(\bm{r},t)\\
w(\bm{r},t)
\end{array}
\right),
\end{equation}
where $u$, $v$ and $w$ are the components of the velocity perturbation in the local azimuthal, latitudinal and radial directions, respectively. Next, we define the fluid buoyancy,
\begin{equation}
b(\bm{r},t) = -g\, \frac{\rho(\bm{r},t)}{\rho_0},
\label{eq:buoyancy}
\end{equation}
where $\rho(\bm{r},t)$ is the density fluctuation field. Then, we define the buoyancy frequency {in the Boussinesq approximation},
\begin{equation}
N^2(z) = -\frac{g}{\rho_0}\frac{\text{d}\bar{\rho}}{\text{d}z},
\label{eq:BruntVaisaleFreqLinear}
\end{equation}
where $\bar{\rho}(z)$ is the resulting background density profile. {We stress that $\rho_0$ is the reference Boussinesq density value, while $\bar{\rho}$ is the background density distribution associated with the layered density profile. We assume $\displaystyle \max_{|z|<L_z/2} \bar{\rho}(z) - \min_{|z|<L_z/2} \bar{\rho}(z) \ll \rho_0$.}
In Section \ref{SEC:layeredCase}, we will model layered semi-convection by a succession of convective layers, in which we assume $N^2=0$, separated by infinitesimally thin stably stratified interfaces, in which $N^2>0$.

The three linearised components of the momentum equation are given by
\begin{align}
\frac{\uppartial u}{\uppartial t} - fv + \tilde{f}w &= -\frac{1}{\rho_0} \frac{\uppartial p}{\uppartial x},\label{eq:momx}\\
\frac{\uppartial v}{\uppartial t} + fu &= -\frac{1}{\rho_0} \frac{\uppartial p}{\uppartial y}, \label{eq:momy}\\
\frac{\uppartial w}{\uppartial t} - \tilde{f}u &= -\frac{1}{\rho_0} \frac{\uppartial p}{\uppartial z} + b, \label{eq:momz}
\end{align}
where $p(\bm{r},t)$ is the pressure fluctuation, and $\tilde{f}$ and $f$ are expressed by Eqs. (\ref{eq:ftilde}) and (\ref{eq:f}), respectively. Then, we write the continuity equation,
\begin{equation}
\frac{\uppartial u}{\uppartial x} + \frac{\uppartial v}{\uppartial y} + \frac{\uppartial w}{\uppartial z} = 0.
\label{eq:cont}
\end{equation}

Finally, we write the {thermal energy equation} {in the adiabatic limit},
\begin{equation}
\frac{\uppartial b}{\uppartial t} + N^2w = 0.
\label{eq:energy}
\end{equation}

\subsection{Propagation of gravito-inertial waves}\label{subsec:GIW:propagation}
\subsubsection{Dispersion relation in a uniformly stratified medium}
We now introduce some key aspects to analyse the linear propagation of GIWs. We consider monochromatic plane wave solutions of the form
\begin{equation}
x(\bm{r},t) = \Re\left\{ X \exp\left[\text{i}(\bm{k}\cdot\bm{r}-\omega t)\right] \right\},
\label{eq:transfo_planeWave}
\end{equation}
where $\omega$ is the wave frequency, and $\bm{k}$ the wave vector. Here, $x$ stands for either $\rho$, $p$, $b$, $u$, $v$ or $w$.

Substituting the above solution into Eqs. (\ref{eq:momx})--(\ref{eq:energy}), we obtain the dispersion relation for GIWs in a uniformly stratified ($N(z)=N_0$) and rotating medium:
\begin{equation}
\omega^2 = N_0^2\frac{k_{\perp}^2}{k^2} + \frac{(2\bm{\Omega}\cdot \bm{k})^2}{k^2}.
\label{eq:GIWdispersionRelation}
\end{equation}
One can then derive the frequency domain for which GIWs propagate, corresponding to a real frequency $\omega$. This will be done in Section \ref{subsubsec:GIW_propagation}.

In a convective region, in which we assume isentropy ($N_0=0$), we recover the dispersion relation for pure inertial waves (IWs) given by
\begin{equation}
\omega^2 = \dfrac{(2\bm{\Omega}\cdot\bm{k})^2}{k^2}.
\label{eq:disprelIW}
\end{equation}
Similarly, when rotation is absent, we recover the dispersion relation for pure internal gravity waves given by
\begin{equation}
\omega^2 = N_0^2\dfrac{k^2_{\perp}}{k^2}.
\label{eq:disprelGW}
\end{equation}

\subsubsection{Group velocities}
From the dispersion relation given by Eq. (\ref{eq:GIWdispersionRelation}), we can obtain that the group velocity of GIWs is
\begin{equation}
\bm{v}_{\text{g}}^{\text{(GIW)}} = \frac{1}{\omega}\left[
N_0^2\left( \frac{k_z}{k} \right) \frac{\bm{k} \times (-\hat{\bm{e}}_z \times \bm{k})}{k^3}
+ \left( \frac{2\bm{\Omega}\cdot\bm{k}}{k} \right) \frac{\bm{k} \times (2\bm{\Omega} \times \bm{k})}{k^3}
\right].
\label{eq:groupvelocityGIW}
\end{equation}
Thus, the energy carried by GIWs can propagate along two directions, corresponding to the two signs of $\omega$.

\paragraph{The case of pure inertial waves.}\hfill\break
Let us focus more precisely on the case of pure inertial waves, which propagate in convective regions. Setting $N_0=0$ in Eq. (\ref{eq:groupvelocityGIW}) and using Eq. (\ref{eq:disprelIW}), we get that the group velocity of pure inertial waves is given by 
\begin{equation}
\bm{v}_{\text{g}} = \text{sgn}(\omega) \frac{\bm{k} \times (2\bm{\Omega} \times \bm{k})}{k^3}.
\label{eq:IWgroupVelocity} 
\end{equation}

It is worth considering along which direction the energy of a pure inertial wave propagates, because in a region of layered semi-convection, we expect the volume to be mostly convective (the interfaces are very thin). This can be done with a little algebra, which yields that the group velocity of pure inertial waves makes an angle
\begin{equation}
\lambda = \pm\sin^{-1}\left(\dfrac{\omega}{2\Omega}\right)
\end{equation}
with respect to the rotation vector.

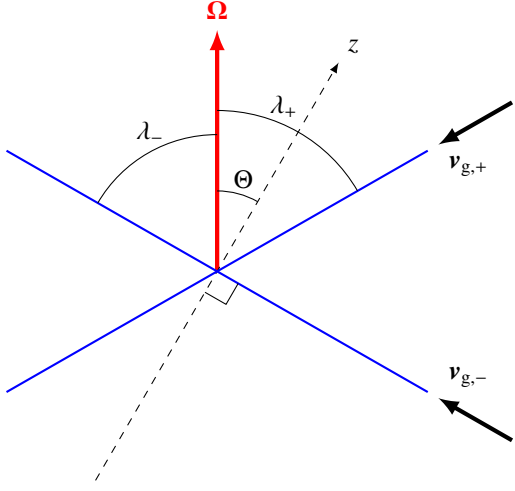
\begin{figure}
\centering
\begin{tikzpicture}[scale=0.8]
\pgfmathsetmacro{\size}{4}
\pgfmathsetmacro{\cpsix}{sqrt(3)/2}
\pgfmathsetmacro{\spsix}{1/2}
\pgfmathsetmacro{\cptroi}{1/2}
\pgfmathsetmacro{\sptroi}{sqrt(3)/2}
\pgfmathsetmacro{\perp}{\size/10}
\draw[load,red] (0,0) -- ++ (0,\size) node[above]{$\bm{\Omega}$};
\draw[axis,dashed] (-\size*\cptroi,-\size*\sptroi) -- (\size*\cptroi,\size*\sptroi) node[above right]{$z$};
\draw[thick,blue] (-\size*\cpsix,-\size*\spsix) -- (\size*\cpsix,\size*\spsix);
\draw[thick,blue] (-\size*\cpsix,\size*\spsix) -- (\size*\cpsix,-\size*\spsix);
\draw (-\perp*\cptroi,-\perp*\sptroi) -- ++ (\perp*\cpsix,-\perp*\spsix) -- ++ (\perp*\cptroi,\perp*\sptroi);
\draw (0,\size/3) arc (90:60:\size/3);
\node at (0.65*\size/6,1.15*\size/3) {$\Theta$};
\draw (0,2*\size/3) arc (90:30:2*\size/3);
\draw (0,0.85*2*\size/3) arc (90:150:0.85*2*\size/3);
\node at (1.1*\size/4,2.05*\size/3) {$\lambda_+$};
\node at (-1.1*\size/4,1.73*\size/3) {$\lambda_-$};
\draw[load] (1.4*\size*\cpsix,1.4*\size*\spsix) -- (1.05*\size*\cpsix,1.05*\size*\spsix) node[below right]{$\bm{v}_{\text{g},+}$};
\draw[load] (1.4*\size*\cpsix,-1.4*\size*\spsix) -- (1.05*\size*\cpsix,-1.05*\size*\spsix) node[above  right]{$\bm{v}_{\text{g},-}$};
\end{tikzpicture}
\caption[Direction of propagation of energy carried by pure inertial waves at the inertial frequency]{Propagation of energy carried by pure inertial waves at the frequency $\omega=f$. The two allowed group velocity vectors, $\bm{v}_{\text{g},\pm}$, form an angle $\lambda_{\pm} = \pm (\pi/2-\Theta)$ with respect to the rotation axis, represented in red. Thus, the energy propagates along two characteristics, represented in blue. At the frequency $\omega=f$, one of the characteristics is perpendicular to the vertical axis, so that the energy propagates in the local horizontal direction.}
\label{fig:IWgroupVelocityDirection}
\end{figure}

At the frequency $\omega = f = 2\Omega\cos\Theta$, we have 
\begin{equation}
\lambda_{\pm} = \pm \left(\dfrac{\uppi}{2}-\Theta\right).
\end{equation}
The solutions are displayed on Fig. \ref{fig:IWgroupVelocityDirection}. At the frequency $\omega=f$, one of the energy propagation directions is perpendicular to the vertical axis, so that the energy propagates along the local horizontal and thus does not propagate toward deeper regions of the giant planet.

Recalling the simplified physical model to be adopted (showed on Fig. \ref{fig:modelOverview}), this means that at this particular frequency, one of the directions of energy propagation is parallel to the stably stratified interfaces (which lie in the local horizontal plane).

\subsubsection{Poincar\'e equation}
Now, we introduce a key partial differential equation (PDE) to study the dynamics of GIWs. By reducing the system (\ref{eq:momx})--(\ref{eq:energy}), we can derive a PDE solely for the vertical velocity $w$ {(see Appendix \ref{app:PoincareEq})},
\begin{equation}
\frac{\uppartial^2}{\uppartial t^2} \nabla^2 w + (\bm{f} \cdot \bm{\nabla})^2 w + N^2 \nabla_{\perp}^2 w
= 0,
\label{eq:Poincare}
\end{equation}
where $\nabla_{\perp}^2 = \dfrac{\uppartial^2}{\uppartial x^2} + \dfrac{\uppartial^2}{\uppartial y^2}$ is the horizontal Laplacian, and we recall that $\bm{f} = (0, \tilde{f}, f)$. 

We want to study the propagation of a given monochromatic GIW with a frequency $\omega$, that propagates in the direction $(\cos \alpha, \sin \alpha)$ in the $(\text{M}xy)$ plane (see Fig. \ref{fig:cartesianBox}). Substituting $w = W(\chi,z)\exp[\text{i}\omega t]$ where $\chi = x \cos\alpha + y \sin\alpha$ is the reduced horizontal coordinate, we obtain the Poincar\'e equation for GIWs:

\begin{equation}
\left[ N^2(z) - \omega^2 + \tilde{f}_{\text{s}}^2 \right] \dfrac{\uppartial^2 W}{\uppartial\chi^2}
+ 2f\tilde{f}_{\text{s}}\dfrac{\uppartial^2 W}{\uppartial\chi\uppartial z} + \left[ f^2 - \omega^2 \right]\dfrac{\uppartial^2 W}{\uppartial z^2}  = 0,
\end{equation}
where
\begin{equation}
\tilde{f}_{\text{s}} = \tilde{f}\sin\alpha.
\label{eq:fs}
\end{equation}

\paragraph{The case $\omega=f$.}\hfill\break\
Let us focus first on the case $\omega=f$. Substituting 
\begin{equation}
W(\chi,z) = \widetilde{W}(z) \exp\left[\text{i}k_{\perp}\chi \right],
\end{equation}
where $k_{\perp}$ the wave number in the $\chi$ direction, we obtain
\begin{equation}
2f \tilde{f}_{\text{s}} \dfrac{\text{d}\widetilde{W}}{\text{d}z} + \text{i}k_{\perp}\left[N^2 - f^2 + \tilde{f}^2_{\text{s}}\right]\widetilde{W} = 0.
\label{eq:Schrodinger_omEqf} 
\end{equation}
This equation being of first order in $z$, it has only one solution, which we can easily calculate in the case of a uniformly stratified medium ($N=N_0$):
\begin{equation}
\widetilde{W}(z) \propto \exp\left[ \text{i}\left( \dfrac{N_0^2 - f^2 + \tilde{f}_{\text{s}}^2}{2f\tilde{f}_{\text{s}}} \right) k_{\perp}z \right].
\end{equation}

\paragraph{The case $\omega \neq f$.}\hfill\break
For $\omega \neq f$, following \cite{GerkemaShrira2005} we introduce the transformation
\begin{equation}
\label{eq:transformation}
w = \hat{W}(z) \exp\left[\text{i}(k_{\perp}(\chi + \tilde{\delta}z) - \omega t)\right],
\end{equation}
where
\begin{equation}
\label{eq:delta}
\tilde{\delta} = \frac{f\tilde{f}_{\text{s}}}{\omega^2 - f^2}.
\end{equation}
Substituting Eq. (\ref{eq:transformation}) into Eq. (\ref{eq:Poincare}) leads to
\begin{equation}
\label{eq:Schrodinger}
\frac{\text{d}^2 \hat{W}}{\text{d}z^2} + k_z^2 \hat{W} = 0,
\end{equation}
where
\begin{equation}
k_z^2 = k_{\perp}^2 \left[ \frac{N^2-\omega^2}{\omega^2 - f^2} + \left( \frac{\omega \tilde{f}_{\text{s}}}{\omega^2 - f^2}\right)^2\right].
\label{eq:k_z}
\end{equation}
In a uniformly stratified or in a convective medium (which both have $N(z) = \mathrm{const}$), solutions of Eq. (\ref{eq:Schrodinger}) have the form $\hat{W}(z) \propto \exp(\pm \text{i} k_z z)$. {We have $k_z^2>0$ in the propagative regime and $k_z^2<0$ in the evanescent regime.}

We stress that the transformation given by Eq. (\ref{eq:transformation}) has the effect of splitting the vertical wave number into two parts: $\pm k_z$, contained in the $z$-dependence of the $\hat{W}$ function, and $k_{\perp}\tilde{\delta}$, contained in the exponential factor of Eq. (\ref{eq:transformation}). Thus, we define the total vertical wave number as
\begin{equation}
\tilde{k}_z = \pm k_z + k_{\perp}\tilde{\delta},
\label{eq:total_kz}
\end{equation}
so that $w \propto \exp\left[\text{i}(k_{\perp}\chi+\tilde{k}_zz-\omega t) \right]$. The non-traditional component $k_{\perp}\tilde{\delta}$ corresponds to the intrinsic 2D behaviour of GIWs when taking the complete Coriolis acceleration into account. {It vanishes at the pole or for a null rotation, for which the problem is separable.}

\subsubsection{Frequency spectrum}
\label{subsubsec:GIW_propagation}
We now describe GIWs propagation as a function of their frequency $\omega$, along the lines of \cite{MathisNeinerTranminh2014}. At a given $z$, GIWs are propagative if $k_z^2 > 0$, where $k_z^2$ is defined by Eq. (\ref{eq:k_z}), which occurs when
\begin{equation}
\omega_- < \omega < \omega_+,
\end{equation}
with
\begin{equation}
\omega_{\pm} = \frac{1}{\sqrt{2}} \sqrt{\left[N^2+4\widetilde{\Omega}^2\right] \pm \sqrt{\left[N^2+4\widetilde{\Omega}^2\right]^2 - (2fN)^2}},
\label{eq:freq_spectrum}
\end{equation}
where we have defined a modified rotation rate of the planet
\begin{equation}
\widetilde{\Omega} \equiv \frac{1}{2} \sqrt{f^2 + \tilde{f}_{\text{s}}^2} = \Omega \sqrt{1 - \sin^2\Theta\cos^2\alpha}.
\label{eq:Omega_s}
\end{equation}
In convective regions, the local vertical wave number given by Eq. (\ref{eq:k_z}) is obtained by setting $N = 0$ and becomes
\begin{equation}
k_{z,\mathrm{c}}^2 \equiv k_{\perp}^2 \frac{\omega^2(4\widetilde{\Omega}^2 - \omega^2)}{\left(\omega^2 - f^2\right)^2},
\label{eq:kCZ}
\end{equation}

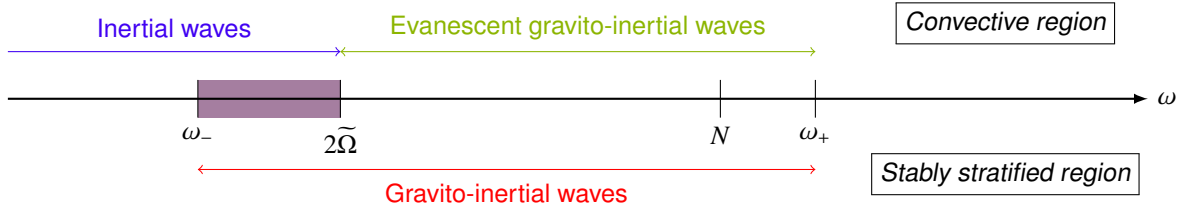
\begin{figure*}
\centering
\begin{tikzpicture}[scale=1.25]
\fill[fill=palatinatepurple!60] (2,-0.2) -- (2, 0.2) -- (3.5,0.2) -- (3.5,-0.2) -- cycle;
\draw[axis,thick] (0,0)--(12,0) node[right] {\normalsize $\omega$};
\node at (10.5,0.8) {\normalsize \it \fbox{\textsf{Convective region}}};
\node at (10.5,-0.8) {\normalsize \it \fbox{\textsf{Stably stratified region}}};
\draw (2,0.2) -- (2,-0.2) node[below] {\normalsize $\omega_{-}$};
\draw (3.5,0.2) -- (3.5,-0.2) node[below] {\normalsize $2\widetilde{\Omega}$};
\draw (7.5,0.2) -- (7.5,-0.2) node[below] {\normalsize $N$};
\draw (8.5,0.2) -- (8.5,-0.2) node[below] {\normalsize $\omega_{+}$};
\draw[color=electricultramarine, ->] (0,0.5) -- (3.5,0.5);
\node[color=electricultramarine] at (1.75,0.75) {\normalsize \textsf{Inertial waves}};
\draw[color=applegreen, <->] (3.5,0.5) -- (8.5,0.5);
\node[color=applegreen] at (6,0.75) {\normalsize \textsf{Evanescent gravito-inertial waves}};
\draw[color=red, <->] (2,-0.75) -- (8.5,-0.75);
\node[color=red] at (5.25,-1) {\normalsize \textsf{Gravito-inertial waves}};
\end{tikzpicture}
\caption[Propagation of internal waves depending on their frequency]{Low-frequency spectrum for internal waves in a rotating giant planet. Waves in the convective and stably stratified regions are indicated at the top and bottom, respectively. The purple box corresponds to sub-inertial gravito-inertial waves that are propagative in both convective and stably stratified regions. Adapted from \cite{MathisNeinerTranminh2014}.
}
\label{fig:propagationSpectrum}
\end{figure*}

On Fig. \ref{fig:propagationSpectrum}, we illustrate the spectrum of internal waves, depending on their frequency $\omega$ and on the type of layer they propagate in: convective or stably stratified. In a stably stratified layer, in which both rotation and stratification are present, gravito-inertial waves propagate for $\omega_- < \omega < \omega_+$, as explained above. In a convective layer, we have pure inertial waves propagative for $\omega<2\widetilde{\Omega}$, and evanescent for $\omega>2\widetilde{\Omega}$.

\subsubsection{Properties of the reflected wave upon an interface}
\cite{GoodmanLackner2009} have considered the reflection of a monochromatic plane pure inertial wave upon a rigid wall, with normal $\hat{\bm{e}}_z$. In this section, we extend their work to the more general case of a gravito-inertial wave, which obeys the dispersion relation given by Eq. (\ref{eq:GIWdispersionRelation}). The wave vector {is expanded as} $\bm{k} = k_{\perp}\hat{\bm{e}}_{\chi} + \tilde{k}_z\hat{\bm{e}}_{z}$. 

The goal of the following calculation is to understand how the vertical wave vector of the reflected wave is expected to change as a result of the non-specular reflection from an interface. In our case, the interfaces are in the horizontal plane, with $\hat{\bm{e}}_{z}$ being the vector normal to the interfaces. We thus consider incident and reflected waves of the form
\begin{align}
\hat{W}_{\text{in}} &\propto \exp \left[\text{i}\left(\bm{k}\cdot\bm{r}-\omega t \right)\right],\\
\hat{W}_{\text{re}} &\propto \exp \left[ \text{i}\left(\bm{k}'\cdot\bm{r}-\omega' t \right)\right],
\end{align}
where the scattered (outgoing) wave vector $\bm{k}' = k'_{\perp}\hat{\bm{e}}_{\chi} + \tilde{k}_z'\hat{\bm{e}}_{z}$ is determined by two conditions. First, in order that the incident and reflected waves have the same relative phase at all points along the interface, as required by the impermeability condition, it is necessary that $\bm{k}$ and $\bm{k}'$ have the same components parallel to the interface. That means that $\hat{\bm{e}}_{z}\times\left(\bm{k}'-\bm{k}\right) = \bm{0}$, or equivalently
$k'_{\perp} = k_{\perp}$.
Thus, 
\begin{equation}
\bm{k}' = k_{\perp}\hat{\bm{e}}_{\chi} + \tilde{k}'_z\hat{\bm{e}}_{z},
\label{eq:kz_reflected}
\end{equation}
where $\tilde{k}'_z$ (the total vertical wave number) remains to be determined. Similarly, in order that the relative phase stays constant in time, the two waves must have the same frequency, i.e
$\omega'=\omega$,
so that (see Eq. (\ref{eq:GIWdispersionRelation}))
\begin{equation}
\omega^2 = \frac{k_{\perp}^2}{k'^2} N^2 + \frac{(2\bm{\Omega}\cdot \bm{k}')^2}{k'^2}.
\label{eq:disprel_reflected}
\end{equation}
Substituting (\ref{eq:kz_reflected}) into (\ref{eq:disprel_reflected}) leads to
\begin{equation}
\left(\omega^2-f^2\right)\tilde{k}_z'^2 -2f\tilde{f}_{\text{s}}k_{\perp}\tilde{k}'_z + \left[\omega^2-\left(\tilde{f}_{\text{s}}^2+N^2\right)\right]k_{\perp}^2 = 0,
\label{eq:kz_reflected_polynome}
\end{equation}
where we have used that $(2\bm{\Omega}\cdot\bm{k}) = \tilde{f}_{\text{s}}k_{\perp} + f\tilde{k}_z$.
From Eq. (\ref{eq:kz_reflected_polynome}), it is possible to get the two roots, which are
\begin{align}
\tilde{k}'_z &= {\left(\frac{f\tilde{f}_{\text{s}}}{\omega^2-f^2}\right)} k_{\perp} \pm \frac{\sqrt{(f\tilde{f}_{\text{s}})^2 - (\omega^2-f^2)[\omega^2-(\tilde{f}_{\text{s}}^2+N^2)]}}{\omega^2-f^2}k_{\perp} \nonumber\\
&= \tilde{\delta}k_{\perp} \pm k_z, \label{eq:kz_reflected_twoRoots}
\end{align}
where $\tilde{\delta}$ and $k_z$ are given by Eqs. (\ref{eq:delta}) and (\ref{eq:k_z}), respectively.
We recognize that the expression of $\tilde{k}'_z$ given above matches the definition of the total vertical wave number defined by Eq. (\ref{eq:total_kz}), the plus sign corresponding to the incident (ingoing) wave, and the minus sign corresponding to the reflected (outgoing) wave. Therefore, this calculation provides an independent check that the formalism introduced with Eq. (\ref{eq:transformation}), which splits the vertical wave number into two parts, is correct.

\subsubsection{Polarization relations}
The Poincar\'e equation given by Eq. (\ref{eq:Poincare}) is a PDE solely for the vertical velocity $w$. In order to study the dynamics of the other fields: the horizontal components of the velocity, $u$ and $v$, the pressure $p$, and the buoyancy $b$, we can derive analytic formulae to get $u$, $v$, $p$ and $b$ in term of $w$ and its derivatives.
To do that, we first express each field $x(\chi,z,t)$ describing the perturbed flow using the transformation introduced in Eq. (\ref{eq:transformation}) for the vertical velocity,
\begin{equation}
x = \Re\left\{X(\chi,z)\text{e}^{-\text{i}\omega t}\right\},
\label{eq:transformation_time}
\end{equation}
where
\begin{equation}
X = \hat{X}(z) \exp\left[\text{i}k_{\perp}(\chi + \tilde{\delta}z)\right].
\label{eq:transformation_allfields}
\end{equation}
We stress again that $\hat{X}$ carries only part of the vertical dependence, the other part being included in $\exp(\text{i}k_{\perp}\tilde{\delta}z)$. The system of equations in Section \ref{sec:linearSetOfEquations} thus becomes
\begin{align}
-\text{i}\omega U - fV + \tilde{f}W &= -\frac{\text{i}k_{\perp}\cos\alpha}{\rho_0}P,\\
-\text{i}\omega V + fU &= -\frac{\text{i}k_{\perp}\sin\alpha}{\rho_0} P,\\
-\text{i}\omega W - \tilde{f}U &= -\frac{1}{\rho_0}\dfrac{\uppartial P}{\uppartial z} + B,
\end{align}
for the three components of the equation of momentum, and
\begin{gather}
\text{i}k_{\perp}\cos\alpha U + \text{i}k_{\perp}\sin\alpha V + \dfrac{\uppartial W}{\uppartial z} = 0,\\[2mm]
-\text{i}\omega B + N^2 W = 0,
\end{gather}
for the equations of conservation of mass and energy, respectively.
Noting that 
\begin{equation*}
\dfrac{\uppartial W}{\uppartial z} = \left(\hat{W}' + \text{i}k_{\perp}\tilde{\delta}\hat{W}\right)\exp\left[\text{i}k_{\perp}(\chi + \tilde{\delta}z)\right],
\end{equation*}
all the fields can be expressed in term of $\hat{W}$ as follows:
\begin{align}
\hat{U} &= \frac{\tilde{f}_{\text{s}}(f\cos\alpha - \text{i}\omega\sin\alpha)}{f^2-\omega^2}\hat{W} + \frac{f\sin\alpha + \text{i}\omega\cos\alpha}{\omega k_{\perp}}\hat{W}', \label{eq:polU}\\
\hat{V} &= \frac{\tilde{f}_{\text{s}}(f\sin\alpha + \text{i}\omega\cos\alpha)}{f^2-\omega^2}\hat{W} - \frac{f\cos\alpha - \text{i}\omega\sin\alpha}{\omega k_{\perp}}\hat{W}',\label{eq:polV}\\
\hat{P} &= \text{i}\frac{\rho_0\tilde{f}_{\text{c}}}{k_{\perp}}\hat{W} + \text{i}\frac{\rho_0(f^2-\omega^2)}{\omega k_{\perp}^2}\hat{W}',\label{eq:polP}\\
\hat{B} &= \text{i}\frac{N^2}{\omega}\hat{W}, \label{eq:polB}
\end{align}
where the prime denotes differentiation with respect to $z$, and $\tilde{f}_{\text{c}}=\tilde{f}\cos\alpha$. Then, each field has to be multiplied by $\exp \left[\text{i}\left(k_{\perp}(\chi + \tilde{\delta}z) - \omega t\right)\right]$ to get the complete solution. The physical solution is then the real part of the complete complex solution.

\subsection{Energetical aspects}\label{subsec:GIW:energy}
One of our motivations is to predict energy transport by internal waves in regions of layered semi-convection. In this section, we thus focus on the energetics of GIWs propagation.

\subsubsection{Expression of the kinetic and potential energies in term of $\hat{W}$}
Kinetic and potential energy {densities}, averaged over a wave period $2\uppi/\omega$, can be written as
\def\mean#1{\left< #1 \right>}
\begin{align}
E_{\text{k}} &= \frac{1}{2}\rho_0 \mean{\Re(u)^2 + \Re(v)^2 + \Re(w)^2} \nonumber\\[1mm]
&= \frac{1}{4}\rho_0\left(UU^* + VV^* + WW^* \right), \label{eq:kin_averaged_temp}
\end{align}
and
\begin{align}
E_{\text{p}} &= \frac{1}{2}\rho_0 \frac{\displaystyle \mean{\Re(b)^2}}{N^2} \nonumber \\[1mm]
&= \frac{1}{4}\rho_0\frac{BB^*}{N^2} \label{eq:pot_averaged_temp},
\end{align}
where the brackets $\mean{\cdot}$ denotes time averaging, and the asterisks, the complex conjugate. To get equations (\ref{eq:kin_averaged_temp}) and (\ref{eq:pot_averaged_temp}), we have used the identity $\Re(A)\Re(B) = \frac{1}{2}\Re\left(AB+AB^*\right)$, so that provided the transformation given by Eq. (\ref{eq:transformation_time}), $\mean{\Re(x)^2} = \frac{1}{2}\left<{\Re(X^2\text{e}^{-2\text{i}\omega t} + XX^*)}\right> = \frac{1}{2}XX^*$. Then, using the transformation given by Eq. (\ref{eq:transformation_allfields}), we have $XX^*=\hat{X}\hat{X}^*$, so that the kinetic and potential energies can be written as
\begin{align}
E_{\text{k}} &= \frac{1}{4}\rho_0\left(\hat{U}\hat{U}^* + \hat{V}\hat{V}^* + \hat{W}\hat{W}^*\right) \label{eq:kinetic_energy_functionPrimitiveVar}
\end{align}
and
\begin{align}
E_{\text{p}} &= \frac{1}{4}\rho_0 \frac{\hat{B}\hat{B}^*}{N^2}. \label{eq:potential_energy_functionPrimitiveVar}
\end{align}
Then, using the polarization relations given by Eqs. (\ref{eq:polU}), (\ref{eq:polV}) and (\ref{eq:polB}), we can express the kinetic and potential energies in term of $\hat{W}$ only as
\begin{align}
E_{\text{k}} = \frac{1}{4}\rho_0\left\{\left[\frac{\tilde{f}_{\text{s}}^2(f^2+\omega^2)}{(\omega^2-f^2)^2}+1\right]\right.&\hat{W}\hat{W}^* + \frac{1}{k_{\perp}^2}\left[\left(\frac{f}{\omega}\right)^2 + 1\right]\hat{W}'\left(\hat{W}'\right)^* \nonumber\\
& ~ ~ ~ ~ ~ + \left.\frac{4f\tilde{f}_{\text{s}}}{k_{\perp}(f^2-\omega^2)}\Im\left[\hat{W}\left(\hat{W}'\right)^*\right] \right\}
\end{align}
and
\begin{equation}
E_{\text{p}} = \frac{1}{4}\rho_0\left(\frac{N}{\omega}\right)^2 \hat{W}\hat{W}^*,
\end{equation}
where we recall that $\hat{W}$ is a function of $z$ only. Assuming a plane wave form in the vertical direction also (as suggested by Eq. (\ref{eq:Schrodinger})),
\begin{equation}
\hat{W} = \mathcal{A}\exp(\text{i}k_zz),
\label{eq:WhatPlaneWave}
\end{equation}
we find that the total energy is
\begin{align}
E_{\text{k}} + E_{\text{p}} = \frac{1}{4}\rho_0\left\{\left[\frac{\tilde{f}_{\text{s}}^2(f^2+\omega^2)}{(\omega^2-f^2)^2} \right. \right. &\left.+ \left(\frac{N}{\omega}\right)^2 + 1\right] + \left[\left(\frac{f}{\omega}\right)^2 + 1\right]\left(\frac{k_z}{k_{\perp}}\right)^2\nonumber \\
 & ~~~~~~~~~~~~~~~~~~ \left. + \frac{4f\tilde{f}_{\text{s}}}{(\omega^2-f^2)}\frac{k_z}{k_{\perp}}\right\}\left|\mathcal{A}\right|^2.
\label{eq:sumPotentialKineticEnergies}
\end{align}

\subsubsection{Expression of the vertical energy flux density in term of $\hat{W}$}
In our setup, the energy flux density is $\mean{p\bm{u}}$ in real variables, so that the vertical energy flux density is given by
\begin{equation}
F_z = \left<\Re(w)\Re(p)\right> = \frac{1}{2}\Re\left(\hat{W}\hat{P}^*\right). \label{eq:verticalFluxOfEnergyDef}
\end{equation}
Using Eq. (\ref{eq:polP}) for $\hat{P}$, we get
\begin{equation}
F_z = \frac{1}{2}\rho_0\left(\frac{f^2-\omega^2}{\omega k_{\perp}^2}\right) \Im\left[\hat{W}\left(\hat{W}'\right)^*\right].
\label{eq:verticalFluxOfEnergyInTermW}
\end{equation}
Finally, using Eq. (\ref{eq:WhatPlaneWave}) for $\hat{W}$, we obtain that the vertical energy flux density is given by
\begin{equation}
F_z = \frac{1}{2}\rho_0\left(\frac{\omega^2-f^2}{\omega k_{\perp}^2}\right) k_z|\mathcal{A}|^2,
\label{eq:verticalFluxOfEnergyInTermAmplitudeOfW}
\end{equation}
where we recall that $\mathcal{A}$ is the amplitude of the vertical component of the velocity, and $k_z$ the vertical wave number.

It is interesting to note that the vertical energy flux density can be expressed in an alternative manner, noting that the energy is transported vertically {with} the vertical component of the group velocity, $\bm{v}_{\text{g}}\cdot\hat{\bm{e}}_z$. That suggests that the vertical energy flux density can be expressed as
\begin{equation}
F_z = (\bm{v}_{\text{g}}\cdot\hat{\bm{e}}_z)\,(E_{\text{k}} + E_{\text{p}}),
\label{eq:def_verticalFluxOfEnergy}
\end{equation}
where the group velocity is defined by Eq. (\ref{eq:disprelIW}) and the sum of the mean kinetic and potential energies has been derived in the case of a plane wave in Eq. (\ref{eq:sumPotentialKineticEnergies}). We stress that this expression only makes sense when $E_{\text{k}}$ and $E_{\text{p}}$ are mean quantities, averaged over a wavelength and period, as it has been done to obtain Eq. (\ref{eq:sumPotentialKineticEnergies}). 
It has been checked that both expressions are equal, providing that $k_z$ takes the expression given by Eq. (\ref{eq:k_z}).

\subsubsection{Expression of the transmission coefficient}
\label{subsubsec:GIW:transmission}
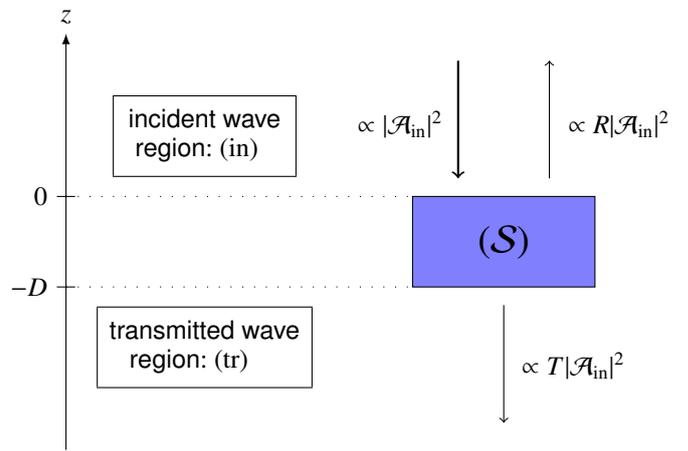
\begin{figure}
\centering
\begin{tikzpicture}[scale=0.8]
\pgfmathsetmacro{\sizex}{3}
\pgfmathsetmacro{\sizey}{\sizex/2}
\pgfmathsetmacro{\axisposx}{1.9*\sizex}
\draw[black,fill=blue!50] (0,0) -- ++ (\sizex,0) -- ++ (0,\sizey) -- ++ (-\sizex,0) -- cycle;
\node at (\sizex/2,\sizey/2) {\Large $(\mathcal{S})$};
\draw[->, thick] (\sizex/4,2.5*\sizey) -- (\sizex/4,1.2*\sizey);
\node at (-0.2,1.8*\sizey) {$\propto |\mathcal{A}_{\text{in}}|^2$};
\draw[<-] (3*\sizex/4,2.5*\sizey) -- (3*\sizex/4,1.2*\sizey);
\node at (\sizex+0.4,1.8*\sizey) {$\propto R |\mathcal{A}_{\text{in}}|^2$};
\draw[->] (\sizex/2,-0.2*\sizey) -- (\sizex/2,-1.5*\sizey);
\node at (3*\sizex/4+0.4,-\sizey+0.2) {$\propto T |\mathcal{A}_{\text{in}}|^2$};
\draw[axis] (-\axisposx,-1.8*\sizey) -- ++ (0,4.6*\sizey) node[above] {$z$};
\draw[loosely dotted] (-\axisposx,0) -- ++ (\axisposx,0);
\draw[loosely dotted] (-\axisposx,\sizey) -- ++ (\axisposx,0);
\draw (-\axisposx+0.15,\sizey) -- ++ (-0.3,0) node[left] {$0$};
\draw (-\axisposx+0.15,0) -- ++ (-0.3,0) node[left] {$-D$};
\node at (-1.2*\axisposx/2,\sizey+1) {\centering \framebox(68,30){\minibox{\textsf{incident wave}\\ ~\textsf{region}: $\text{(in)}$}}};
\node at (-1.2*\axisposx/2,-1) {\framebox(80,30){\minibox{\textsf{transmitted wave}\\ ~~~\textsf{region}: $\text{(tr)}$}}};
\end{tikzpicture}
\caption[Transmission and reflection of an internal wave]{Illustration of the system to be considered in Section \ref{SEC:layeredCase}, consisting of a density staircase $(\mathcal{S})$ of vertical extent $D$. An incident wave of amplitude $\mathcal{A}_{\text{in}}$, thus carrying an energy density $\propto |\mathcal{A}_{\text{in}}|^2$, comes from above $(\mathcal{S})$. The reflected wave has an amplitude $\mathcal{A}_{\text{re}}$ and thus carries an energy density $\propto |\mathcal{A}_{\text{re}}|^2 = R |\mathcal{A}_{\text{in}}|^2$, while the transmitted wave has an amplitude $\mathcal{A}_{\text{tr}}$ and thus carries an energy density $\propto |\mathcal{A}_{\text{tr}}|^2 = T |\mathcal{A}_{\text{in}}|^2$. $R$ and $T$ are the reflection and transmission coefficients, respectively. The arrows indicate the schematic vertical direction of propagation of energy. The horizontal components are not shown.}
\label{fig:Tcoeff}
\end{figure}

In Section \ref{SEC:layeredCase}, we will study in detail the transmission of internal waves through a portion of a density staircase produced by semi-convection: an incident  (downward propagating) wave carrying an energy density $\propto |\mathcal{A}_{\text{in}}|^2$ enters the staircase $(\mathcal{S})$, of vertical extent $D$. At the top of the staircase, a reflected (upward propagating) wave is created, carrying an energy density $\propto |\mathcal{A}_{\text{re}}|^2 = R|\mathcal{A}_{\text{in}}|^2$. At the bottom of the staircase, the transmitted (downward propagating) wave comes out, carrying an energy density $\propto |\mathcal{A}_{\text{tr}}|^2 = T|\mathcal{A}_{\text{in}}|^2$. Here, $R$ and $T$ are the reflection and transmission coefficients, respectively; this is illustrated on Fig. \ref{fig:Tcoeff}.

The transmission coefficient itself is defined to be the ratio of transmitted (labelled by $(\text{tr})$) to incident (labelled by $(\text{in})$) energy flux densities, for which we have found an analytic expression in the previous section:
\begin{equation}
T = \frac{F_z^{(\text{tr})}}{F_z^{(\text{in})}}.
\label{eq:def_Tcoeff}
\end{equation}
In our case, using Eq. (\ref{eq:verticalFluxOfEnergyInTermAmplitudeOfW}) for the vertical energy flux density gives
\begin{equation}
T = \frac{k_z^{(\text{tr})}}{k_z^{(\text{in})}} \left|\frac{\mathcal{A}_{\text{tr}}}{\mathcal{A}_{\text{in}}}\right|^2.
\label{eq:Tcoeff_expression}
\end{equation}
This is the expression we will use to calculate transmission coefficients. The regions above and below $(\mathcal{S})$, respectively defined by $z>0$ and $z<-D$ (see Fig. \ref{fig:Tcoeff}), are {a priori} different: in particular, the stratification {(stable or unstable)} in those regions is not necessarily the same, so that the vertical wave numbers $k_z$ will not be either. 
Note that here, $k_z$ is not the total vertical wavenumber, that we denote by $\tilde{k}_z$ (see Eq. (\ref{eq:total_kz})), but only the part defined through Eq. (\ref{eq:Schrodinger})--(\ref{eq:k_z}).

In the absence of energy sources such as background shear, we expect that the energy is conserved. Therefore, the reflection coefficient is given by
\begin{equation}
R = 1 - T.
\label{eq:Rcoeff}
\end{equation}

\section{Propagation of internal waves in layered semi-convection}\label{SEC:layeredCase}

Armed with the results of the previous section, we can now study the propagation of internal waves in a idealised model describing layered semi-convection. 

\subsection{Physical set up}\label{sec:FREE:model}
We continue to work in the local Cartesian box described in Section \ref{sec:MainAssumptions}.
The background state is assumed to be in hydrostatic equilibrium with constant gravity pointing in the $-\hat{\bm{e}}_z$ direction, so that $\mathrm{d}P/\text{d}z = -\rho_0 g$.

We consider a region of a giant planet envelope in which double-diffusive convection has produced a layered density profile, as described in the introduction. Thus, the density profile is close to a density staircase in which convective layers of size $d$ are separated by infinitesimally thin stably stratified interfaces. This idealised reference model is similar to that considered by \citetalias{BelyaevQuataertFuller2015} and \citetalias{Sutherland2016}. The distance between adjacent interfaces, $d$, is assumed constant. {We will later discuss the effects of relaxing some of these assumptions in Section \ref{subsec:FREE:t_NUstepSizes} and \ref{sec:FREE:t_FiniteSizeInterface}}. At each interface, the density undergoes a discontinuous jump by a value $\Delta\rho > 0$. This is illustrated on Fig. \ref{fig:modele_globalView}, on which are displayed the physical quantities introduced so far.

\begin{figure}
\centering
\begin{tikzpicture}[scale=0.8]
	\draw[thick, color=gray!50] (0,0) -- (2.7,6.48);
	\draw[thick, color=gray!50] (0,0) -- (6.48,2.7);
	\draw[thick, color=gray!50] (6.48, 2.7) arc (22.62: 67.38: 7.02);
	\draw[thick, color=gray!50] (1.92, 0.8) arc (22.62: 67.38: 2.08);
	\fill[fill=red!50]
		(0,0) -- (1.92, 0.8) arc (22.62: 67.38: 2.08) -- (0.8,1.92) -- cycle;
	\draw[load, color=red!80] (0,0) -- ++ (0,7) node[above] {\large $\bm{\Omega}$};
	\draw[load, color=blue!80] (3.2,3.84) -- (2.2,2.64) node[below right] {\hspace{-1mm}\large $\bm{g}$};
	\draw[axis, color=orange!60] (1.6,2.4) -- (3.6,4.8) node[above] {\vspace{-5mm} \normalsize $z$};
	\draw[axis, color=orange!60] (3.1,4.65) -- (4.8,3.2333) node[right] {\normalsize $\bar{\rho}$};	
	\draw[thick, color=orange!100] (3.8,4.067) -- (3.4,3.5867);
	\draw[thick, color=orange!100] (3.4,3.5867) -- (3.6,3.42);
	\draw[thick, color=orange!100] (3.6,3.42) -- (3.2, 2.94);
	\draw[thick, color=orange!100] (3.2,2.94) -- (3.4,2.7733);
	\draw[thick, color=orange!100] (3.4,2.7733) -- (3,2.2933);
	\draw[<->, color=black!80] (3.9,4.197) -- (4.1,4.03033);
	\node[black!80] at (4.28,4.32) {\footnotesize $\Delta\rho$};
	\draw[<->, color=black!80] (3.5,2.69) -- (3.9,3.17);
	\node[black!80] at (3.89,2.85) {\small $d$};	
	\draw[axis, color=black!80] (0.02, 4.2175) arc (90:50.65:4.2175);
	\node[black!80] at (1.2,4.36) {\large $\Theta$};
	\node[black!80] at (2.92,4.43) {\footnotesize M};	
\end{tikzpicture}
\caption[Internal structure of a giant planet]{Model of the internal structure of a giant planet hosting layered semi-convection. The red area represents {the core}, while the white one represents the gaseous envelope. Within the latter, double-diffusive convection acts to create a staircase-like profile for the density (in orange). Convective layers of vertical extent $d$ are separated by infinitesimally thin stably stratified interfaces. The density undergoes a density jump by a value $\Delta \rho$ at each interface. The rotation axis $\bm{\Omega}$ (in red) forms an angle $\Theta$ with respect to the gravity $g$ (in blue), directed along the (M$z$) direction.}
\label{fig:modele_globalView}
\end{figure}
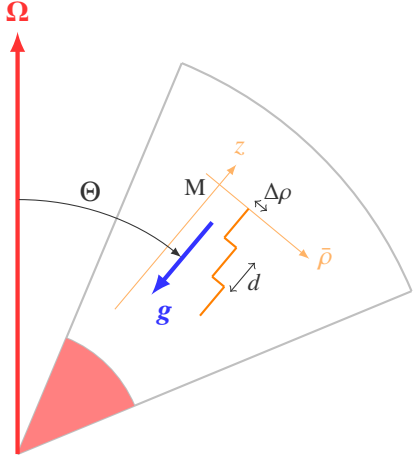

\begin{figure*}
\centering
\begin{subfigure}[b]{0.3\textwidth}
\caption{General scheme}
\label{fig:Model_generalScheme}
\centering
\begin{tikzpicture}[scale=0.8]
	\draw (-0.15,8.25) -- ++ (-0.2,0) node[left] {\footnotesize $0$};
	\draw (-0.15,6.75) -- ++ (-0.2,0) node[left] {\footnotesize $-d$};
	\draw (-0.15,5) -- ++ (-0.2,0) node[left] {\footnotesize $-(m-1)d$};
	\draw (-0.15,3.5) -- ++ (-0.2,0) node[left] {\footnotesize $-md$};
	\fill[bottom color=orange!40, top color=white]
	(0,8.25) -- (4.5,8.25) -- (4.5,10) -- (0,10);
	\fill[color=orange!40]
	(0,8.25) -- (4.5,8.25) -- (4.5,6.75) -- (0,6.75);
	\fill[top color=orange!40, bottom color=white]
	(0,6.75) -- (4.5,6.75) -- (4.5,5.95) -- (0,5.95);
	\fill[bottom color=orange!40, top color=white]
	(0,5.8) -- (4.5,5.8) -- (4.5,5) -- (0,5);
	\fill[color=orange!40]
	(0,5) -- (4.5,5) -- (4.5,3.5) -- (0,3.5);
	\fill[bottom color=white, top color=orange!40]
	(0,3.5) -- (4.5,3.5) -- (4.5,1.75) -- (0,1.75);
	\draw[axis] (-0.25,6.3) -- ++ (0,4) node[above] {$z$};
	\draw (-0.25,1.75) -- (-0.25,5.45);
	\draw[dashed] (-0.25,5.45) -- (-0.25,6.3);
	\draw[thick, color=red!80] (0,8.25) -- (4.5,8.25);
	\draw[thick, color=red!80] (0,6.75) -- (4.5,6.75);
	\draw[thick, color=red!80] (0,5) -- (4.5,5);
	\draw[thick, color=red!80] (0,3.5) -- (4.5,3.5);
	\draw[->, color=black, thick] (1,9.5) -- (2,8.5); \node at (1.1,8.9) {\small $\mathcal{A}_0$};
	\draw[->, color=black] (2.5,8.5) -- (3.5,9.5); \node at (3.4,8.9) {\small $\mathcal{B}_0$};
	\draw[->, color=black] (1,8) -- (2,7); \node at (1.1,7.4) {\small $\mathcal{A}_1$};
	\draw[->, color=black] (2.5,7) -- (3.5,8); \node at (3.4,7.4) {\small $\mathcal{B}_1$};
	\node at (1.5,6) {$\vdots$}; \node at (3,6) {$\vdots$};
	\draw[->, color=black] (1,4.75) -- (2,3.75); \node at (1.1,4.15) {\small $\mathcal{A}_m$};
	\draw[->, color=black] (2.5,3.75) -- (3.5,4.75); \node at (3.4,4.15) {\small $\mathcal{B}_m$};
	\draw[->, color=black] (2.25,3.25) -- (3.25,2.25); \node at (2.15,2.65) {\small $\mathcal{A}_{m+1}$};
\end{tikzpicture}
\end{subfigure}
\begin{subfigure}[b]{0.26\textwidth}
\caption{Density profile}
\label{fig:Model_densityProfile}
\centering
\begin{tikzpicture}[scale=0.8]
	\draw[axis,color=black!40] (5.75,6.3) -- ++ (0,4) node[above] {$z$};
	\draw[color=black!40] (5.75,1.75) -- (5.75,5.45);
	\draw[dashed,color=black!40] (5.75,5.45) -- (5.75,6.3);
	\draw[color=black!40] (5.5,8.25) -- (7,8.25);
	\draw[axis,color=black!40] (7.5,8.25) -- (10,8.25) node[right,color=black] {$\bar{\rho}$};
	\draw[color=black!40] (5.85,8.25) -- ++ (-0.2,0);
	\draw[color=black!40] (5.85,6.75) -- ++ (-0.2,0);
	\draw[color=black!40] (5.85,5) -- ++ (-0.2,0);
	\draw[color=black!40] (5.85,3.5) -- ++ (-0.2,0);
	\draw[thick, color=black] (7,9.5) -- (7,8.25);
	\draw[densely dotted, color=red!50, thick] (7,8.25) -- (7.5,8.25);
	\draw[thick, color=black] (7.5,8.25) -- (7.5,6.75);
	\draw[densely dotted, color=red!50, thick] (7.5,6.75) -- (8,6.75);
	\draw[thick, color=black] (8,6.75) -- (8,6.5);
	\draw[thick, color=black] (8.0833,5.25) -- (8.0833,5);
	\draw[densely dotted, color=red!50, thick] (8.0833,5) -- (8.5833,5);
	\draw[thick, color=black] (8.5833,5) -- (8.5833,3.5);
	\draw[densely dotted, color=red!50, thick] (8.5833,3.5) -- (9.0833,3.5);
	\draw[thick, color=black] (9.0833,3.5) -- (9.0833,2.25);
	\draw[densely dashed] (6.75,9.75) -- (9.3333,2);
	\draw[<-] (7.25,8.5) -- ++ (0.5,0.5) node[right] {\footnotesize 
		$\displaystyle \left|\frac{\text{d}\bar{\rho}}{\text{d}z}\right| = \frac{\Delta\rho}{d}$};
\end{tikzpicture}
\end{subfigure}
\begin{subfigure}[b]{0.18\textwidth}
\caption{$N^2$ profile}
\label{fig:Model_Nprofile}
\centering
\begin{tikzpicture}[scale=0.8]
	\draw[axis,color=black!40] (11.25,6.3) -- ++ (0,4) node[above] {$z$};
	\draw[color=black!40] (11.25,1.75) -- (11.25,5.45);
	\draw[dashed,color=black!40] (11.25,5.45) -- (11.25,6.3);
	\draw[axis,color=black!40] (11,8.25) -- (14.5,8.25) node[right,color=black] {$N^2$};
	\draw[color=black!40] (11.35,8.25) -- ++ (-0.2,0);
	\draw[color=black!40] (11.35,6.75) -- ++ (-0.2,0);
	\draw[color=black!40] (11.35,5) -- ++ (-0.2,0);
	\draw[color=black!40] (11.35,3.5) -- ++ (-0.2,0);
	\draw[axis, thick, color=red!80] (11.25,8.25) -- ++ (2.5,0);
	\draw[axis, thick, color=red!80] (11.25,6.75) -- ++ (2.5,0);
	\draw[axis, thick, color=red!80] (11.25,5) -- ++ (2.5,0);
	\draw[axis, thick, color=red!80] (11.25,3.5) -- ++ (2.5,0);
	\draw[dashed] (13,2.75) -- (13,8.7) node[above]
	{\footnotesize $\displaystyle \bar{N}^2=\frac{g\Delta\rho}{\rho_0 d}$};
\end{tikzpicture}
\end{subfigure}
\caption[Local physical model with discrete interfaces]{Summary of our physical model: $m$ convective steps of constant size $d$ and indexed by the integer $n=\{1, \dots, m\}$ are separated by discrete interfaces. 
(a) General scheme: the incident, reflected and transmitted waves have amplitudes $\mathcal{A}_0$, $\mathcal{B}_0$ and $\mathcal{A}_{m+1}$, respectively. In the $n$-th convective step, the ingoing wave have an amplitude $\mathcal{A}_n$, and the outgoing has an amplitude $\mathcal{B}_n$. 
(b) The corresponding density profile: at each interface, the density undergoes a discontinuous jump by a value $\Delta\rho>0$. In the convective steps, it follows an adiabatic gradient.  This creates a mean density gradient $|\text{d}\bar{\rho}/\text{d}z|=\Delta\rho/d$.
(c) The corresponding buoyancy frequency profile: $N=0$ everywhere, except at the location of the interfaces, where it is a Dirac distribution, which creates a mean stratification $\bar{N}^2=g\Delta\rho/\rho_0 d$.}
\label{fig:tcoeff_modele_CC}
\end{figure*}
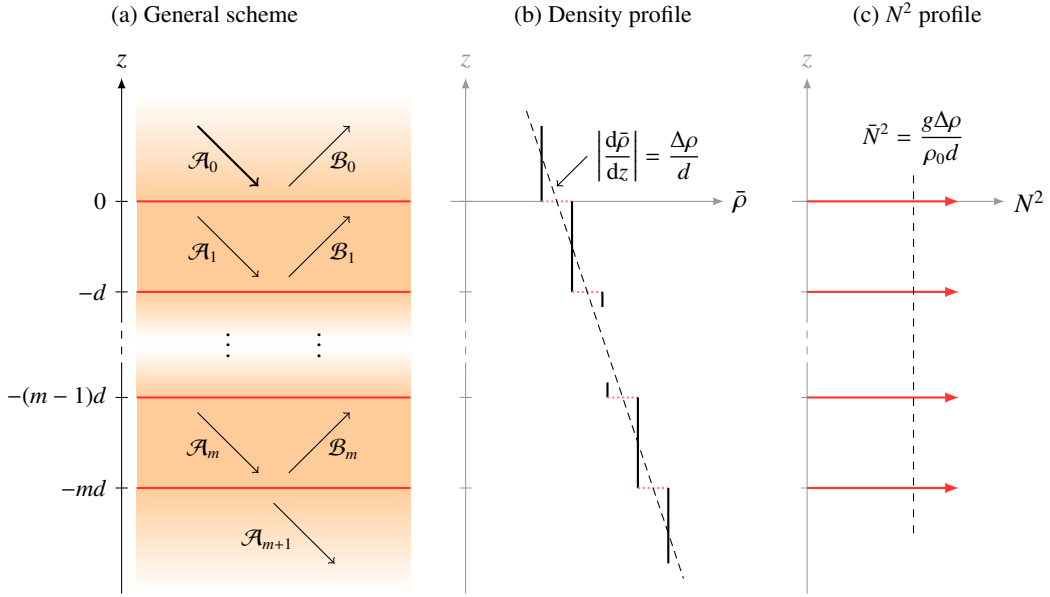

The Boussinesq approximation restricts us to have $d \ll H$, where $H$ is the characteristic length over which the background quantities vary. Under this condition, the magnitude of the density jump, $\Delta\rho$, between adjacent steps (which we obtain by integrating Eq. (\ref{eq:BruntVaisaleFreqLinear}) over one step including the interface) is given by
\begin{equation}
\bar{N}^2 = \frac{g\Delta\rho}{\rho_0 d}
\label{eq:meanN}
\end{equation}
\citepalias{BelyaevQuataertFuller2015}. Here, $\bar{N}$ corresponds to a mean stratification: typically, an internal wave with a large wavelength compared to the size of the steps will see the staircase as a continuously stratified medium characterised by the buoyancy frequency $\bar{N}$. Since $N=0$ within a convective layer, the only contribution to $\bar{N}$ comes from the density jump across the discrete interfaces. Also, because $d\ll H$ by assumption, we generally have $\Delta\rho\ll\rho_0$, so that the background density profile $\bar{\rho}(z)$ does not depart significantly from the constant reference density value $\rho_0$.

On Fig. \ref{fig:tcoeff_modele_CC}, three panels are displayed to summarize our reference model. Panel \ref{fig:Model_generalScheme} shows the general scheme: $m$ convective steps, each of vertical extent $d$, are separated by infinitesimally thin stably stratified interfaces. The incident (ingoing) wave -- whose vertical velocity has an amplitude $\mathcal{A}_0$ -- enters the staircase from above (region $z>0$); a reflected wave with amplitude $\mathcal{B}_0$ is created, and a transmitted (outgoing) wave with amplitude $\mathcal{A}_{m+1}$ comes out of the staircase (region $z<-md$). By causality, there is no upward propagating wave below the staircase \citepalias{Sutherland2016}. The regions above and below the staircase can either be convective ($N=0$), or stably stratified ($N>0$). We will denote the stratification above and below the staircase by $N_{\text{a}}$ and $N_{\text{b}}$, respectively. Note that Fig. \ref{fig:tcoeff_modele_CC} corresponds to the particular case where $N_{\text{a}}=N_{\text{b}}=0$, i.e. a staircase embedded in a convective medium, but we also consider the general case.

In each convective step, labelled by the integer $n$ ranging from $0$ to $m$, the solution is the sum of an upward and a downward propagating (or evanescent) wave, whose vertical velocities have amplitudes $\mathcal{A}_n$ and $\mathcal{B}_n$, respectively. 

We recall that the propagative or evanescent behaviour of a pure inertial wave depends on the value of $|\omega/2\Omega|$ ($<1$ for propagative and $>1$ for evanescent). Panel \ref{fig:Model_densityProfile} shows the corresponding density profile: the density undergoes a discontinuous jump by a value $\Delta\rho$ across each interface, and in between is uniform. This creates a mean density gradient 
\begin{equation}
\left|\frac{\text{d}\bar{\rho}}{\text{d}z}\right| = \frac{\Delta\rho}{d}.
\end{equation}
Finally, the buoyancy frequency profile $N^2(z)$ is displayed on panel \ref{fig:Model_Nprofile}. It consists of a sum of Dirac distributions centred on the interfaces, each with an integrated value of $\bar{N}^2$, such as to create a profile
\begin{equation}
N^2(z) = \left\{
\begin{array}{cl}
N^2_{\text{a}} & ~ \text{for} ~ z>0,\\
\displaystyle \bar{N}^2 \sum_{n=0}^{m}\updelta(z+nd) & ~ \text{for} ~ 0 > z > -md,\\
N^2_{\text{b}} & ~ \text{for} ~ z<-md,
\end{array}
\right.
\end{equation}
where $\updelta$ denotes the Dirac distribution. We stress that the vertical wave number, $k_z$, is a function of the buoyancy frequency $N$. This means that the incident gravito-inertial wave (or pure inertial wave if $N_{\text{a}}=0$) has a vertical wave number $k_{z,\text{a}}$ which is in general different from the one of the transmitted wave, $k_{z,\text{b}}$, which are both {a priori} different from the wave number of purely inertial waves inside the convective steps, $k_{z,\text{c}}$ ($N_{\text{c}}=0$). Remembering Eq. (\ref{eq:k_z}), we have defined
\begin{equation}
k_{z,\upalpha}^2 = -k_{\perp}^2 \left[ \frac{N_{\upalpha}^2-\omega^2}{\omega^2 - f^2} + \left( \frac{\omega \tilde{f}_{\text{s}}}{\omega^2 - f^2}\right)^2\right].
\label{eq:kz_minus}
\end{equation}
Here, $\upalpha$ stands for either $\text{a}$, $\text{b}$ or $\text{c}$. Note that this definition is minus the one given by Eq. (\ref{eq:k_z}), a choice we will explain in the following section. Once again, we stress that the total vertical wave number is given by $\tilde{k}_z = \pm k_z + \tilde{\delta}k_{\perp}$.

In the following sections, we will mostly consider a finite staircase constituted of $m$ steps, as described above, but we will also consider as a first approach in Section \ref{subsec:FREE:disprelInfiniteStaircase} an infinite staircase, for which $m \gg 1$.

\subsection{Mathematical statement of the problem}\label{sec:FREE:maths}
The quantity we will manipulate is the GIW's vertical velocity $w$, whose dynamics is governed by the Poincar\'e equation (see Section \ref{subsec:GIW:propagation}). 
Above and below the staircase (where $N=N_{\text{a}}, N_{\text{b}}$ respectively) and within each step (where $N=0$), the differential equation given by Eq. (\ref{eq:Schrodinger_omEqf}) for $\omega=f$ and by Eq. (\ref{eq:Schrodinger}) otherwise (modified by the definition of $k_z$ given by Eq. (\ref{eq:kz_minus})), can be solved explicitly. 

\paragraph{The case $\omega=f$.}\hfill\break
First, let us focus on the case $\omega=f$. In that case, the solution of the equation governing the vertical dependence of the vertical velocity is given by (\ref{eq:Schrodinger_omEqf}).
The solution for prescribed $k_{\perp}$ and $\omega$ is
\begin{equation}
\widetilde{W}(z) = \left\{
\begin{array}{l}
\mathcal{E}_0\text{e}^{\text{i}\gamma_{\text{a}}z} ~~~ \text{for} ~~ z>0,\\[3mm]
\mathcal{E}_n\text{e}^{\text{i}\gamma(z+nd)} ~~~ \text{for} ~~ -nd<z<-(n+1)d,
n=\{1,\dots,m\}\\[3mm]
\mathcal{E}_{m+1}\text{e}^{\text{i}\gamma_{\text{b}}(z+md)} ~~~ \text{for} ~~ z<-md,
\end{array}
\right.
\label{eq:w_n_omEqf}
\end{equation}
where we have defined
\begin{equation}
\gamma_{\upalpha} = \left( \dfrac{N_{\upalpha}^2 - f^2 + \tilde{f}_{\text{s}}^2}{2f\tilde{f}_{\text{s}}} \right) k_{\perp}
\end{equation}
(we have omitted the subscript 'c' in the convective steps to make notations lighter). {Equation} (\ref{eq:Schrodinger_omEqf}) being of first order in $z$, the only boundary condition we need to get the full solution comes from the requirement that the two fluids above and below the interface stay in contact: $w_n=w_{n+1}$, for $n=\{0, \dots, m+1\}$. It is then straightforward that the amplitudes of the incident and transmitted waves, $\mathcal{E}_0$ and $\mathcal{E}_{m+1}$ respectively, are related by
\begin{equation}
\mathcal{E}_0 = \text{e}^{\text{i}(m-1)\gamma d}\mathcal{E}_{m+1}.
\end{equation}
The coefficient $\gamma$ being real, we conclude that
\begin{equation}
|\mathcal{E}_0| = |\mathcal{E}_{m+1}|.
\label{eq:E0EqEmp1}
\end{equation}

\paragraph{The case $\omega \neq f$.}\hfill\break
The solution for prescribed $k_{\perp}$ and $\omega$ is then
\begin{equation}
\hat{W}(z) = \left\{
\begin{array}{l}
\mathcal{A}_0\text{e}^{k_{z,\text{a}}z} + \mathcal{B}_0\text{e}^{-k_{z,\text{a}}z} ~~~ \text{for} ~~ z>0,\\[3mm]
\mathcal{A}_n\text{e}^{k_z[z+(n-1/2)d]} + \mathcal{B}_n\text{e}^{-k_z[z+(n-1/2)d]} \\
\hspace{0.8cm} \text{for} ~~ -nd<z<-(n+1)d, n=\{1,\dots,m\}\\[3mm]
\mathcal{A}_{m+1}\text{e}^{k_{z,\text{b}}[z+md]} + \mathcal{B}_{m+1}\text{e}^{-k_{z,\text{b}}[z+md]} ~~~ \text{for} ~~ z<-md,
\end{array}
\right.
\label{eq:w_n}
\end{equation}
where 
\begin{equation}
k_z \equiv \sqrt{k_{z,\text{c}}^2} = \text{i}\sqrt{-k_{z,\text{c}}^2},
\end{equation}
and the addition of $(n-1/2)d$ in the exponents have been included to take advantage of symmetry in finding analytic solutions \citepalias[][]{Sutherland2016}. Thus, defining $k_z^2$ as in Eq. (\ref{eq:kz_minus}) allows us to treat the propagative {versus} evanescent behaviour of the ingoing and outgoing waves self consistently.

We note that under the traditional approximation, $\tilde{f} = 0$ so that $\tilde{\delta} = 0$. In this sense, the term $\tilde{\delta}$ accounts for non-traditional effects, {together with the other non-traditional terms contained in Eq. (\ref{eq:kz_minus})}. In the general case, these effects vanish when $f=0$ or $\tilde{f}=0$, i.e. at the equator ($\Theta=\uppi/2$) and at the pole ($\Theta=0$), respectively.

Equation (\ref{eq:Schrodinger}) being of second order, we need a second boundary condition to get the complete solution. It arises from the requirement that the momentum flux is continuous across the interface. It can be obtained by integrating Eq. (\ref{eq:Poincare}) across the interface, situated at $z_n = -nd$ {for the example case that follows:}
\begin{align}
\label{eq:int1}
\underbrace{\int_{-nd-\epsilon d}^{-nd+\epsilon d}\left(f^2-\omega^2\right) \frac{\uppartial^2w}{\uppartial z^2}\,\text{d}z}_{\textstyle \alpha_1}
+ \underbrace{\int_{-nd-\epsilon d}^{-nd+\epsilon d} 2\text{i}f\tilde{f}_{\text{s}}k_{\perp}\frac{\uppartial w}{\uppartial z}\,\text{d}z}_{\textstyle \alpha_2} & \nonumber\\
- \underbrace{\int_{-nd-\epsilon d}^{-nd+\epsilon d} k_{\perp}^2\left(N^2-\omega^2+\tilde{f}_{\text{s}}^2\right)w\,\text{d}z}_{\textstyle \alpha_3}
& = 0,
\end{align}
where we have introduced a dimensionless parameter $\epsilon < 1$. We have
\begin{equation}
\label{eq:alp1}
\alpha_1 = \left(f^2-\omega^2\right)\left(w_n' - w_{n+1}'\right),
\end{equation}
where the primes denote differentiation with respect to $z$, and
\begin{equation}
\label{eq:alp2}
\alpha_2 = 2\text{i}f\tilde{f}_{\text{s}}k_{\perp} (w_n - w_{n+1}) = 0,
\end{equation}
where we have used the first boundary condition $w_n = w_{n+1}$. Finally, we have
\begin{align}
\alpha_3 &= \int_{-nd-\epsilon d}^{-nd+\epsilon d} k_{\perp}^2\left(N^2-\omega^2+\tilde{f}_{\text{s}}^2\right)w\,\text{d}z\nonumber\\
 &= k_{\perp}^2 w_n d \left[ \underbrace{\int_{-\epsilon}^{+\epsilon}N^2\,\text{d}\zeta}_{\textstyle \bar{N}^2} + \left(\tilde{f}_{\text{s}}^2-\omega^2\right)\times (2\epsilon)\right],
\end{align}
where we used the change of variable $\zeta = (z+nd)/d$, and took $w$ out of the integral as it does not vary across the interface. Replacing $\bar{N}^2$ by its expression given by Eq. (\ref{eq:meanN}), and taking the limit $\epsilon \rightarrow 0$ lead to
\begin{equation}
\label{eq:alp3}
\alpha_3 = k_{\perp}^2 g \frac{\Delta \rho}{\rho_0} w_n.
\end{equation}
Thus, we finally obtain a set of two boundary conditions involving the vertical velocity $w_n$ and its $z$-derivative $w_n'$,
\begin{align}
w_{n+1} &= w_n,\label{eq:BC1}\\
w_{n+1}' &= w_n' + \frac{g\bar{k}^2}{\omega^2} \frac{\Delta \rho}{\rho_0}w_n,\label{eq:BC2}
\end{align}
where we have defined
\begin{equation}
\label{eq:kbar}
\bar{k} = k_{\perp} \frac{\omega}{\sqrt{\omega^2 - f^2}}.
\end{equation}
These results agree with those of \citetalias{BelyaevQuataertFuller2015} {for the two cases they consider of the pole and the equator}.

Let us consider the interior of the staircase, i.e. consider $n=\{1, \dots, m\}$. Substituting the expression of $w_n$ given by Eq. (\ref{eq:w_n}) into the two boundary conditions given by Eqs. (\ref{eq:BC1}) and (\ref{eq:BC2}), we obtain two recurrence relations between the coefficients $(\mathcal{A}_n, \mathcal{B}_n)$,
\begin{align}
\mathcal{A}_{n} &= \text{e}^{-\text{i}\tilde{\varphi}}\left[\Delta\left(1-\Gamma\right)\mathcal{A}_{n+1} - \Gamma\mathcal{B}_{n+1}\right],\label{eq:RI1}\\
\mathcal{B}_{n} &= \text{e}^{-\text{i}\tilde{\varphi}}\left[\Gamma\mathcal{A}_{n+1} +\Delta^{-1} \left(1+\Gamma\right)\mathcal{B}_{n+1}\right],\label{eq:RI2}
\end{align}
where the following dimensionless quantities have been defined:
\begin{align}
\tilde{\varphi} &\equiv \tilde{\delta}k_{\perp}d,\\[2.5mm]
\Delta &\equiv \text{e}^{k_zd},\\
\Gamma &\equiv \frac{1}{2}\frac{g\bar{k}^2}{k_z\omega^2}\frac{\Delta\rho}{\rho_0} = \frac{1}{2} \left(\frac{\bar{N}}{\omega}\right)^2 \frac{\left(\bar{k}d\right)^2}{k_zd}.
\end{align}
The coefficient $\Gamma$ can further be expressed as a function of more appropriate variables for our problem, as
\begin{equation}
\Gamma = \frac{1}{2} \left(k_{\perp}d\right)\left(\frac{\bar{N}}{\omega}\right)^2\left[1-\left(\frac{2\widetilde{\Omega}}{\omega}\right)^2\right]^{-1/2}.
\end{equation}
Inside the staircase, the coefficients of adjacent convective steps are then related by
\begin{equation}
\left[
\begin{array}{c}
      \mathcal{A}_{n}\\
      \mathcal{B}_{n}
\end{array}
\right]
=
\widetilde{\mathbfsf{T}}
\left[
\begin{array}{c}
      \mathcal{A}_{n+1}\\
      \mathcal{B}_{n+1}
\end{array}
\right],
\label{eq:matrixform}
\end{equation}
where the transfer matrix $\widetilde{\mathbfsf{T}}$ is defined by
\begin{equation}
\widetilde{\mathbfsf{T}} = \text{e}^{-\text{i}\tilde{\varphi}}
\left[
\begin{array}{cc}
      \Delta\left(1-\Gamma\right) & -\Gamma\\[3mm]
      \Gamma & \Delta^{-1}\left(1+\Gamma\right)
\end{array}
\right].
\end{equation}

Above and below the staircase, we have to perform a separate calculation because the stratification is not the same {a priori}. This will give us the boundary conditions of the entire staircase. 

\subsection{Dispersion relation for gravito-inertial waves in a staircase density profile}\label{sec:FREE:disprel}
The dispersion relation for pure gravity modes in a staircase modelled as described in Section \ref{sec:FREE:model} has been derived by \citetalias{BelyaevQuataertFuller2015}. This was done in detail in the case without rotation, and they discussed the effects of rotation for the two particular cases where the spin axis is parallel or perpendicular to $\hat{\bm{e}}_z$ (respectively $\Theta=0$ at the pole and $\Theta=\uppi/2$ at the equator). In this section, we extend their calculation to the general case where the spin axis makes an arbitrary angle $\Theta$ to the local radial direction, in order to obtain the free modes of a density staircase at any latitude.

\subsubsection{Infinite staircase}\label{subsec:FREE:disprelInfiniteStaircase}
{Following \citetalias{BelyaevQuataertFuller2015}, we first consider an infinite staircase. The local model strictly loses} validity in this case, but this {idealized} calculation will provide us useful insights. We first consider strictly periodic boundary conditions. That is, we assume that there is an integer $m$ so that $\mathcal{A}_{n}=\mathcal{A}_{n+m}$ and $\mathcal{B}_{n}=\mathcal{B}_{n+m}$, so that $m$ is the periodicity of the infinite staircase. Recalling Eq. (\ref{eq:matrixform}), we obtain
\begin{equation}
\left[
\begin{array}{c}
      \mathcal{A}_{n}\\
      \mathcal{B}_{n}
\end{array}
\right]
=
\widetilde{\mathbfsf{T}}^m
\left[
\begin{array}{c}
      \mathcal{A}_{n}\\
      \mathcal{B}_{n}
\end{array}
\right].
\end{equation}
Non trivial solutions of this equation exist only if
\begin{equation}
\det\left(\widetilde{\mathbfsf{T}}^m - \bm{\updelta}\right)=0,
\end{equation}
where $\bm{\updelta}$ is the $2\times 2$ identity matrix. With some algebra, following \citetalias{BelyaevQuataertFuller2015}, we obtain the dispersion relation for periodic solutions with rotation,
\begin{equation}
\omega^2 = \bar{N}^2 \left(
\frac{(\bar{k}d)^2/k_zd}{2\coth (k_zd)-2\cos \theta \csch (k_zd)}
\right),
\label{eq:BQFdisprel_infinite}
\end{equation}
where
\begin{equation}
\theta = \frac{2\uppi n}{m} \pm \tilde{\delta}(k_{\perp}d),
\end{equation}
$n$ being an integer that ranges from $0$ to $m-1$, and $\bar{N}^2$ is the background buoyancy frequency that corresponds to the averaged density gradient, defined by Eq. (\ref{eq:meanN}). Equation (\ref{eq:BQFdisprel_infinite}) gives the frequencies for the modes of the staircase.

In the case without rotation, $\tilde{\delta}=0$ and $k_zd=\bar{k}d=k_{\perp}d$, so that we recover \citetalias{BelyaevQuataertFuller2015}'s dispersion relation (see their Eq. (18)):
\begin{equation}
\omega^2 = \bar{N}^2 \left(
\frac{k_{\perp}d}{2\coth (k_{\perp}d)-2\cos \theta \csch (k_{\perp}d)}
\right).
\label{eq:BQFdisprel_infiniteWithourRotation}
\end{equation}
Equation (\ref{eq:BQFdisprel_infinite}) also agrees with \citetalias{BelyaevQuataertFuller2015} for the specific cases with rotation that they study, i.e. at the pole and the equator.

\subsubsection{Finite staircase embedded in a convective medium}\label{subsec:FREE:disprelFiniteStaircase}
In this section, we {extend} the calculation made in \citetalias{BelyaevQuataertFuller2015} for the case of a finite staircase embedded in a convective medium, including rotation at any colatitude $\Theta$. This is done in order to get the free modes of oscillation of a finite staircase in that set-up. The perturbations are assumed to decay as $z \rightarrow \pm\infty$. The boundary conditions are then $\mathcal{A}_0=0$ at the top, and $\mathcal{B}_{m+1}=0$ at the bottom. 

Applying those boundary conditions and using Eq. (\ref{eq:matrixform}), we can write
\begin{equation}
\left[
\begin{array}{c}
      \mathcal{A}_{m+1}\\
      0
\end{array}
\right]
=
\widetilde{\mathbfsf{T}}^m
\left[
\begin{array}{c}
      0\\
      \mathcal{B}_{0}
\end{array}
\right].
\end{equation}
The equation above is true in general if the lower-right corner of the $2\times 2$ matrix $\widetilde{\mathbfsf{T}}^m$ is zero. The dispersion relation thus becomes
\begin{equation}
\left( \widetilde{\mathbfsf{T}}^m \right)_{22} = 0
\end{equation}
We can diagonalise this matrix, and with some algebra \citepalias[following][]{BelyaevQuataertFuller2015}, we can obtain a similar dispersion relation:
\begin{equation}
\omega^2 = \bar{N}^2 \left(
\frac{(\bar{k}d)^2/k_zd}{2\coth (k_zd)-2\cos\theta\csch (k_zd)}
\right),
\label{eq:BQFdisprel_finite}
\end{equation}
but now $\cos\theta$ is one of the roots of the polynomial
\begin{equation}
T_m(\cos\theta) + [\cos\theta \coth (k_zd) - \csch (k_zd)]U_{m-1}(\cos\theta) = 0,
\end{equation}
where $T_m$ and $U_m$ are Chebyshev polynomials of the first and second kinds, respectively, defined by
\begin{align}
\cos(m\theta) &= T_m(\cos\theta),\\
\sin(m\theta) &= U_{m-1}(\cos\theta)\sin\theta.
\end{align}
We refer the reader to \citetalias{BelyaevQuataertFuller2015} to see the details of the calculation {in the case without rotation}. Here, the same lines are reproduced. However, taking into account rotation at any colatitude somewhat complicates the analysis.

\subsection{Transmission of an incident (gravito-)inertial wave}\label{sec:FREE:transmission}
In this section, we aim to answer the following question: what is the effect of a density staircase on the transmission of an incident internal wave with amplitude $\mathcal{A}_0$ {taking into account the complete Coriolis acceleration}? This should help us to understand how the presence of layered semi-convection and associated density staircases would affect the fate of tidally excited waves launched in one region as it propagates towards another. {It could also be of great interest when studying the seismology of giant planets, to predict what modes can be observed when looking at the oscillations of their surfaces \citep[][]{GaulmeEtal2011,Fuller2014}.}

We thus analyse the properties of the transmission of an internal wave upon a finite-length density staircase. We will use the same formalism as previously, except for the boundary conditions. Indeed, because we consider the transmission of an internal wave upon the staircase, we refer to Fig. \ref{fig:Tcoeff}: an incident (ingoing) wave with amplitude $\mathcal{A}_0$ enters the staircase, creating a reflected (outgoing) wave with amplitude $\mathcal{B}_0$. {Only a downward propagating wave is assumed to exist below the staircase}. This is the transmitted wave, with amplitude $\mathcal{A}_{m+1}$ ($\mathcal{B}_{m+1}=0$).

The first work of this type was S16, who studied internal wave transmission through a density staircase in the ocean using the traditional approximation. If this is often appropriate for a thin oceanic layer on the Earth, this is not suitable to model low-frequency internal waves in the deep envelopes of giant planets \citep{OgilvieLin2004}. In this section, we generalise \citetalias{Sutherland2016} to arbitrary top and bottom layer properties, and include the complete Coriolis acceleration.

\subsubsection{Analytic expression of the transmission coefficient}\label{subsec:FREE:t_analytic}
{Our aim is now to calculate a transmission coefficient in the general case of arbitrary boundary conditions and number of steps}. This will allow us to determine what fraction of the incident wave energy makes its way through the staircase and propagates to deeper regions of a giant planet. Recalling the expression we derived in Section \ref{subsubsec:GIW:transmission}, the transmission coefficient is defined by
\begin{equation}
T = \frac{k_{z,\text{b}}}{k_{z,\text{a}}}
\left| \frac{\mathcal{A}_{m+1}}{\mathcal{A}_0} \right|^2,
\end{equation}
which simply reduces to $T=\left|\mathcal{A}_{m+1}/\mathcal{A}_0 \right|^2$ when the stratification is the same above and below the staircase, e.g. in the case of a staircase embedded in a stably stratified medium \citepalias{Sutherland2016}.

Substituting Eq. (\ref{eq:w_n}) (including the boundary condition discussed above, namely $\mathcal{B}_{m+1}$) into the interface conditions given by Eqs. (\ref{eq:BC1}) and (\ref{eq:BC2}), we obtain after some algebra the following set of equations:
\begin{align}
\mathcal{A}_0 &= \textstyle\frac{1}{2}\Delta^{1/2}\left[1-K^{\text{c}}_{\text{a}}(1-\Gamma)\right]\,\mathcal{A}_1 +\\ 
& ~ ~ ~ ~ ~ ~ ~ \textstyle\frac{1}{2}\Delta^{-1/2}\left[1+K^{\text{c}}_{\text{a}}(1+\Gamma)\right]\,\mathcal{B}_1\\[2mm]
\mathcal{A}_n &= \Delta(1-\Gamma)\,\mathcal{A}_{n+1} - \Gamma\,\mathcal{B}_{n+1}\\
\mathcal{B}_n &= \Gamma\,\mathcal{A}_{n+1} + \Delta^{-1}(1+\Gamma)\,\mathcal{B}_{n+1}\\[2mm]
\mathcal{A}_m &= \textstyle\frac{1}{2}\Delta^{1/2}\left[1-\Gamma+K^{\text{b}}_{\text{c}}\right]\,\mathcal{A}_{m+1}\\
\mathcal{B}_m &= \textstyle\frac{1}{2}\Delta^{-1/2}\left[1+\Gamma-K^{\text{b}}_{\text{c}}\right]\,\mathcal{A}_{m+1},
\end{align}
where $n=\left\{ 1, \dots, m-1\right\}$, and we have defined
\begin{equation}
K^{\upalpha}_{\upbeta} = \frac{k_{z,\upalpha}}{k_{z,\upbeta}}.
\end{equation}
We recall that it has been defined that $k_z\equiv k_{z,\text{c}}$ in convective steps, to make notations lighter. Those equations can be combined to express $\mathcal{A}_0$ in term of solely $\mathcal{A}_{m+1}$,
\begin{equation}
\mathcal{A}_0 = \left[\bm{\text{b}}_{\text{a}}^{\text{T}}\,\widetilde{\mathbfsf{T}}^{m-1} \bm{\text{b}}_{\text{b}} \right] \mathcal{A}_{m+1},
\end{equation}
where $^{\text{T}}$ denotes transposition, and the left and right vectors are defined by
\begin{align}
\bm{\text{b}}_{\text{a}} &=\frac{1}{2}\left[
\begin{array}{c}
\Delta^{1/2}\left[1-K^{\text{c}}_{\text{a}}(1-\Gamma)\right]\\
\Delta^{-1/2}\left[1+K^{\text{c}}_{\text{a}}(1+\Gamma)\right]
\end{array}
\right],\\[2mm]
\bm{\text{b}}_{\text{b}} &=\frac{1}{2}\left[
\begin{array}{c}
\Delta^{1/2}\left[1-\Gamma+K^{\text{b}}_{\text{c}}\right]\\
\Delta^{-1/2}\left[1+\Gamma-K^{\text{b}}_{\text{c}}\right]
\end{array}
\right].
\end{align}
Therefore, the transmission coefficient is given by
\begin{equation}
T = \frac{k_{z,\text{b}}}{k_{z,\text{a}}} \left| \bm{\text{b}}_{\text{a}}^{\text{T}}\,\widetilde{\mathbfsf{T}}^{m-1} \bm{\text{b}}_{\text{b}} \right|^{-2},
\label{eq:Tcoeff_msteps_analytic}
\end{equation}
with
\begin{equation}
\frac{k_{z,\text{b}}}{k_{z,\text{a}}} = \left(
\frac{\omega^4+\omega^2[N^2_{\text{b}}-4\widetilde{\Omega}^2] + (N_{\text{b}}f)^2}{\omega^4+\omega^2[N^2_{\text{a}}-4\widetilde{\Omega}^2] + (N_{\text{a}}f)^2}
\right)^{1/2},
\label{eq:kzb/kza}
\end{equation}
where we recall that $2\widetilde{\Omega} \equiv \sqrt{f^2+f_{\text{s}}^2}$ (i.e. $2\Omega$ when $\alpha=\uppi/2$).

Eq. (\ref{eq:Tcoeff_msteps_analytic}) is a general analytic expression of the transmission coefficient, that can now be used in order to analyse the behaviour of the transmission in the parameter space (wave frequency, wavelength, number of steps, boundary conditions, {etc}).

\subsubsection{Transmission across one step}\label{subsec:FREE:t_onestep}

\begin{figure}
\centering
\begin{tikzpicture}[scale=1]
\node[anchor=south west,inner sep=0] at (0,0)
    {\includegraphics[height=6.875cm]{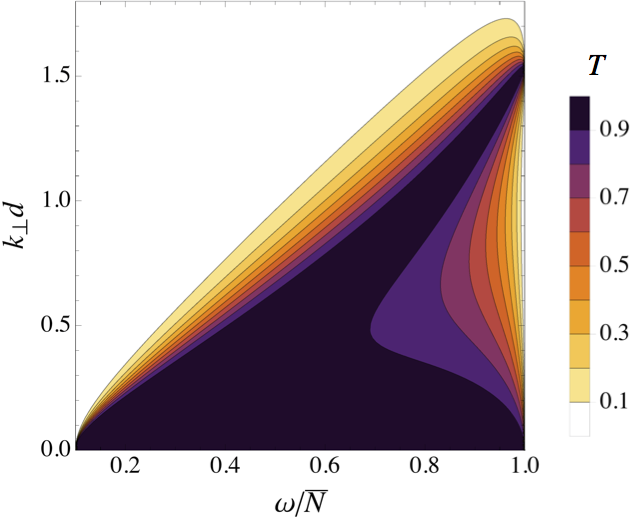}};
\draw[densely dotted] (1,1) -- ++ (0,-0.5) node[below]{$\boxed{\omega_-}$};
\draw[densely dotted] (6.93,1) -- ++ (0,-0.5) node[below]{$\boxed{\omega_+}$};
\end{tikzpicture}
\caption[Transmission coefficient -- comparison with Sutherland (2016)]{Transmission coefficient as a function of normalised frequency, $\omega/\bar{N}$ and horizontal wave number, $k_{\perp}d$, with $\Omega=0.05\bar{N}$ and $\Theta=0$ ($f=0.1\bar{N}$), for comparison with \citetalias{Sutherland2016}. The transmission coefficient, $T$, lies between $0$ and $1$ and is calculated for the range of frequency over which the incident wave is propagative, which we denote by $\omega_-<\omega<\omega_+$.}
\label{fig:OneStep_CompS16}
\end{figure}

\paragraph{Staircase embedded in a stably stratified medium.}\label{subsec:FREE:t_rad}\hfill\break
\citetalias{Sutherland2016} have considered the transmission of an internal wave through a density  staircase embedded in a stably stratified medium ($N_{\text{a}}, N_{\text{b}} > 0$). Because the aim of his work is to predict energy transport by internal waves incident upon observed density staircases in the strongly stratified ocean, the traditional approximation is adopted \citep[even though][have demonstrated that this is not always appropriate, even for the ocean]{GerkemaShrira2005}. We begin by recovering their results in the case of the transmission across one step (see his Section II.B) under the traditional approximation, which corresponds to a colatitude $\Theta = 0$, where the rotation axis and gravity are aligned. This will provide us a first check that our mathematical formalism, in which the traditional approximation is {not} assumed, is correct. The result is shown on Fig. \ref{fig:OneStep_CompS16}.

The transmission coefficient $T$ is displayed as a function of the wave frequency normalized by the mean buoyancy frequency, $\omega/\bar{N}$, and of the horizontal wave number rendered dimensionless by multiplication by the size of the convective layers, $k_{\perp}d$. In this and subsequent figures, the range of frequency over which the transmission coefficient is calculated is chosen so that both the incident {and} reflected waves (carrying energy densities $\propto |\mathcal{A}_0|^2$ and $\propto |\mathcal{A}_{m+1}|^2$, respectively) are {propagative in both regions $z>0$ and $z<-D$}, respectively. This gives a range $\omega_- < \omega < \omega_+$, where
\begin{align}
\omega_- &= \max\left(\omega_-(N_{\text{a}}),\omega_-(N_{\text{b}})\right),\\
\omega_+ &= \min\left(\omega_+(N_{\text{a}}),\omega_+(N_{\text{b}})\right),
\end{align}
where $\omega_{\pm}(N_{\upalpha})$ is given by Eq. (\ref{eq:freq_spectrum}). Fig. \ref{fig:OneStep_CompS16} can be directly compared to Fig. 2a of \citetalias{Sutherland2016} for parameters $\Omega=0.05\bar{N}$, $\Theta=0$, giving  $f=0.1\bar{N}$. No transmission occurs for $\omega<f$ because the incident wave would be evanescent in that case, as explained above {(this occurs in the traditional approximation)}. 

We can see that the qualitative behaviour of the transmission coefficient is the same: transmission is enhanced for large wavelengths (low wave numbers) compared to the size of the staircase ($D=d$ in this section because there is only one step), and there is a branch of enhanced transmission near $\omega_+$, where transmission can be large even with $k_{\perp}d$ of order unity. 

However, a quantitative discrepancy is found between the two figures: perfect transmission for incident waves with $\omega \approx \omega_+$ occur for $k_{\perp}d \approx 2.4$ in the case of \citetalias{Sutherland2016}, and $k_{\perp}d \approx 1.5$ in our case. This difference is due to the fact that the density jumps at the first and the last interfaces is taken to be $\Delta\rho/2$ in \citetalias{Sutherland2016}, and $\Delta\rho$ everywhere in our work in order to treat all cases self-consistently, regardless of the boundary conditions applied to the staircase. However, it has been checked that the model of \citetalias{Sutherland2016} with our formalism yields his results.
The origin of this band of enhanced transmission is a resonance between the ingoing wave with interfacial gravity waves on either side of the convective step. The dispersion relation for these waves on an interface with density jump $\Delta \rho$ (assuming infinitely deep layer) is
\begin{equation}
\omega^2 = \frac{1}{2} \frac{k_{\perp} g \Delta\rho}{\rho_0} = \frac{1}{2} (k_{\perp} d) \bar{N}^2.
\end{equation} 
So when $\omega\approx \bar{N}$, we expect enhanced transmission when $k_\perp d \approx 2$, which is what we (and \citetalias{Sutherland2016}) observe, at least approximately. However, given that our layer is not infinitely deep, we expect quantitative discrepancies from this simple estimate.

Note that waves with low wave numbers, and thus large wavelengths
\begin{equation}
\lambda \gg d
\end{equation}
are expected to be unaffected by the staircase. This is confirmed by the fact that we get perfect transmission at low wave numbers, for any frequency. Those waves will see the staircase as a continuously stratified medium with $N_0=\bar{N}$. However, shorter wavelength waves, with $\lambda \sim d$, are strongly affected by the presence of a staircase, only being efficiently transmitted in certain regions of parameter space.
\\

\begin{figure*}
\centering
\begin{subfigure}{0.31\textwidth}
\centering
\caption{$\Omega = 0$}
\includegraphics[height=5.55cm]{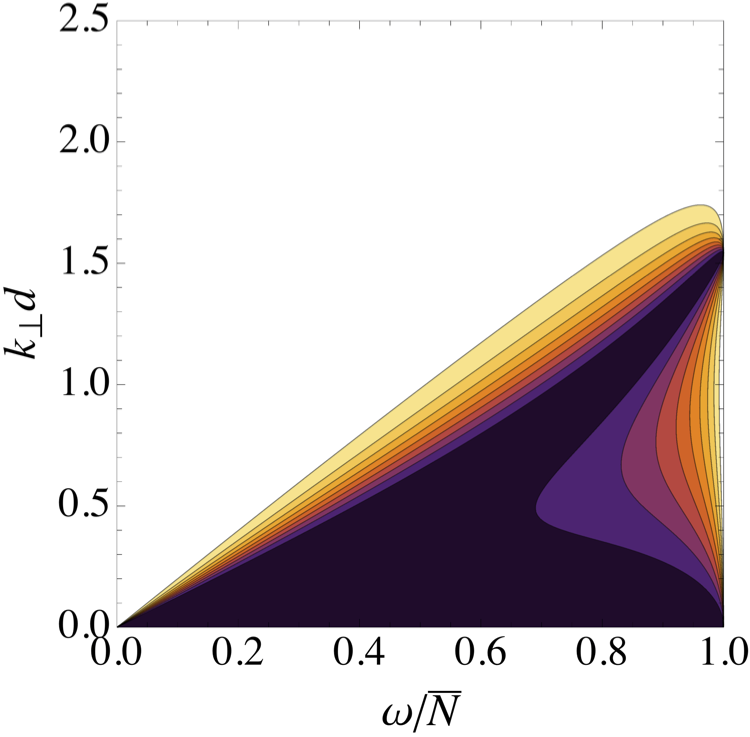}
\end{subfigure}
\begin{subfigure}{0.31\textwidth}
\centering
\caption{$\Omega = 0.4\bar{N}$}
\includegraphics[height=5.55cm]{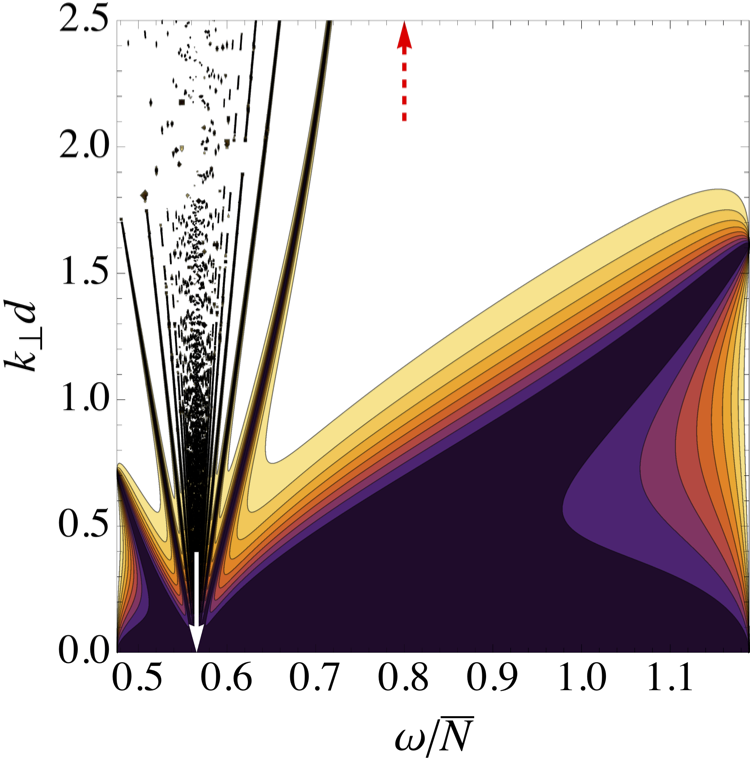}
\end{subfigure}
\begin{subfigure}{0.36\textwidth}
\centering
\caption{$\Omega = 0.6\bar{N}$}
\includegraphics[height=5.55cm]{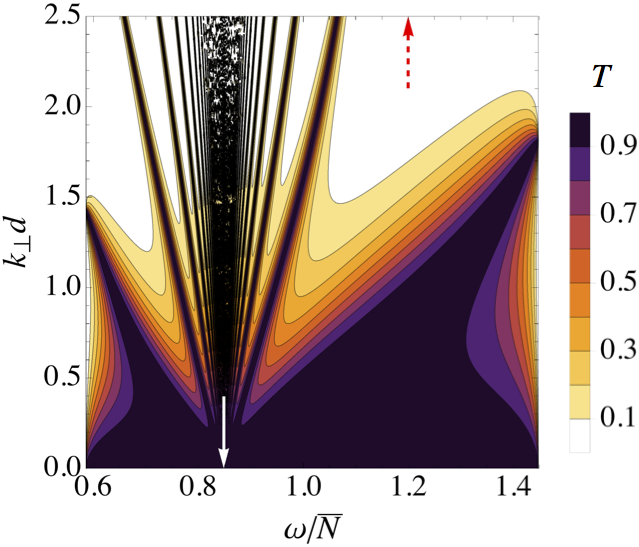}
\end{subfigure}
\caption[Transmission coefficient -- one step embedded in a stably stratified medium with rotation]{Transmission coefficient for one step embedded in a stably stratified medium, in an inclined box with $\Theta=\uppi/4$, for different rotation frequencies (a) $\Omega=0$, (b) $\Omega=0.4\bar{N}$ and (c) $\Omega=0.6\bar{N}$, as a function of frequency and perpendicular wave number. The white and dashed red arrows indicate frequencies $\omega=f$ and $\omega=2\Omega$, respectively. A set of bands depart from $f=2\Omega\cos\Theta$ ($=\sqrt{2}\Omega$ here) corresponding with a resonance with short-wavelength inertial waves. The transmission coefficients are calculated over a frequency range $\omega_- < \omega < \omega_+$, calculated such that both the incident and transmitted wave are propagative.}
\label{fig:onestep_RRrotation}
\centering
\begin{subfigure}{0.31\textwidth}
\centering
\caption{$\Omega = 0$}
\includegraphics[height=5.55cm]{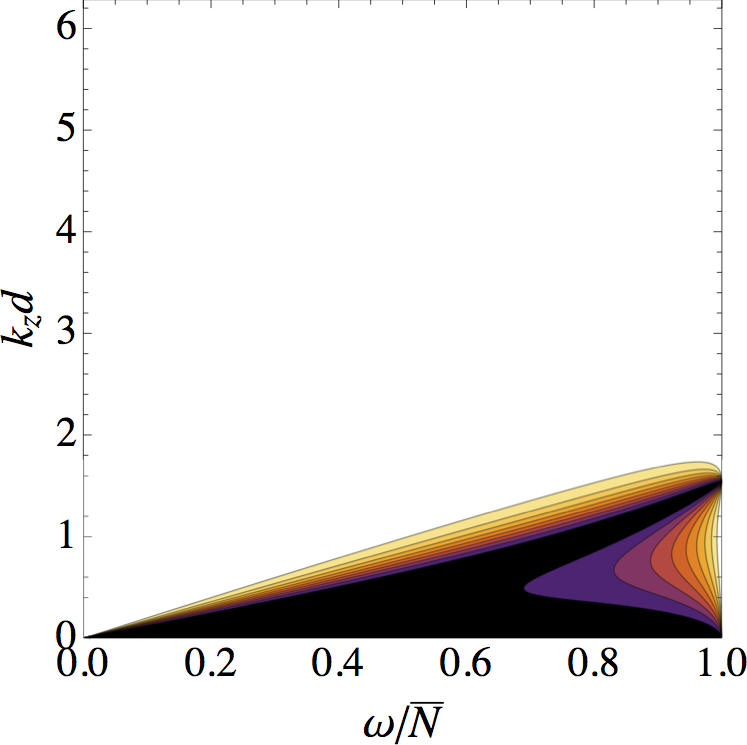}
\end{subfigure}
\begin{subfigure}{0.31\textwidth}
\centering
\caption{$\Omega = 0.4\bar{N}$}
\includegraphics[height=5.55cm]{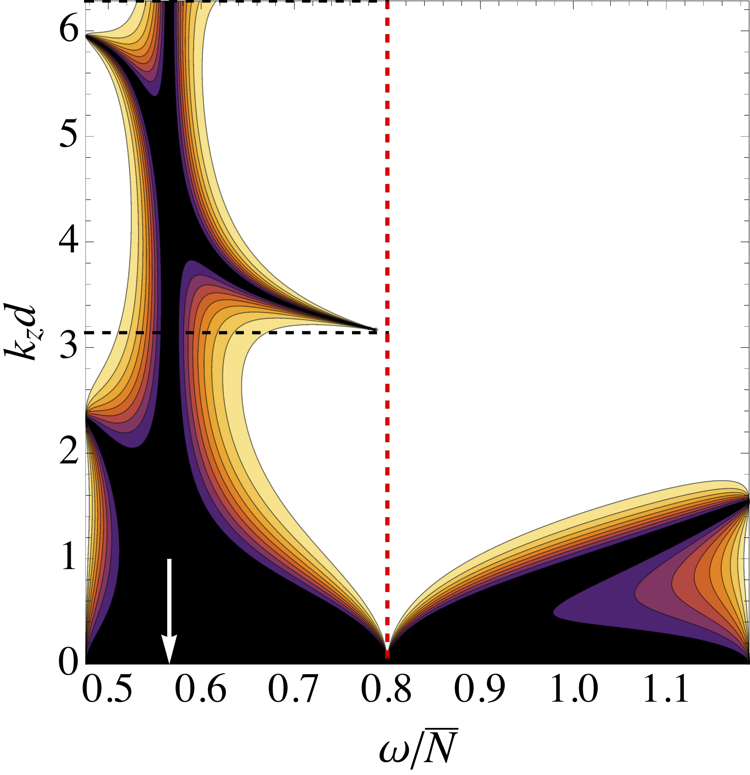}
\end{subfigure}
\begin{subfigure}{0.36\textwidth}
\centering
\caption{$\Omega = 0.6\bar{N}$}
\includegraphics[height=5.55cm]{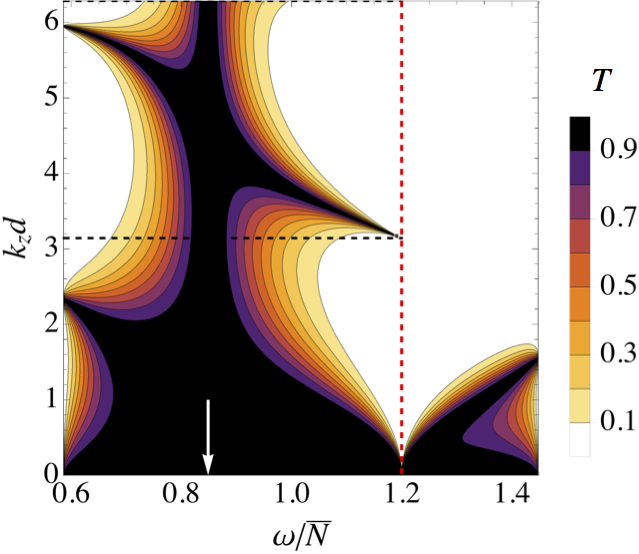}
\end{subfigure}
\caption[Transmission coefficient -- one step embedded in a stably stratified medium with rotation]{Transmission coefficient for one step embedded in a stably stratified medium, in an inclined box with $\Theta=\uppi/4$, for different rotation frequencies (a) $\Omega=0$, (b) $\Omega=0.4\bar{N}$ and (c) $\Omega=0.6\bar{N}$, as a function of frequency and of the vertical wave number. The white arrow indicates the frequency $\omega=f$, while the vertical dashed red line indicates the frequency $\omega=2\Omega$. The horizontal dashed black lines indicates multiple of $\uppi$.}
\label{fig:onestep_RRrotationKZ}
\end{figure*}

Now, to quantify the importance of the non-traditional effects, we move the box away from the pole and choose a colatitude $\Theta=\uppi/4$. We calculate the transmission coefficient embedded in a stably stratified medium with $N_{\text{a}} = N_{\text{b}} = \bar{N}$ for various rotation rates to quantify the influence of rotation. The results are displayed on Fig. \ref{fig:onestep_RRrotation} for $\Omega=0$, $0.4\bar{N}$ and $0.6\bar{N}$. Thus, in an astrophysical context we mostly consider fast rotators, so that the effects of the complete Coriolis acceleration are clearer, especially near $\omega = f$. 
On Fig. \ref{fig:onestep_RRrotation}, we see that new features arise compared to the traditional case (see Fig. \ref{fig:OneStep_CompS16}), in the form of bands of perfect transmission departing from $\omega = f$ ($\approx 0.57\bar{N}$ for $\Omega=0.4\bar{N}$, and $\approx 0.85\bar{N}$ for $\Omega=0.6\bar{N}$), these getting thicker with increasing rotation (but not more numerous, as we will see later). Note that the discontinuous (scattered, even) appearance of some of these branches is a plotting issue, because the branches are very narrow. However, the underlying branches are physical and not due to numerical errors, as we will discuss below. 

Fig. \ref{fig:onestep_RRrotationKZ} shows the behaviour of the transmission coefficient as a function of the vertical wave number $k_z$ (instead of $k_{\perp}$), rendered dimensionless by muliplication by $d$. It will be seen that as soon as rotation {is non-zero} (see Figs. \ref{fig:onestep_RRrotationKZ}b,c), we obtain perfect transmission around the Coriolis frequency $\omega=f$, which is indicated by a white arrow. Furthemore, two regions can be identified: 
\begin{itemize}
\item the region $\omega_- < \omega < 2\widetilde{\Omega}$ (delimited by the vertical dashed red line) -- in which transmission is enhanced even for large vertical wave-lengths (meaning low vertical wave numbers) -- shows bands of perfect transmission departing from the vertical transmission line at $\omega=f$, to reach $k_zd$ equalling multiple of $\uppi$ lines (indicated by the horizontal dashed black lines) near $2\widetilde{\Omega}$.
\item the region $2\widetilde{\Omega} < \omega < \omega_+$, which shows a similar behaviour to Fig. \ref{fig:onestep_RRrotation} as a function of the horizontal wave number $k_{\perp}$.
\end{itemize}
We note that the region $\omega_- < \omega < 2\widetilde{\Omega}$ corresponds to the frequency window for which both inertial waves in the convective layers, and GIWs in the stably stratified ones, are propagative. Thus, it makes sense to obtain enhanced transmission for those frequencies.
Note that $\widetilde{\Omega}=\Omega$ for $\alpha=\uppi/2$ that is adopted throughout this and subsequent sections.

Thus, the transmission differs mostly for GIWs, which have frequency close to $f$. We find that the qualitative behaviour remains the same for $\omega > 1.5f$. However, because including the complete Coriolis acceleration {extends} the range of admissible frequencies below $f$, there is a new band of perfect transmission for $k_{\perp}d$ of order unity arising for sub-inertial waves (i.e. for $\omega_- < \omega < f$). Similarly, the range of admissible frequencies is extended above $N_{\text{a,b}}=\bar{N}$, so that the band of perfect transmission for $\omega \approx \omega_+$ reaches higher values of $k_{\perp}d$ with increasing rotation rates.

The features described above only arise when the complete Coriolis force is taken into account, and therefore they are intrinsically caused by non-traditional effects. We will give them a physical interpretation in Section \ref{subsec:FREE:t_physicalInterpretation}.

\begin{figure*}
\centering
\begin{subfigure}{0.31\textwidth}
\centering
\caption{$\Omega = 0.2\bar{N}$}
\includegraphics[height=5.55cm]{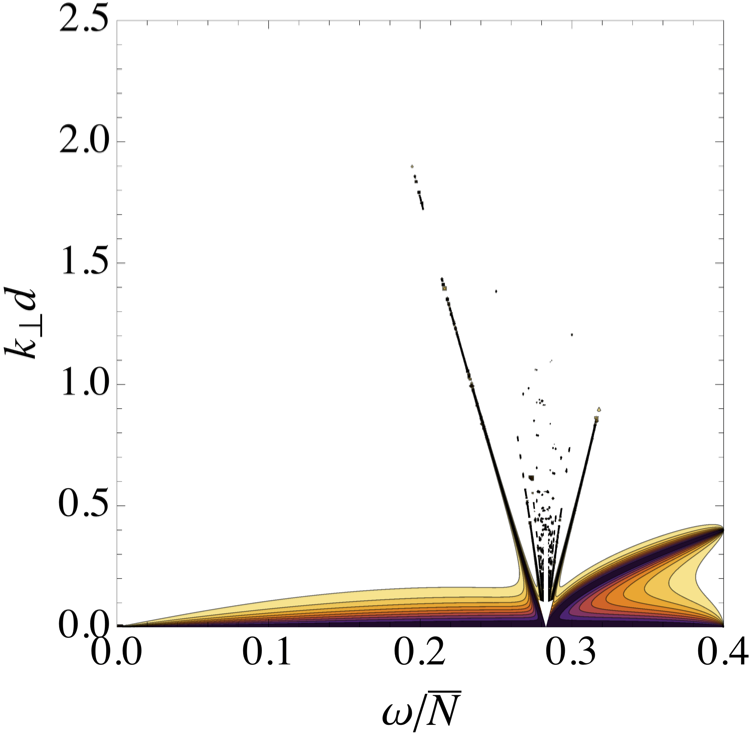}
\end{subfigure}
\begin{subfigure}{0.31\textwidth}
\centering
\caption{$\Omega = 0.4\bar{N}$}
\includegraphics[height=5.55cm]{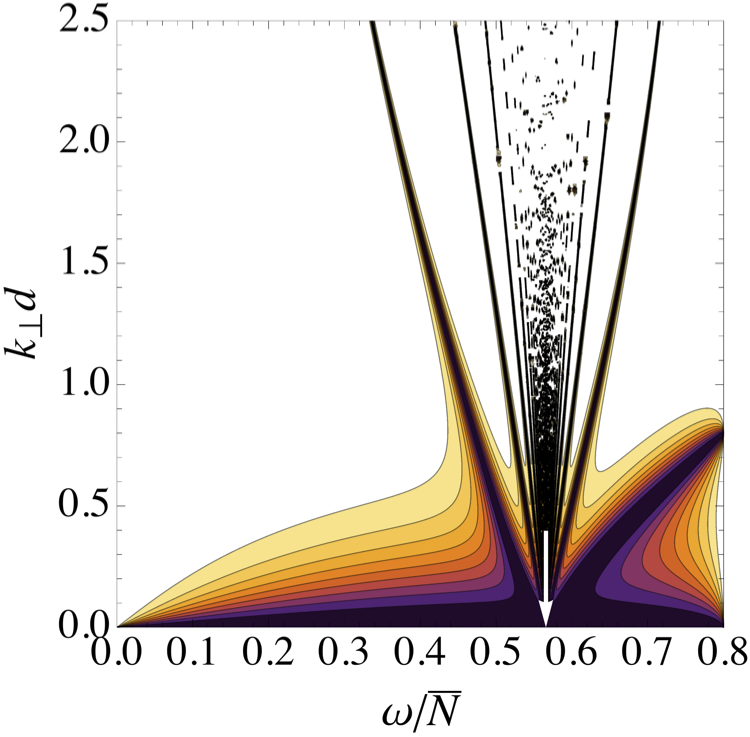}
\end{subfigure}
\begin{subfigure}{0.36\textwidth}
\centering
\caption{$\Omega = 0.6\bar{N}$}
\includegraphics[height=5.55cm]{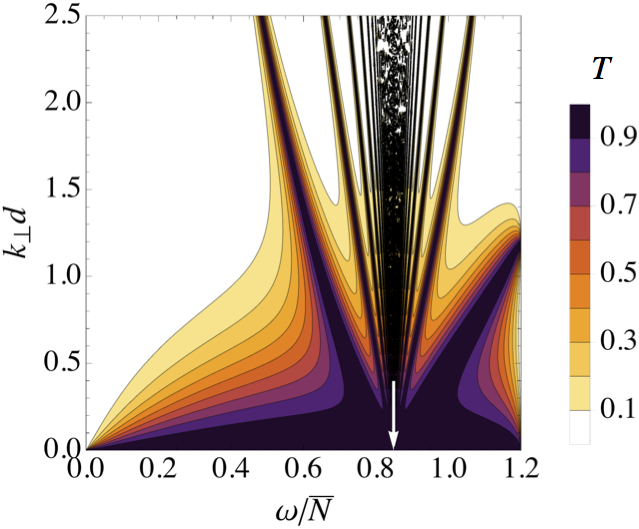}
\end{subfigure}
\caption[Transmission coefficient -- one step embedded in a convective medium with rotation]{Same as Fig. \ref{fig:onestep_RRrotation}, but for one step embedded in a convective medium. The frequency domain extends from $0$ to $2\Omega$.}
\label{fig:onestep_CCrotation}
\centering
\begin{subfigure}{0.31\textwidth}
\centering
\caption{$\Omega = 0.2\bar{N}$}
\includegraphics[height=5.55cm]{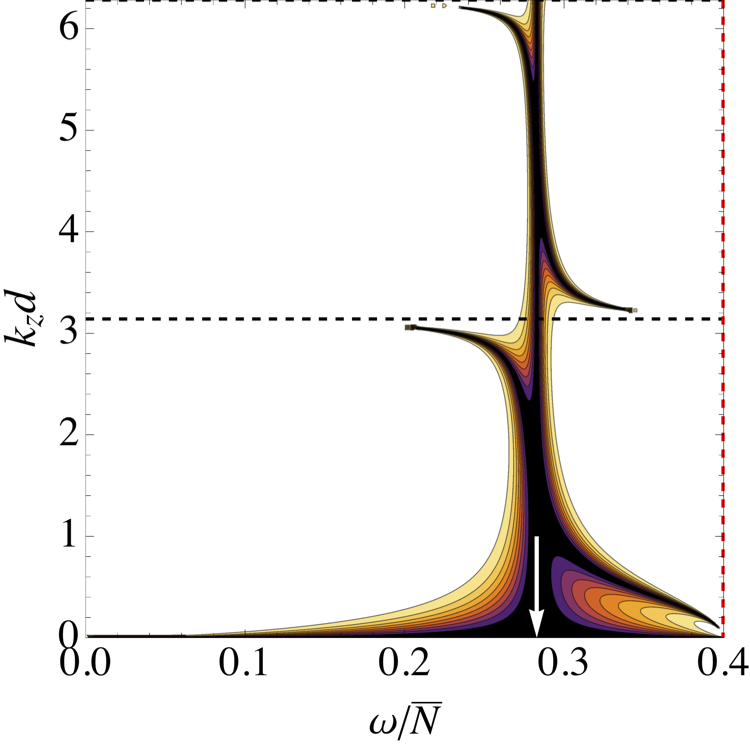}
\end{subfigure}
\begin{subfigure}{0.31\textwidth}
\centering
\caption{$\Omega = 0.4\bar{N}$}
\includegraphics[height=5.55cm]{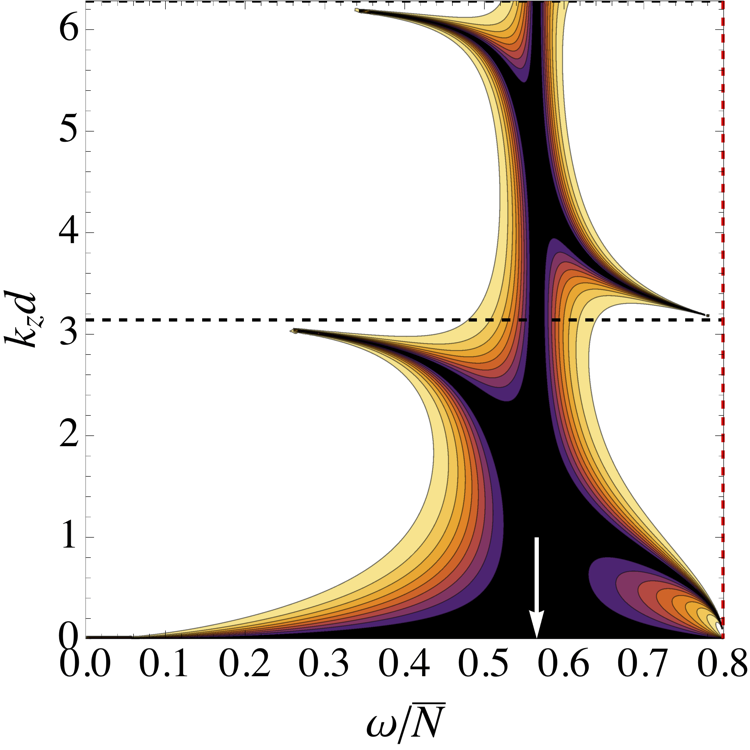}
\end{subfigure}
\begin{subfigure}{0.36\textwidth}
\centering
\caption{$\Omega = 0.6\bar{N}$}
\includegraphics[height=5.55cm]{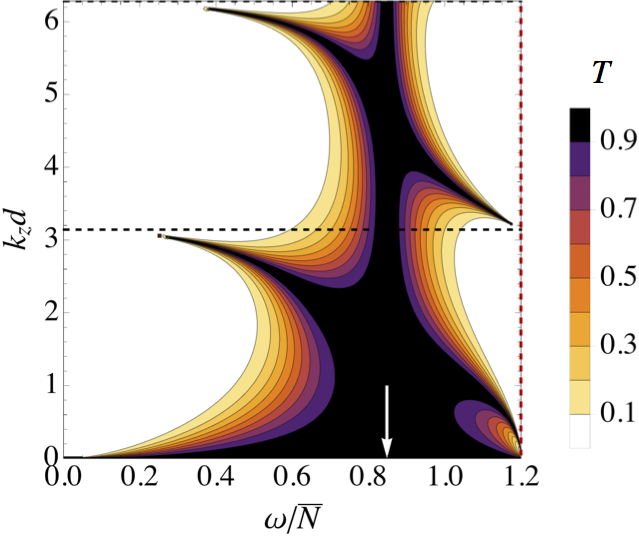}
\end{subfigure}
\caption[Transmission coefficient -- one step embedded in a convective medium with rotation]{Same as Fig. \ref{fig:onestep_RRrotationKZ}, but for one step embedded in a convective medium. The frequency domain extends from $0$ to $2\Omega$.}
\label{fig:onestep_CCrotationKZ}
\end{figure*}

\begin{figure*}
\centering
\begin{subfigure}{0.31\textwidth}
\centering
\caption{$\Omega = 0.2\bar{N}$}
\includegraphics[height=5.55cm]{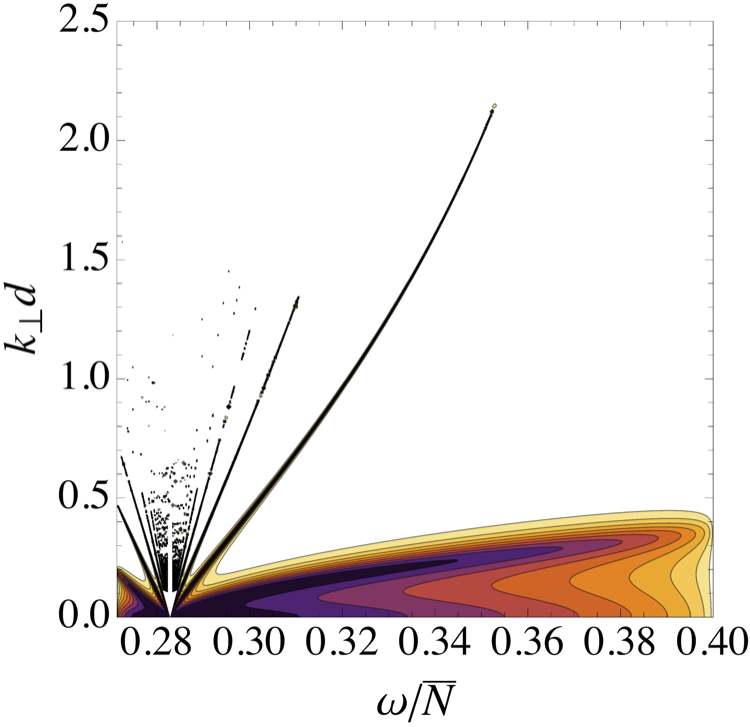}
\end{subfigure}
\begin{subfigure}{0.31\textwidth}
\centering
\caption{$\Omega = 0.4\bar{N}$}
\includegraphics[height=5.55cm]{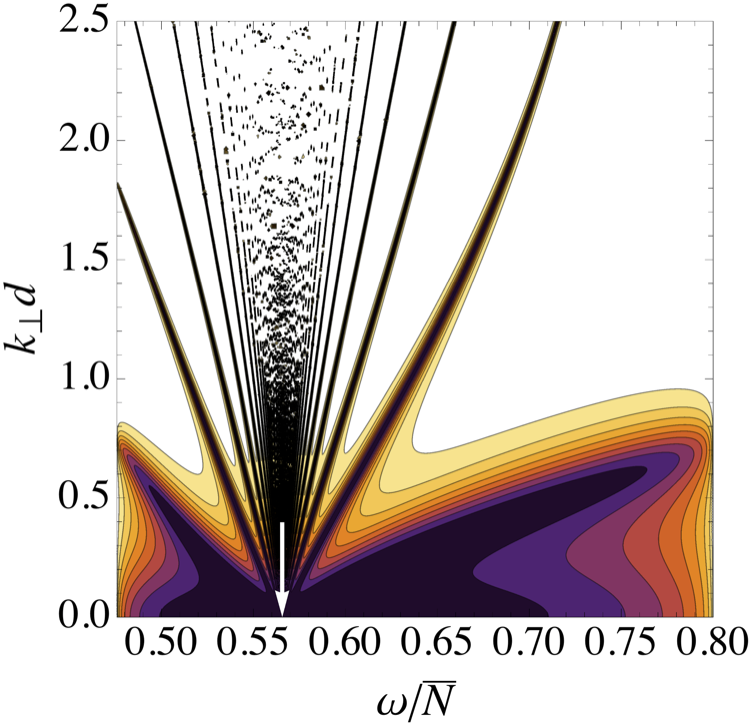}
\end{subfigure}
\begin{subfigure}{0.36\textwidth}
\centering
\caption{$\Omega = 0.6\bar{N}$}
\includegraphics[height=5.55cm]{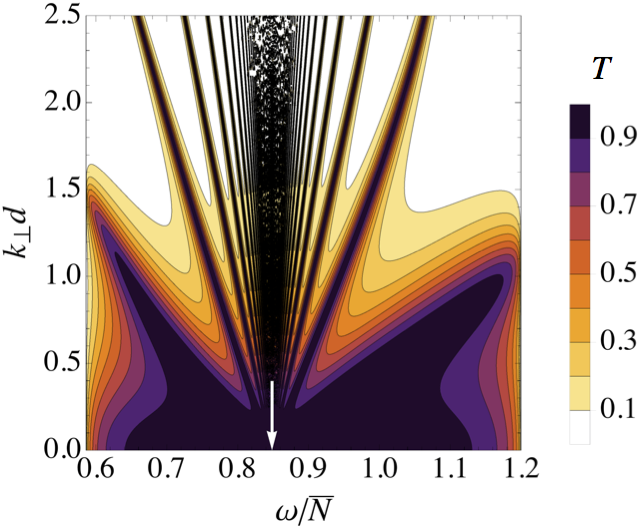}
\end{subfigure}
\caption[Transmission coefficient -- one step embedded in a stably stratified medium at the bottom and a convective medium at the top, with rotation]{Same as Fig. \ref{fig:onestep_RRrotation}, but for one step embedded in a convective medium at the top and a stably stratified medium at the bottom.
}
\label{fig:onestep_CRrotation}
\centering
\begin{subfigure}{0.31\textwidth}
\centering
\caption{$\Omega = 0.2\bar{N}$}
\includegraphics[height=5.55cm]{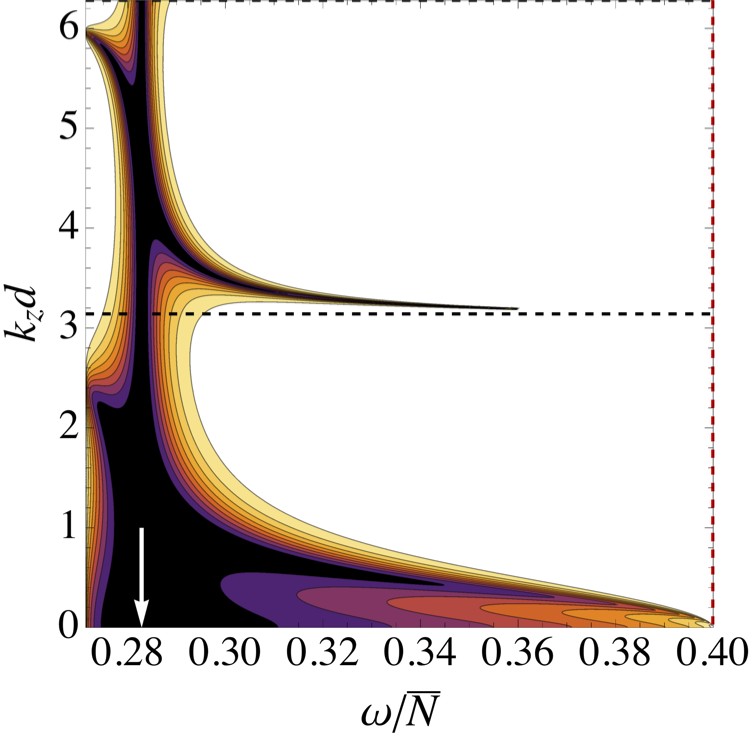}
\end{subfigure}
\begin{subfigure}{0.31\textwidth}
\centering
\caption{$\Omega = 0.4\bar{N}$}
\includegraphics[height=5.55cm]{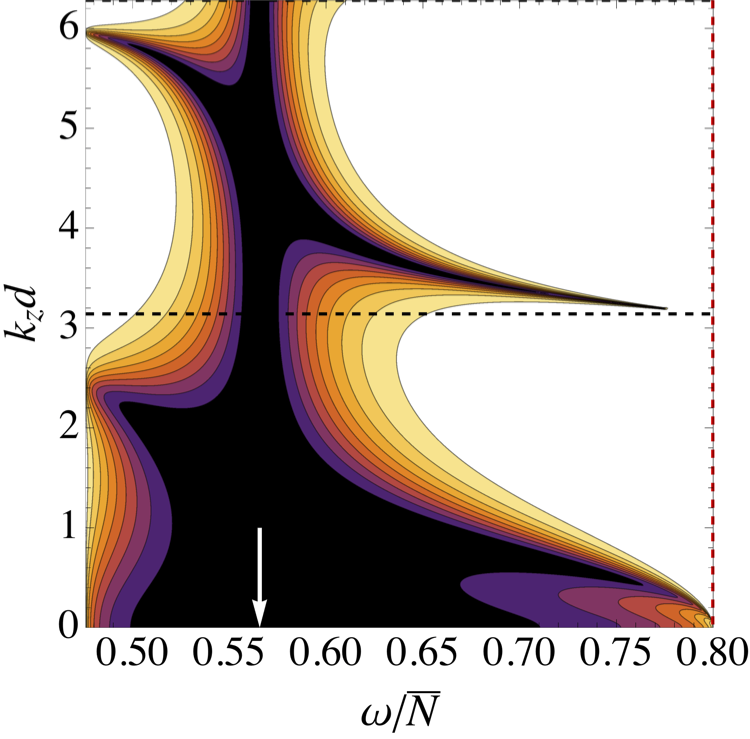}
\end{subfigure}
\begin{subfigure}{0.36\textwidth}
\centering
\caption{$\Omega = 0.6\bar{N}$}
\includegraphics[height=5.55cm]{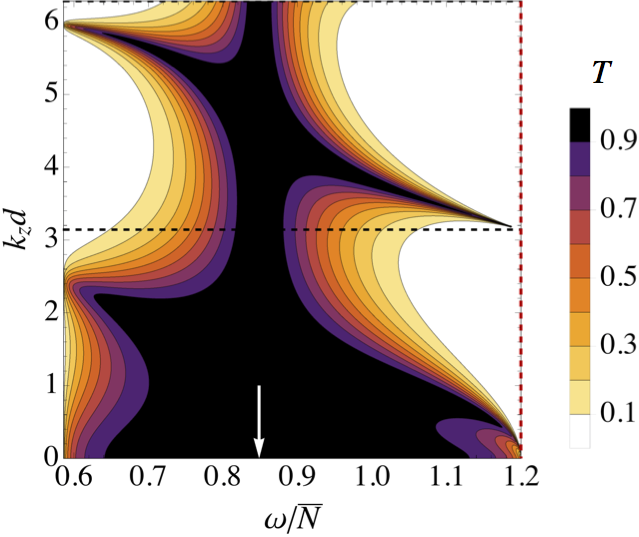}
\end{subfigure}
\caption[Transmission coefficient -- one step embedded in a stably stratified medium at the bottom and a convective medium at the top, with rotation]{Same as Fig. \ref{fig:onestep_RRrotationKZ}, but for one step embedded in a convective medium at the top and a stably stratified medium at the bottom. 
}
\label{fig:onestep_CRrotationKZ}
\end{figure*}

\paragraph{Staircase embedded in a convective medium.}\label{subsec:FREE:t_conv}\hfill\break
Now, we consider a staircase embedded in a convective medium, which is probably more relevant to a portion of a giant planet interior in which double-diffusive convection operates. In this case, the incident and transmitted waves are pure inertial waves. Considering a staircase embedded in a convective medium can also be seen as considering a portion of a vertically (more) extended staircase.  

In the special case of a single convective layer embedded in convective medium ($m=1, N_{\text{a}}=N_{\text{b}}=0$), the expression of the transmission coefficient, $T$ given by Eq. (\ref{eq:Tcoeff_msteps_analytic}) takes the following analytical expression:
\begin{equation}
T = \dfrac{1}{1+4\Gamma^2\left[\cos(k_zd) - \Gamma\sin(k_zd)\right]^2},
\label{eq:Tanalytical_CC}
\end{equation}
where $k_z$ is the vertical wave number of inertial waves,
\begin{equation}
k_z = k_{\perp}\sqrt{\frac{\omega^2(4\widetilde{\Omega}^2-\omega^2)}{\left(\omega^2 - f^2\right)^2}}.
\end{equation}
From this formula, it is obvious that $T$ lies between $0$ and $1$, as we expect for a transmission coefficient.

In this case, the range of frequency over which the transmission coefficient is calculated is the range of frequencies over which inertial waves are propagative in a convective medium, i.e. between $0$ and $2\widetilde{\Omega}$ (see Eq. (\ref{eq:Omega_s})). In the case where $\alpha=\uppi/2$ that we consider here, $\widetilde{\Omega}=\Omega$. The results are displayed on Fig. \ref{fig:onestep_CCrotation} for different rotation rates $\Omega=0.2\bar{N}$, $\Omega=0.4\bar{N}$, $\Omega=0.6\bar{N}$, and the colatitude $\Theta=\uppi/4$.
Again, we get perfect transmission for large enough wavelengths (small wave numbers), and a set of bands depart from $\omega=f$ ($\approx 0.28\bar{N}$ for $\Omega=0.2\bar{N}$, $\approx 0.57\bar{N}$ for $\Omega=0.4\bar{N}$, and $\approx 0.85\bar{N}$ for $\Omega=0.6\bar{N}$). What differs from the previous case is that transmission stays close to unity for higher and higher wave numbers as rotation is increased, which made us choose a range in $k_{\perp}d$ from $1.0$ to $2.5$ for $\Omega=0.2\bar{N}$ and $\Omega=0.6\bar{N}$, in order to observe this effect.

Fig. \ref{fig:onestep_CCrotationKZ} shows the behaviour of the transmission coefficient as a function of the vertical wave number, $k_z$. Again, we find perfect transmission near $\omega=f$ (white arrow), and a set of bands of perfect transmission are departing from this frequency to reach lines of $k_zd=0,\uppi,2\uppi,\dots$ (black dashed lines) for any vertical wave-length.

\paragraph{Staircase embedded in convective medium at the top and stably stratified medium at the bottom.}\hfill\break
The boundary conditions of the staircase are changed once again to consider a staircase embedded in a convective medium at the top and a stably stratified medium at the bottom. Thus, the incident wave is a pure inertial wave, and the transmitted wave is a gravito-inertial wave. This situation is interesting to consider for astrophysical applications because density staircases might develop in the deep interiors of giant planets where there could be a stably stratified region just outside the core \citep{Fuller2014}, on top of which could sit a region of layered semi-convection and associated density staircases.

This is found to be equivalent to the opposite situation where $N_{\text{a}}$ and $N_{\text{b}}$ are interchanged, which could correspond to a stably stratified layer near the surface of a hot Jupiter matching the convective envelope of deeper regions through a region of layered semi-convection. The fact that both cases are identical likely results from the symmetry of the Boussinesq equations under the transformation $\{z \rightarrow -z, \Theta \rightarrow -\Theta, b \rightarrow -b, w \rightarrow -w\}$ (i.e. 'up $\rightarrow$ down', 'hot $\rightarrow$ cold').

The results are displayed on Fig. \ref{fig:onestep_CRrotation} and \ref{fig:onestep_CRrotationKZ}, as a function of $k_{\perp}d$ and $k_zd$, respectively, for different rotation rates $\Omega=0.2\bar{N}$, $\Omega=0.4\bar{N}$, $\Omega=0.6\bar{N}$. The behaviour of the transmission coefficient is similar to the two previous cases. One difference that arises, however, concerns the band of perfect transmission corresponding to resonance with interfacial gravity waves on either side of the convective step, which do not extend to $\omega_+$ like in the two previous cases.

\subsubsection{Transmission across $m$ steps}
\label{subsec:FREE:t_msteps}
Density staircases that might exist in giant planet interiors are expected to have a large number of steps, {with each step being} much smaller than the planet’s radius \citep{LeconteChabrier2012}. Therefore, we now study the effects of having more than one step. We choose again to perform our calculations at a colatitude $\Theta=\uppi/4$, for illustration and comparison with the results of the previous section. The transmission coefficient is given by Eq. (\ref{eq:Tcoeff_msteps_analytic}), and we checked that our results agreed with \citetalias{Sutherland2016} assuming the traditional approximation, and for similar parameters. We recall that here, unlike in \citetalias{Sutherland2016}, the complete Coriolis acceleration is taken into account.

\begin{figure*}
\centering
\begin{subfigure}{0.31\textwidth}
\centering
\caption{$m = 2$}
\includegraphics[height=5.6cm]{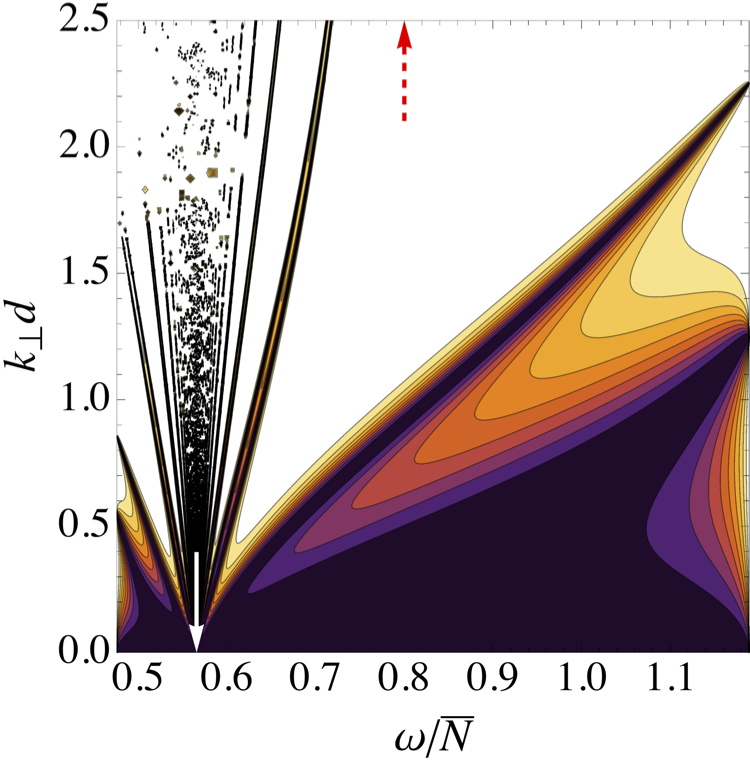}
\end{subfigure}
\begin{subfigure}{0.31\textwidth}
\centering
\caption{$m=5$}
\includegraphics[height=5.6cm]{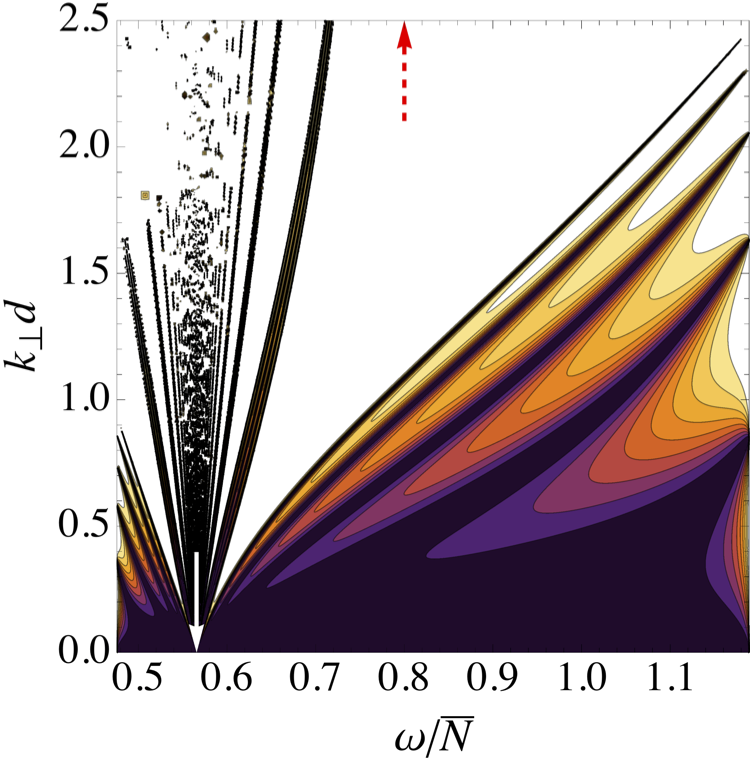}
\end{subfigure}
\begin{subfigure}{0.36\textwidth}
\centering
\caption{$m=10$}
\includegraphics[height=5.6cm]{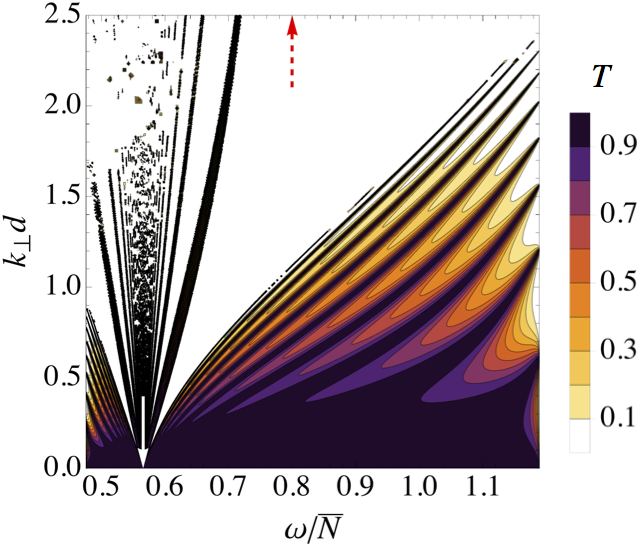}
\end{subfigure}
\centering
\vspace{2mm}
\begin{subfigure}{0.312\textwidth}
\centering
\includegraphics[height=5.7cm]{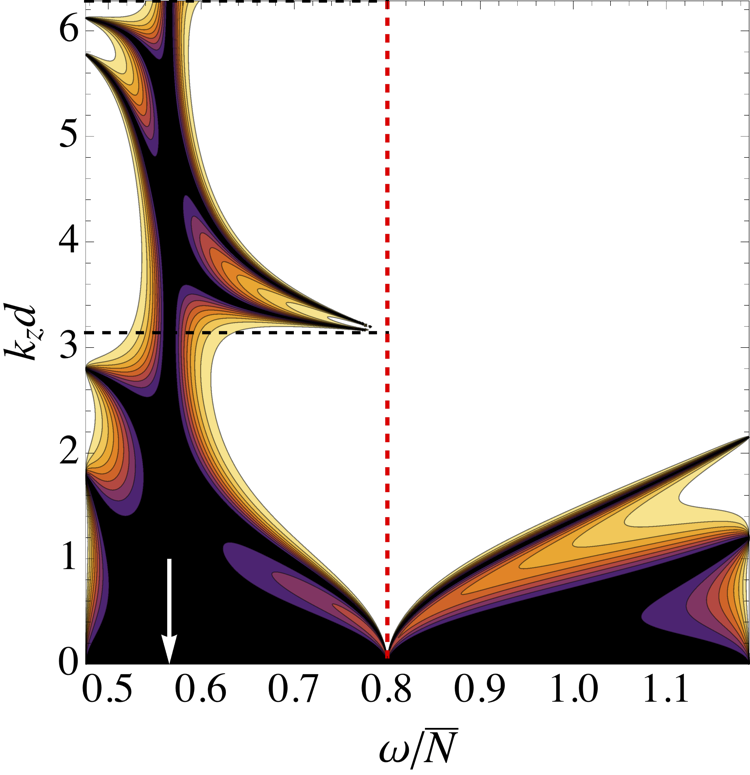}
\end{subfigure}
\begin{subfigure}{0.312\textwidth}
\centering
\vspace{2mm}
\includegraphics[height=5.7cm]{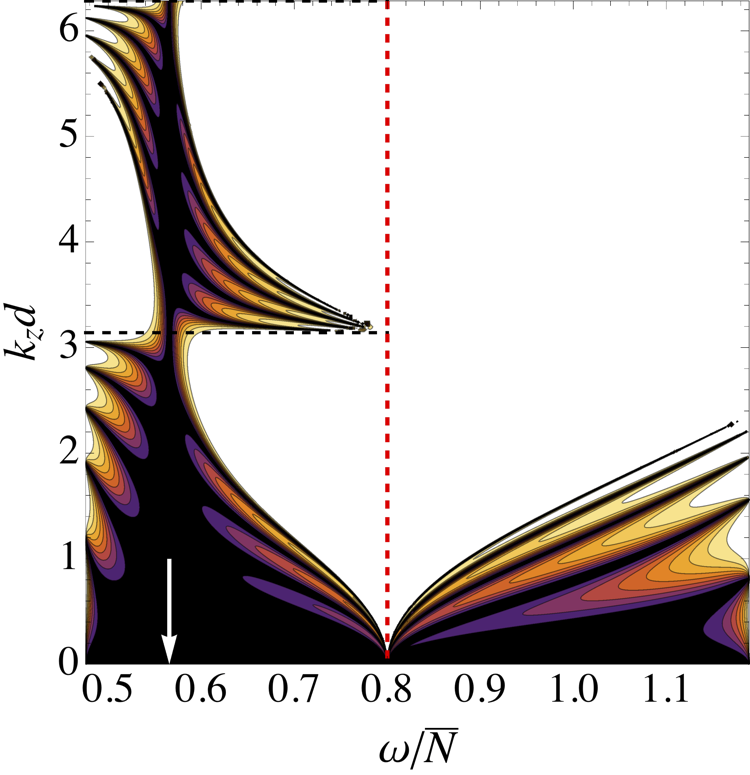}
\end{subfigure}
\begin{subfigure}{0.356\textwidth}
\centering
\vspace{2mm}
\includegraphics[height=5.7cm]{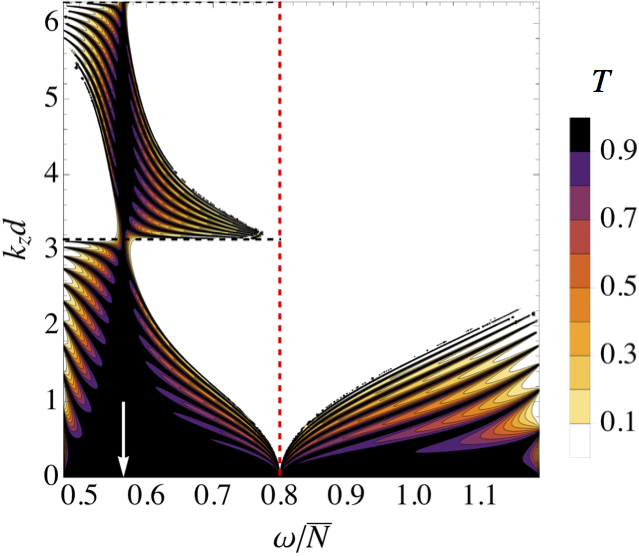}
\end{subfigure}
\caption{Transmission coefficient for a staircase embedded in a stably stratified medium, in an inclined box with $\Omega=0.4\bar{N}$ and $\Theta=\uppi/4$, for different number of steps (a) $m=2$, (b) $m=5$ and (c) $m=10$, as a function of dimensionless frequency and horizontal wave number (upper panel) or vertical wave number (bottom panel). The white arrow indicates the frequency $\omega=f$. A set of bands extending vertically depart from $f=2\Omega\cos\Theta$ ($\approx 0.57\bar{N}$). In addition, $m$ bands of perfect transmission appear in the regions $\omega<f$ and $\omega>f$, each becoming narrower with increasing $m$.}
\label{fig:T_mstepsRR}
\end{figure*}

Fig. \ref{fig:T_mstepsRR} shows the results for $m=2$, $5$ and $10$ steps embedded in a stably stratified medium with $\Omega=0.4\bar{N}$ (upper panels) and in a convective medium with $\Omega=0.6\bar{N}$ (bottom panels). The upper and bottom panels can directly be compared to Figs. \ref{fig:onestep_RRrotation}b and \ref{fig:onestep_CCrotation}c, respectively, on which is displayed the corresponding single step case with same colatitude and rotation rates. 

In the regions $\omega>f$ and $\omega<f$, it will be seen that as the number of steps increases, more and more transmission peaks appear. These are particularly visible for $\omega \lesssim \omega_+$ ($\omega \gtrsim \omega_-$, respectively), and propagate back (forward, respectively) to $\omega = f$. Confirming the results of \citetalias{Sutherland2016} in the case of a staircase embedded in a stably stratified medium, we find that if there are $m$ steps, the number of peaks equals $m$. This will be explained in the following section. Also, as the numbers of steps increases, those bands of perfect transmission become narrower, and transmission becomes inhibited for larger and larger wavelengths ($k \rightarrow 0 \Rightarrow \lambda \rightarrow \infty$).

\begin{figure*}
\centering
\begin{subfigure}{0.31\textwidth}
\centering
\caption{$m = 2$}
\includegraphics[height=5.6cm]{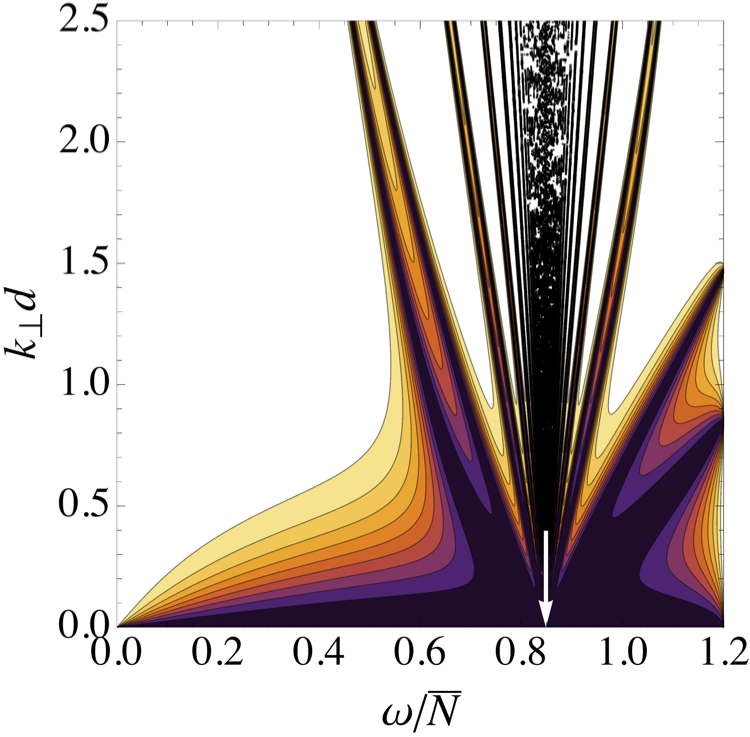}
\end{subfigure}
\begin{subfigure}{0.31\textwidth}
\centering
\caption{$m = 5$}
\includegraphics[height=5.6cm]{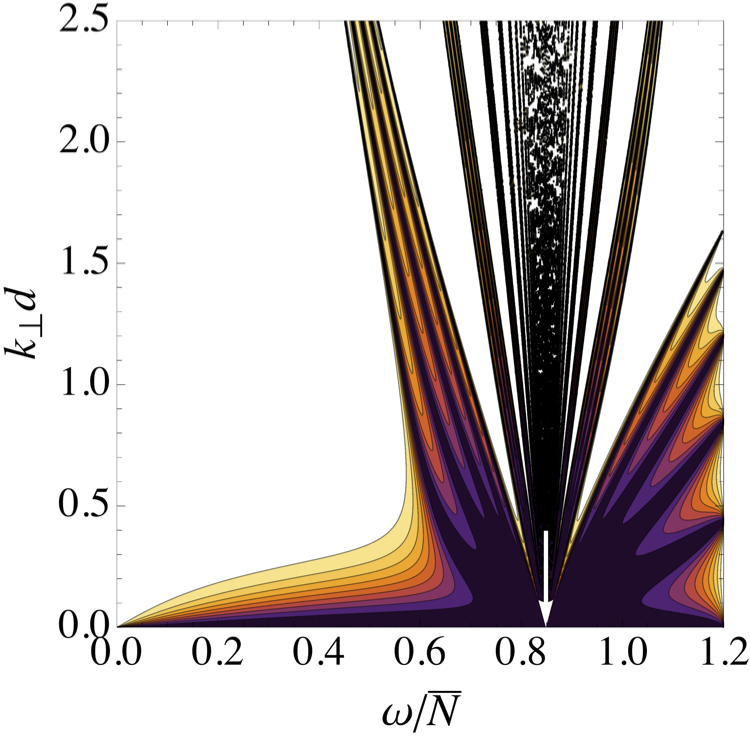}
\end{subfigure}
\begin{subfigure}{0.36\textwidth}
\centering
\caption{$m = 10$}
\includegraphics[height=5.6cm]{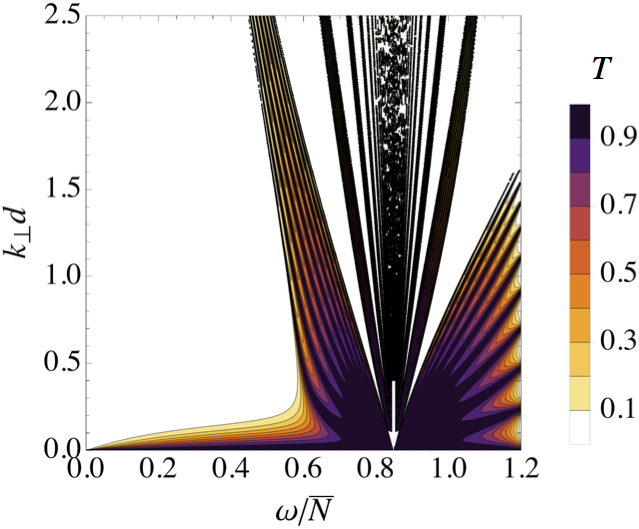}
\end{subfigure}
\centering
\begin{subfigure}{0.312\textwidth}
\centering
\vspace{2mm}
\includegraphics[height=5.7cm]{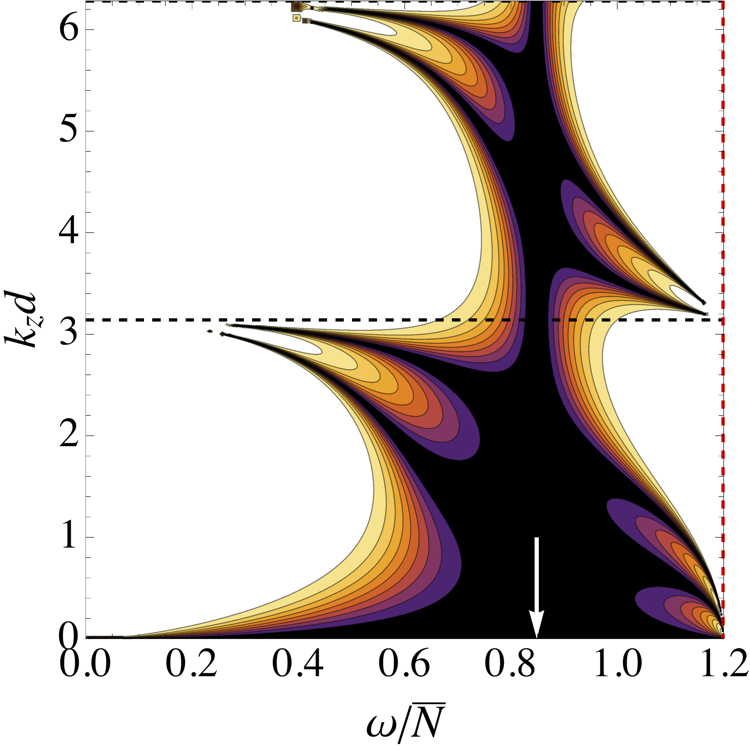}
\end{subfigure}
\begin{subfigure}{0.312\textwidth}
\centering
\vspace{2mm}
\includegraphics[height=5.7cm]{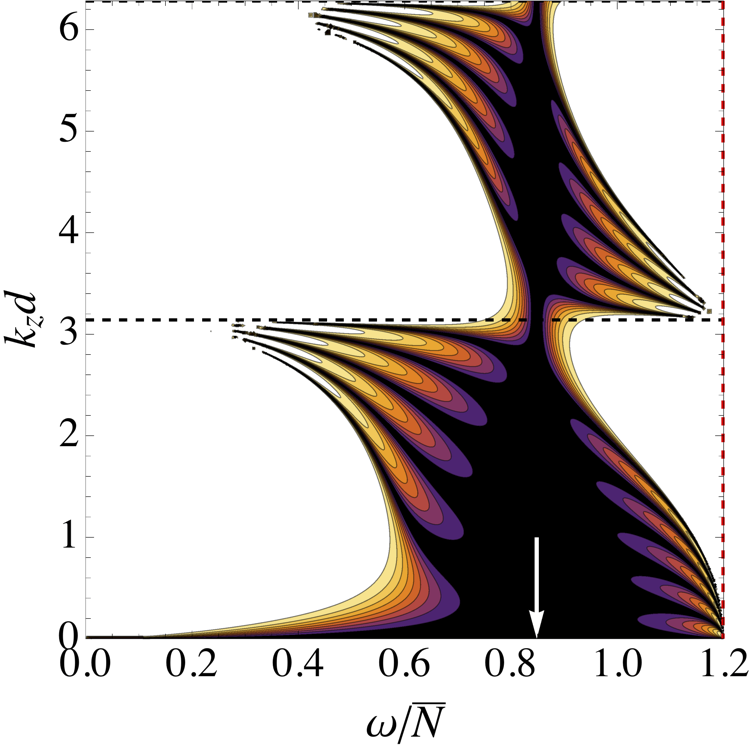}
\end{subfigure}
\begin{subfigure}{0.356\textwidth}
\centering
\vspace{2mm}
\includegraphics[height=5.7cm]{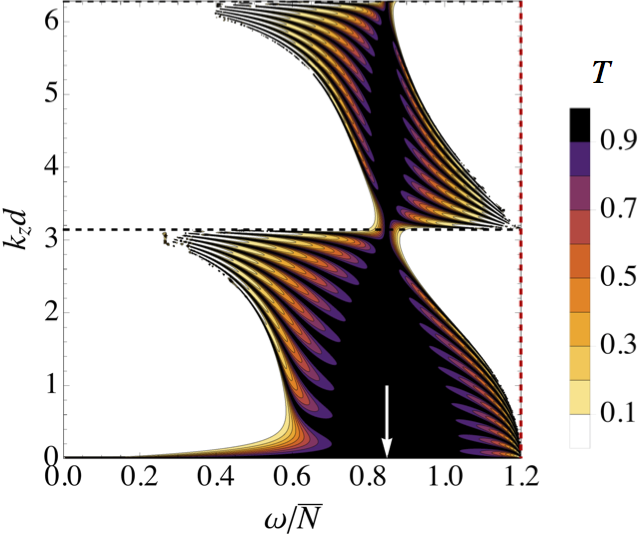}
\end{subfigure}
\caption[Transmission coefficient -- multiples steps embedded in a stably stratified and a convective medium]{Same as Fig. \ref{fig:T_mstepsRR}, but for a staircase embedded in a convective medium, with a rotation rate $\Omega=0.6\bar{N}$ ($f \approx 0.85\bar{N}$).}
\label{fig:T_mstepsCC}
\end{figure*}

As a function of the vertical wave number (see the bottom panel of Fig. \ref{fig:T_mstepsRR}), those observations hold. Namely, in each region of enhanced transmission previously described, one can see $m$ bands of perfect transmission for a staircase with $m$ steps.

\subsubsection{Understanding the transmission coefficient}
\label{subsec:FREE:t_physicalInterpretation}

\paragraph{Perfect transmission at $\omega=f$.}\hfill\break
We have seen that the transmission is perfect at the Coriolis frequency $\omega=f$ for almost any wavelength (see e.g. Fig. \ref{fig:onestep_CCrotationKZ}). This result can be interpreted by noting that when $\omega=f$ there is only one direction for wave propagation that is allowed by the dispersion relation that is not parallel to the interface (see Fig. \ref{fig:IWgroupVelocityDirection}: the group velocity $\bm{v}_{\text{g},-}$ is directed along the perpendicular to the $z$-axis), so there is only one wave solution with non-zero vertical velocity. This means that the boundary condition at the interface that matches the vertical velocity between the incoming and transmitted waves (see Eq. (\ref{eq:BC1_finite})) requires the transmitted wave to have the same amplitude, so that $T=1$.
Furthermore, from the separate calculation we have made in Section \ref{sec:FREE:maths} for $\omega=f$ (see Eq. \ref{eq:E0EqEmp1}), it is straightfoward that $T=|\mathcal{E}_{m+1}/\mathcal{E}_0|^2=1$ (at least when the stratification is the same above and below the staircase).

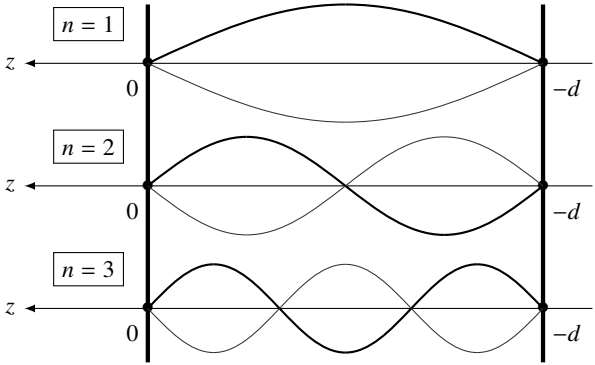
\begin{figure}
\centering
\begin{tikzpicture}[scale=0.65]
\draw[ultra thick] (2,1.2) -- ++ (0,-7.3);
\draw[ultra thick] (10,1.2) -- ++ (0,-7.3);
\pgfmathsetmacro{\axis}{11}
\draw[axis] (\axis,0) -- ++ (-\axis-0.5,0) node[left]{\footnotesize $z$};
\draw (2,0.15) -- ++ (0,-0.3) node[below left]{\footnotesize $0$};
\draw (\axis-1,0.15) -- ++ (0,-0.3) node[below right]{\footnotesize $-d$};
\node at (2,0){$\bullet$};
\node at (\axis-1,0){$\bullet$};
\node at (0.8,0.8){\footnotesize $\boxed{n=1}$};
\draw[thick] (2,0) sin (6,1.2); 
\draw[thick] (6,1.2) cos (10,0);
\draw[color=black!80] (2,0) sin (6,-1.2); 
\draw[color=black!80] (6,-1.2) cos (10,0);
\pgfmathsetmacro{\step}{-2.5}
\node at (0.8,\step+0.8){\footnotesize $\boxed{n=2}$};
\draw[axis] (\axis,\step) -- ++ (-\axis-0.5,0) node[left]{\footnotesize $z$};
\draw (2,\step+0.15) -- ++ (0,-0.3) node[below left]{\footnotesize $0$};
\draw (\axis-1,\step+0.15) -- ++ (0,-0.3) node[below right]{\footnotesize $-d$};
\node at (2,\step){$\bullet$};
\node at (\axis-1,\step){$\bullet$};
\draw[thick] (2,\step) sin (4,\step+1); 
\draw[thick] (4,\step+1) cos (6,\step);
\draw[thick] (6,\step) sin (8,\step-1); 
\draw[thick] (8,\step-1) cos (10,\step);
\draw[color=black!80] (2,\step) sin (4,\step+-1); 
\draw[color=black!80] (4,\step-1) cos (6,\step);
\draw[color=black!80] (6,\step) sin (8,\step+1); 
\draw[color=black!80] (8,\step+1) cos (10,\step);
\pgfmathsetmacro{\step}{-5}
\node at (0.8,\step+0.8){\footnotesize $\boxed{n=3}$};
\pgfmathsetmacro{\hdt}{2.6666666666666667/2}
\draw[axis] (\axis,\step) -- ++ (-\axis-0.5,0) node[left]{\footnotesize $z$};
\draw (2,\step+0.15) -- ++ (0,-0.3) node[below left]{\footnotesize $0$};
\draw (\axis-1,\step+0.15) -- ++ (0,-0.3) node[below right]{\footnotesize $-d$};
\node at (2,\step){$\bullet$};
\node at (\axis-1,\step){$\bullet$};
\draw[thick] (2,\step) sin (2+\hdt,\step+0.9); 
\draw[thick] (2+\hdt,\step+0.9) cos (2+2*\hdt,\step);
\draw[thick] (2+2*\hdt,\step) sin (2+3*\hdt,\step-0.9); 
\draw[thick] (2+3*\hdt,\step-0.9) cos (2+4*\hdt,\step);
\draw[thick] (2+4*\hdt,\step) sin (2+5*\hdt,\step+0.9); 
\draw[thick] (2+5*\hdt,\step+0.9) cos (2+6*\hdt,\step);
\draw[color=black!80] (2,\step) sin (2+\hdt,\step-0.9); 
\draw[color=black!80] (2+\hdt,\step+-0.9) cos (2+2*\hdt,\step);
\draw[color=black!80] (2+2*\hdt,\step) sin (2+3*\hdt,\step+0.9); 
\draw[color=black!80] (2+3*\hdt,\step+0.9) cos (2+4*\hdt,\step);
\draw[color=black!80] (2+4*\hdt,\step) sin (2+5*\hdt,\step-0.9); 
\draw[color=black!80] (2+5*\hdt,\step-0.9) cos (2+6*\hdt,\step);
\end{tikzpicture}
\caption[Illustration of the enhanced transmission for inertial waves whose wavelength fits exactly inside a convective step]{Transmission is enhanced for incident waves for which a multiple of the vertical semi-wavelength $\lambda_z/2=\uppi/k_z$ matches the vertical semi-wavelength of inertial waves that fits inside one step, i.e. $k_zd = n\uppi$ with $n$ an integer. This is shown for $n=\{1, \dots, 3\}$.}
\label{fig:corde}
\end{figure}

\begin{figure*}
\centering
\begin{subfigure}{0.43\textwidth}
\caption{$k_zd=\text{cst}$}
\label{fig:kzdcst}
\centering
\begin{tikzpicture}[scale=1.08]
\node[anchor=south west,inner sep=0] at (0,0)
    {\includegraphics[height=8.1cm]{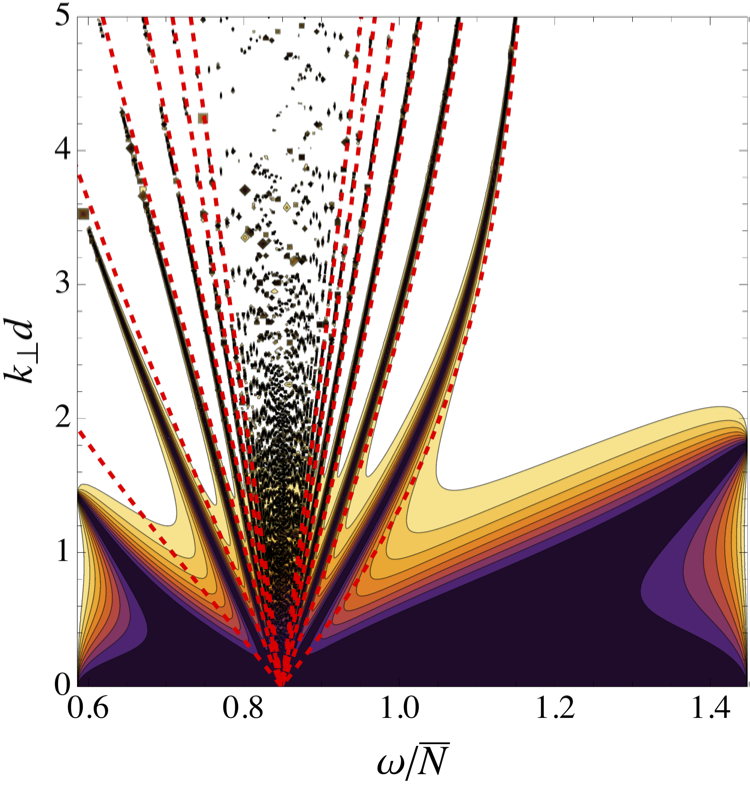}};
\draw[color=red, thick, ->] (5.1,4.5) node[thick, red, right]{$k_zd=\uppi$} --  ++ (-0.56,-0.28);
\draw[color=red, thick, ->] (5.1,5.2) node[thick, red, right]{$k_zd=2\uppi$} --  ++ (-1.12,-0.56);
\draw[color=red, thick, ->] (5.1,5.9) node[thick, red, right]{$k_zd=3\uppi$} --  ++ (-1.36,-0.68);
\node[red] at (6.13,6.5) {$\dots$};
\draw[densely dotted, thick, color=red] (2.69,0.9) -- ++ (0,-0.4) node[color=red, below]{$\boxed{f}$};
\end{tikzpicture}
\end{subfigure}
\begin{subfigure}{0.55\textwidth}
\caption{$m$ steps peaks}
\label{fig:mstepsBQF}
\centering
\includegraphics[height=8.1cm]{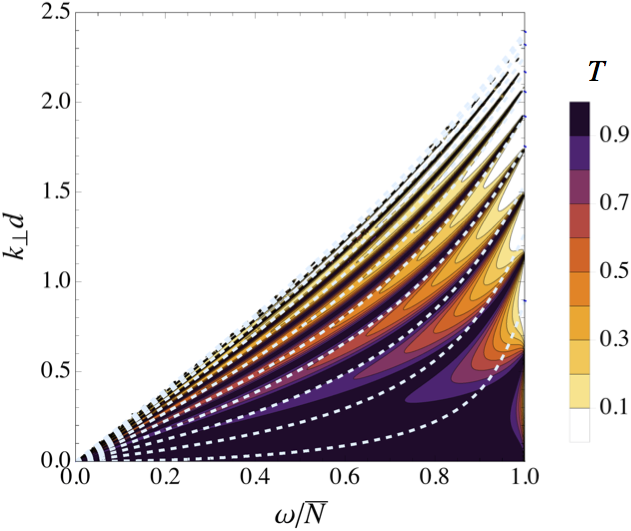}
\vspace{3.5mm}
\end{subfigure}
\caption[Understanding the transmission coefficient]{(a) Same as Fig. \ref{fig:onestep_RRrotation}c, overplotted with curves of equation $k_zd = n\uppi$ for $n=\{1,\dots,6\}$. A set of bands of perfect transmission departs from the frequency $f=2\Omega\cos\Theta$, corresponding with a resonance with short-wavelength inertial waves. (b) The case of 10 steps without rotation, overplotted with the branches of the dispersion relation given by Eq. (\ref{eq:BQFdisprel_finite}). Transmission is enhanced along those branches, corresponding to resonances with free modes of the staircase.}
\end{figure*}

\paragraph{Resonances with short-wavelength inertial modes.}\hfill\break
In Sections \ref{subsec:FREE:t_onestep} and \ref{subsec:FREE:t_msteps}, we have seen that, regardless of the properties of the regions that surround the staircase, we always obtained a set of bands of perfect transmission departing from the inertial frequency $\omega=f$. We interpret those peaks of transmission as being caused by resonances between the incident internal wave and the short-wavelength inertial waves that exist within a convective step. This happens when a multiple of the vertical semi wavelength ($\lambda_z/2$) of the internal waves that propagate inside the step fits inside one step, i.e. for
\begin{equation}
\frac{n\lambda_z}{2} = d,
\end{equation}
where $n$ is an integer. This is illustrated on Fig. \ref{fig:corde} for $n=\{1,2,3\}$.
The condition above can be rewritten in term of the vertical wave number of the inertial wave inside the steps, to read
\begin{equation}
k_zd = n\uppi.
\end{equation}
On Fig. \ref{fig:kzdcst}, curves of equation $k_zd = n\uppi$ (in dashed red) are overplotted to Fig. \ref{fig:onestep_RRrotation}c for $n=\{1, \dots, 6\}$. 
As one can see on this figure, the matching between the bands of transmission and the curves of equation $k_zd = n\uppi$  is satisfying. The largest discrepancy is obtained for the peak of transmission in the bottom left corner of the figure, which corresponds to very low frequency.

\paragraph{Excitation of free gravity modes of the staircase.}\hfill\break
Other features we give a physical interpretation for are the bands of perfect transmission, {more noticeable for the $m>1$ steps case} with $m$ bands of perfect transmission starting from $\omega=\omega_+$ and $k_{\perp}d$ of order unity, and propagating back to $\omega=f$ and $k_{\perp}d \ll 1$ (see Fig. \ref{fig:T_mstepsRR}). In Section \ref{sec:FREE:disprel}, we have derived dispersion relations, whose branches give us the free modes of oscillation of the staircase. Therefore, incident waves with a prescribed frequency can resonate with a free mode of oscillation of the staircase if its frequency is such that it matches {a root of} the dispersion relation given by Eq. (\ref{eq:BQFdisprel_finite}) (in the case of a finite staircase). 

Since it is very challenging to extract the roots of the dispersion relation with rotation (in particular because there are infinitely many solutions near $\omega=f$, which we have just identified), we have focused on the case without rotation, for which the dispersion relation given by (\ref{eq:BQFdisprel_finite}) gives us a direct relation between $\omega/\bar{N}$ and $k_{\perp}d$. On Fig. \ref{fig:mstepsBQF}, the transmission coefficient in the case of $10$ steps without rotation is displayed, together with the branches of the corresponding dispersion relation. As one can see, the bands of perfect transmission are matched satisfyingly well by the branches of the dispersion relation (in light blue). They correspond to incident waves that excite a free mode of oscillation of the staircase. In principle, the same could be done for the case with rotation, the only reason  this has not been done here being that it is hard to extract all of the roots of the dispersion relation, but we have {nevertheless} identified the branches near $\omega=f$ as explained above.

\subsubsection{Non-uniform step sizes}\label{subsec:FREE:t_NUstepSizes}
In this section, we relax the assumption of having equally sized steps.
Observed density staircases in the ocean have approximately equally-sized steps with a typical length-scale, $d \sim 2.5 ~ \text{m}$ \citep[see][]{GhaemsaidiEtal2016}, which determines the vertical extent of the convective steps. We might expect density staircases in giant planet interiors also to have approximately equally-sized steps, though this is not clear from theory or observations. However, we can expect \citep[as is observed in the case of the ocean, see][]{GhaemsaidiEtal2016} some {inherent} variations in the vertical extent of each step around a mean value $d$. To model them, we adopt the same procedure as in \citetalias{Sutherland2016}: inside the staircase, the steps have a vertical extent
\begin{equation}
d_n = d(1+\epsilon\sigma_n),
\end{equation}
where, for $1\leq n<m$, $\sigma_n$ are random numbers between $-1$ and $1$, $\epsilon$ being the amplitude of the fluctuations of the vertical extent of the convective layers about the mean value $d$. Typically, we set $0\leq\epsilon<0.5$, the value $\epsilon=0$ corresponding to equally spaced interfaces, i.e. our reference model. It is interesting to see how the value of the parameter $\epsilon$ affects the results that have been described so far, because having uneven steps is a more realistic situation. The transmission coefficient can be calculated  numerically, and is given by Eq. (\ref{eq:Tcoeff_msteps_analytic}), where $\widetilde{\mathbfsf{T}}^{m-1}$ is now given by
\begin{equation}
\prod_{n=1}^{m-1}\left[
\begin{array}{cc}
      \left(1+\Gamma\right)\Delta^{-1} & \Gamma\Delta^{-2\epsilon\sigma_n}\\[3mm]
      -\Gamma\Delta^{2\epsilon\sigma_n} & \left(1-\Gamma\right)\Delta
\end{array}
\right]\text{e}^{-\text{i}\tilde{\varphi}_n}
,
\label{eq:matrix_uneven}
\end{equation}
where $\tilde{\varphi}_n = \tilde{\delta}k_{\perp}d_n$. The matrix given by Eq. (\ref{eq:matrix_uneven}) reduces to $\widetilde{\mathbfsf{T}}^{m-1}$ when $\epsilon=0$.

\begin{figure*}
\centering
\begin{subfigure}{0.312\textwidth}
\centering
\caption{$\epsilon = 0$}
\includegraphics[height=5.7cm]{RR_ThetaPi4Omega04m5_Over2.png}
\end{subfigure}
\begin{subfigure}{0.312\textwidth}
\centering
\caption{$\epsilon = 0.1$}
\includegraphics[height=5.7cm]{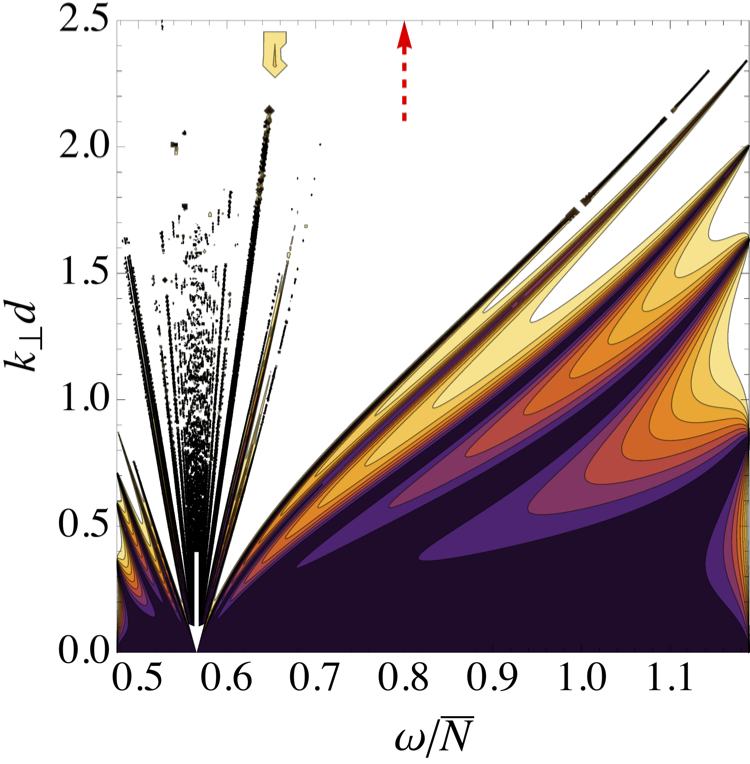}
\end{subfigure}
\begin{subfigure}{0.356\textwidth}
\centering
\caption{$\epsilon = 0.5$}
\includegraphics[height=5.7cm]{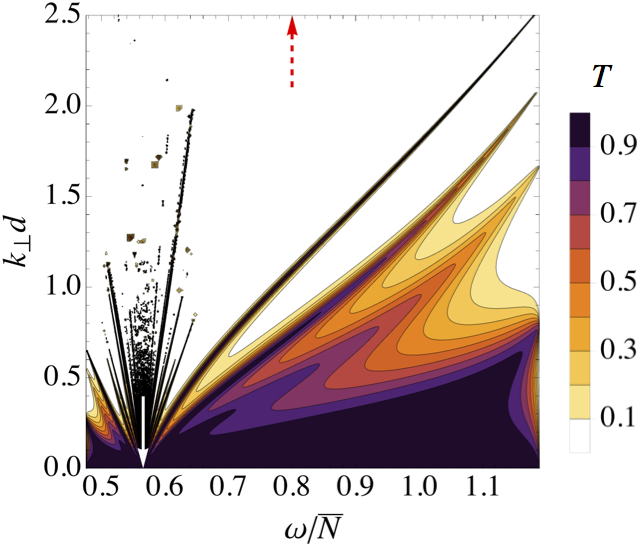}
\end{subfigure}
\caption[Transmission coefficient -- non uniform step sizes]{(a) The even steps case and two uneven steps cases for one particular realisation with (b) $\epsilon=0.1$ and (c) $\epsilon=0.5$. These results have been obtained for 5 steps embedded in a stably stratified medium, with $\Omega=0.4\bar{N}$ and $\Theta=\uppi/4$. {The white and red dashed arrows indicate $\omega=f$ and $\omega=2\Omega$, respectively.}}
\label{fig:Tuneven}
\end{figure*}

The results obtained in the case of five steps ($m=5$) for $\Omega=0.4\bar{N}$, $\Theta=\uppi/4$ and different values of $\epsilon= \{0, 0.1, 0.5\}$ are displayed on Fig. \ref{fig:Tuneven}.
The left panel ($\epsilon=0$) is used as a test of the ability of the method described above to reproduce the analytical solution given by Eq. (\ref{eq:Tcoeff_msteps_analytic}), and is useful for comparison with cases which have $\epsilon>0$. The cases with $\epsilon=0.1$ and $\epsilon=0.5$ show no difference at large wavelengths, so that the perfect transmission for wavelengths $\lambda\gg d$ is a robust result.

However, some differences arise for larger wave numbers {(i.e. smaller wavelengths)}. In particular, the bands of perfect transmission departing from $\omega = f$ weaken with increasing $\epsilon$ (especially the rightmost one). These bands corresponding to half multiple of the wavelength fitting perfectly inside a step (as explained in Section \ref{subsec:FREE:t_physicalInterpretation}), we expect this mechanism to be less effective when the steps are no longer even, since these bands are due to resonances with waves that strongly depend on the step-size. Thus, this observation is perfectly in agreement with the physical interpretation of the bands near the inertial frequency.

The rest of the figure does not differ much from the even step case until the value $\epsilon = 0.5$ is reached (right panel), for which it becomes obvious that there is less transmission near $\omega=\omega_+$. Some modulation of the placement and intensity of the perfect transmission bands between $f$ and $\omega_+$ can also be seen. This means that free modes of the staircase, given by the dispersion relation, are weakened when the steps are no longer even. This is expected to vary for each random realisation, {which we have indeed verified}.

\subsection{Extension of the model: finite size interfaces}\label{sec:FREE:t_FiniteSizeInterface}
\subsubsection{Physical set up}\label{subsec:FREE:FSI_setup}
Now, we relax the assumption of having discrete interfaces. We denote by $l$ the vertical extent of the stably stratified interfaces, and define the {aspect ratio}
\begin{equation}
\varepsilon = \frac{l}{d},
\label{eq:aspect_ratio}
\end{equation}
where $d$ is still the vertical extent of the convective layers. The staircase thus consists of a succession of patterns containing a stably stratified and a convective layer (in red and orange respectively, see Fig. \ref{fig:tcoeff_modele_CC_ThickInterfaces}), of total size $L = l+d$. The total vertical extent of the staircase is $D = mL+l$. This model is of course more realistic since we naturally expect the stably stratified interfaces to have non-zero vertical extent.

\begin{figure*}
\centering
\begin{subfigure}[b]{0.3\textwidth}
\centering
\caption{General scheme}
\begin{tikzpicture}[scale=0.8]
	\fill[bottom color=orange!40, top color=white]
	(0,8.25) -- (4.5,8.25) -- (4.5,10) -- (0,10);
	\fill[color=red!60]
	(0,8.25) -- (4.5,8.25) -- (4.5,7.45) -- (0,7.45);
	\fill[color=orange!40]
	(0,7.45) -- (4.5,7.45) -- (4.5,6) -- (0,6);
	\fill[color=red!60]
	(0,6) -- (4.5,6) -- (4.5,5.2) -- (0,5.2);
	\fill[top color=orange!40, bottom color=white]
	(0,5.2) -- (4.5,5.2) -- (4.5,4.45) -- (0,4.45);
	\fill[bottom color=orange!40, top color=white]
	(0,3.45) -- (4.5,3.45) -- (4.5,4.2) -- (0,4.2);
	\fill[color=red!60]
	(0,3.45) -- (4.5,3.45) -- (4.5,2.65) -- (0,2.65);
	\fill[bottom color=white, top color=orange!40]
	(0,2.65) -- (4.5,2.65) -- (4.5,0.9) -- (0,0.9);
	\draw[axis] (-0.25,4.8) -- (-0.25,10) node[above] {$z$};
	\draw (-0.25,0.75) -- (-0.25,3.85);
	\draw[dashed] (-0.25,3.85) -- (-0.25,4.8);
	\draw (-0.15,8.25) -- ++ (-0.2,0) node[left] {\footnotesize $0$};
	\draw (-0.15,7.45) -- ++ (-0.2,0) node[left] {\footnotesize $-l$};
	\draw (-0.15,6) -- ++ (-0.2,0) node[left] {\footnotesize $-L$};
	\draw (-0.15,5.2) -- ++ (-0.2,0) node[left] {\footnotesize $-(L+l)$};
	\draw (-0.15,3.45) -- ++ (-0.2,0) node[left] {\footnotesize $-mL$};
	\draw (-0.15,2.65) -- ++ (-0.2,0) node[left] {\footnotesize $-(mL+l)$};
	\draw[thick, color=black, dashed] (0,8.25) -- ++ (4.5,0);
	\draw[thick, color=black, dashed] (0,7.45) -- ++ (4.5,0);
	\draw[thick, color=black, dashed] (0,6) -- ++ (4.5,0);
	\draw[thick, color=black, dashed] (0,5.2) -- ++ (4.5,0);
	\draw[thick, color=black, dashed] (0,3.45) -- ++ (4.5,0);
	\draw[thick, color=black, dashed] (0,2.65) -- ++ (4.5,0);
	\draw[->, thick, color=black] (1,9.5) -- (2,8.5); \node at (1.1,8.9) {\small $\mathcal{A}_0$};
	\draw[->, color=black] (2.5,8.5) -- (3.5,9.5); \node at (3.4,8.9) {\small $\mathcal{B}_0$};
	\draw[->, color=black] (1.4,8.1) -- ++ (0.5,-0.5); \node at (1.1,7.8) {\small $\mathcal{C}_0$};
	\draw[->, color=black] (2.55,7.6) -- ++ (0.5,0.5); \node at (3.4,7.8) {\small $\mathcal{D}_0$};
	\draw[->, color=black] (1,7.2) -- ++ (1,-0.95); \node at (1.1,6.6) {\small $\mathcal{A}_1$};
	\draw[->, color=black] (2.5,6.25) -- ++ (1,0.95); \node at (3.4,6.6) {\small $\mathcal{B}_1$};
	\draw[->, color=black] (1.4,5.85) -- ++ (0.5,-0.5); \node at (1.1,5.55) {\small $\mathcal{C}_1$};
	\draw[->, color=black] (2.55,5.35) -- ++ (0.5,0.5); \node at (3.4,5.55) {\small $\mathcal{D}_1$};
	\node at (1.5,4.42) {$\vdots$}; \node at (3,4.42) {$\vdots$};
	\draw[->, color=black] (1.4,3.3) -- ++ (0.5,-0.5); \node at (1.1,3) {\small $\mathcal{C}_m$};
	\draw[->, color=black] (2.55,2.8) -- ++ (0.5,0.5); \node at (3.4,3) {\small $\mathcal{D}_m$};
	\draw[->, thick, color=black] (2.25,2.4) -- ++ (1,-0.95); \node at (2.15,1.8) {\small $\mathcal{A}_{m+1}$};
\end{tikzpicture}
\end{subfigure}
\begin{subfigure}[b]{0.26\textwidth}
\centering
\caption{Density profile}
\begin{tikzpicture}[scale=0.8]
	\draw[axis,color=black!40] (5.75,4.8) -- (5.75,10) node[above] {$z$};
	\draw[color=black!40] (5.75,0.75) -- (5.75,3.85);
	\draw[dashed,color=black!40] (5.75,3.85) -- (5.75,4.8);
	\draw[color=black!40] (5.5,8.25) -- (7,8.25);
	\draw[axis,color=black!40] (7.5,8.25) -- (10,8.25) node[right,color=black] {$\bar{\rho}$};
	\draw[dashed] (6.75,9.75) -- (9.616664,1.15);
	\draw[<-] (7.25,8.5) -- ++ (0.5,0.5) node[right] {\footnotesize 
		$\displaystyle \left|\frac{\text{d}\bar{\rho}}{\text{d}z}\right| = \frac{\Delta\rho}{d}$};
	\draw[thick, color=black] (7,9.5) -- ++ (0,-1.25) -- ++ (0.74999925,-0.8) -- ++ (0,-1.45) -- ++ (0.74999925,-0.8) -- ++ (0,-0.25);
	\draw[thick, color=black] (8.5833,3.7) -- ++ (0,-0.25) -- ++ (0.74999925,-0.8) -- ++ (0,-1.25);
\end{tikzpicture}
\end{subfigure}
\begin{subfigure}[b]{0.18\textwidth}
\centering
\caption{$N^2$ profile}
\begin{tikzpicture}[scale=0.8]
	\draw[axis,color=black!40] (11.25,4.8) -- (11.25,10) node[above] {$z$};
	\draw[color=black!40] (11.25,0.75) -- (11.25,3.85);
	\draw[dashed,color=black!40] (11.25,3.85) -- (11.25,4.8);
	\draw[axis,color=black!40] (11,8.25) -- (14.5,8.25) node[right,color=black] {$N^2$};
	\draw[thick] (13.5,8.25) -- ++ (0,-0.8);
	\draw[thick] (11.25,7.45) -- ++ (0,-1.45);
	\draw[thick] (13.5,6) -- ++ (0,-0.8);
	\draw[thick] (11.25,5.2) -- (11.25,4.8);
	\draw[thick] (11.25,3.85) -- (11.25,3.45);
	\draw[thick] (13.5,8.25) -- ++ (0,-0.8);
	\draw[thick] (13.5,3.45) -- ++ (0,-0.8);
	\draw[dashed, thick] (11.25,8.25) -- ++ (2.25,0);
	\draw[dashed, thick] (11.25,7.45) -- ++ (2.25,0);
	\draw[dashed, thick] (11.25,6) -- ++ (2.25,0);
	\draw[dashed, thick] (11.25,5.2) -- ++ (2.25,0);
	\draw[dashed, thick] (11.25,3.45) -- ++ (2.25,0);
	\draw[dashed, thick] (11.25,2.65) -- ++ (2.25,0);
	\draw (13.5,8.15) -- ++ (0,0.2) node[above] 
		{\footnotesize $\displaystyle N_{\text{i}}^2=\frac{\bar{N}^2}{\varepsilon}$};
	\draw[dashed, densely dotted] (12.2,1.8) -- ++ (0,7.5) node[above]{\footnotesize $\bar{N}^2$};
\end{tikzpicture}
\end{subfigure}
\caption[Local physical model with finite size interfaces]{Summary of our physical model with finite size interfaces: $m$ convective steps of constant size $d$ and indexed by the integer $1<n<m$, are separated by interfaces of size $l$. 
(a) General scheme: the incident, reflected and transmitted waves have amplitudes $\mathcal{A}_0$, $\mathcal{B}_0$ and $\mathcal{A}_{m+1}$, respectively. In the $n$-th convective step, the ingoing wave has an amplitude $\mathcal{A}_n$, and the outgoing has an amplitude $\mathcal{B}_n$, while in the $n$-th stably stratified interface they have amplitudes $\mathcal{C}_n $ and $\mathcal{D}_n$, respectively.
(b) The corresponding density profile: in each interface, the density follows a gradient $|\text{d}\bar{\rho}/\text{d}z|=\Delta\rho/l$, so that the density varies by an ammount $\Delta\rho>0$. In the convective steps, it follows an adiabatic gradient. Thus, the mean density gradient $|\text{d}\bar{\rho}/\text{d}z|=\Delta\rho/L$, where $L=l+d$.
(c) The corresponding buoyancy frequency profiles are $N^2=0$ in the convective layers and $N^2_{\text{i}} = \bar{N}^2/\varepsilon$ in the interfaces to create the mean stratification $\bar{N}^2=g\Delta\rho/\rho_0 L$.}
\label{fig:tcoeff_modele_CC_ThickInterfaces}
\end{figure*}
\begin{figure}
\centering
\begin{tikzpicture}[scale=0.95]
\draw[axis] (-0.3,-0.2) -- ++ (0,4.2) node[above]{$N$};
\draw (-0.5,0) -- ++ (1.5,0); \draw[dashed] (1,0) -- ++ (0,1);
\draw (1,1) -- ++ (2,0); \draw[dashed] (3,1) -- ++ (0,-1);
\draw (3,0) -- ++ (1,0); 
\draw (0-0.5,0) -- ++ (2,0); \draw[dashed] (1.5,0) -- ++ (0,2);
\draw (1.5,2) -- ++ (1,0); \draw[dashed] (2.5,2) -- ++ (0,-2);
\draw (2.5,0) -- ++ (1.5,0);
\draw (-0.5,0) -- ++ (4.5,0);
\draw[load] (2,0) -- ++ (0,3.5);
\draw[axis] (3.3,1.8) arc (200:190:8);
\node at (3.9,2.5){$\varepsilon \rightarrow 0$};
\draw (-0.2,1) -- ++ (-0.2,0) node[left]{$N_{\text{i}}(\varepsilon_1)$};
\draw (-0.2,2) -- ++ (-0.2,0) node[left]{$N_{\text{i}}(\varepsilon_2 < \varepsilon_1)$};
\draw[dotted] (-0.3,1) -- (1,1);
\draw[dotted] (-0.3,2) -- (1.5,2);
\node at (-0.75,0){$0$};
\end{tikzpicture}
\caption[Buoyancy frequency profiles with varying aspect ratios]{Buoyancy frequency profiles with various aspect ratios. As the aspect ratio decreases, $N_{\text{i}}$ increases to maintain the integrated value of $N$ equal to $\bar{N}$. Our model converges towards the reference model (the limit being the Dirac distribution) when $\varepsilon \rightarrow 0$.}
\label{fig:Ni}
\end{figure}
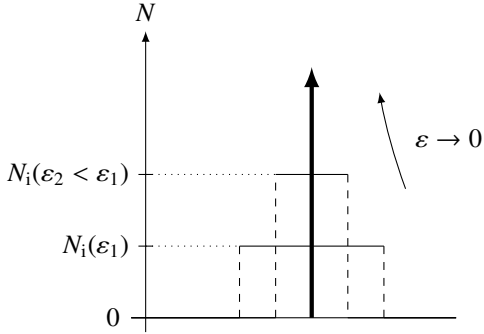

The buoyancy frequency profile is defined step wise (see Fig. \ref{fig:tcoeff_modele_CC_ThickInterfaces}c) through
\begin{equation}
N^2 = \left\{
\begin{array}{cl}
N^2_{\text{i}} & ~ \text{for} ~ -nL<z<-(nL+l),\\
& ~ ~ ~ ~ ~ ~ ~ ~ ~ ~ ~ n = \{0,\dots,m\},\\
0 & ~ \text{otherwise},
\end{array}
\right.
\end{equation}
where $N^2_{\text{i}}$ has been defined in order to keep the mean stratification (and accordingly the mean density gradient, see Fig. \ref{fig:tcoeff_modele_CC_ThickInterfaces}) the same as in the previous model. This gives
\begin{equation}
N^2_{\text{i}} = \frac{\bar{N}^2}{\varepsilon}.
\label{eq:Ni}
\end{equation}
This definition of $N^2_{\text{i}}$ ensures that the integrated value of $N^2$ over one interface is independent of $\varepsilon$ and equals $\bar{N}^2$. Also, for vanishingly small aspect ratios, we recover the reference model with discrete interfaces. This is illustrated on Fig. \ref{fig:Ni}.

The corresponding density profile is shown on Fig. \ref{fig:tcoeff_modele_CC_ThickInterfaces}b. In convective layers (represented in orange on Fig. \ref{fig:tcoeff_modele_CC_ThickInterfaces}a), the density follows an adiabatic gradient, while in the interfaces the density gradient is
\begin{equation}
\left|\frac{\text{d}\bar{\rho}}{\text{d}z}\right| = \frac{\Delta\rho}{l} = \frac{\Delta\rho}{\varepsilon d},
\end{equation}
where the density jump across one interface, $\Delta\rho$, is maintained to be the same as in the reference model of Section \ref{sec:FREE:transmission}, thanks to this procedure. 

\begin{figure*}[t]
\centering
\begin{subfigure}{0.312\textwidth}
\centering
\caption{$\varepsilon = 0.01$}
\includegraphics[height=5.75cm]{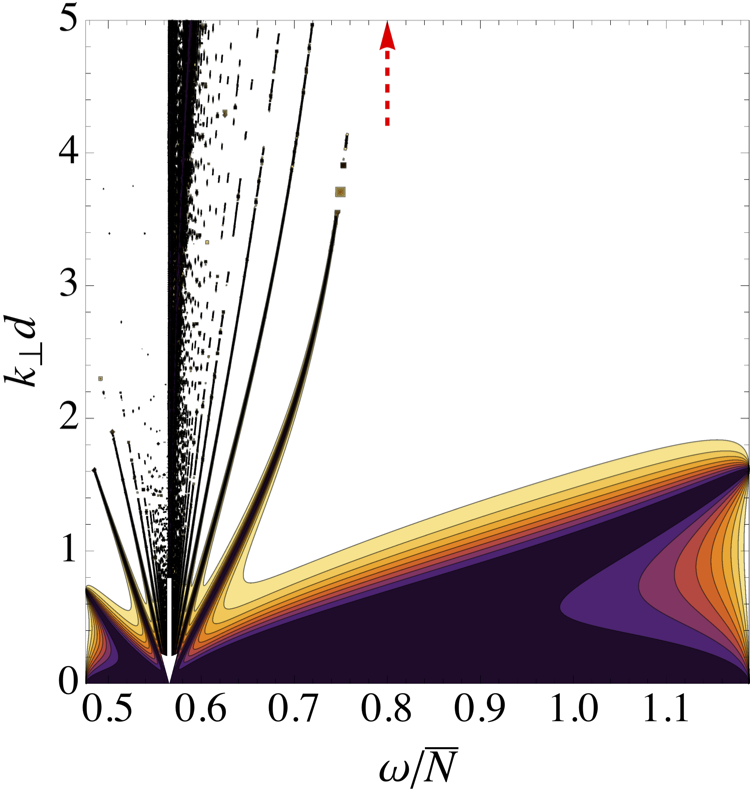}
\end{subfigure}
\begin{subfigure}{0.312\textwidth}
\centering
\caption{$\varepsilon = 0.1$}
\includegraphics[height=5.75cm]{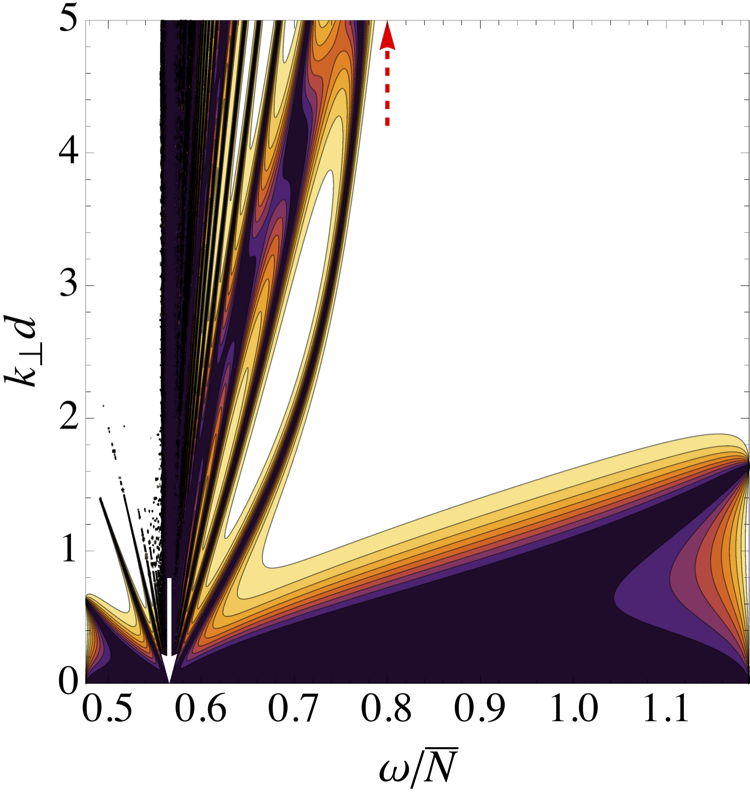}
\end{subfigure}
\begin{subfigure}{0.356\textwidth}
\centering
\caption{$\varepsilon = 0.5$}
\includegraphics[height=5.75cm]{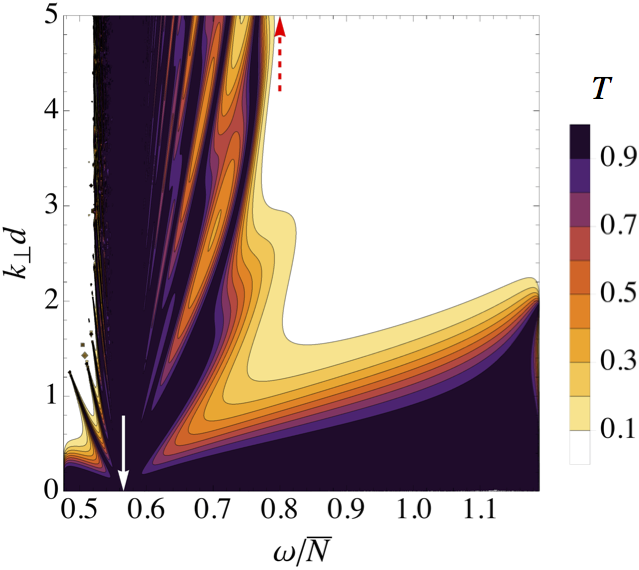}
\end{subfigure}
\caption[Transmission coefficient -- model with finite size interfaces]{Transmission coefficient for one step embedded in a convective medium with $\Omega=0.4\bar{N}$ and $\Theta=\uppi/4$ for different aspect ratios (a) $\varepsilon=0.01$, (b) $\varepsilon=0.1$ and (c) $\varepsilon=0.5$. The white and dashed red arrows indicate frequency $\omega=f$ and $\omega=2\Omega$, respectively. A vertical band of perfect transmission appears near the inertial frequency $\omega=f$, this broadens with increasing $\varepsilon$.}
\label{fig:T_finite}
\end{figure*}

\subsubsection{Analytic expression of the transmission coefficient}\label{subsec:FREE:FSI_Tanalytic}
Since the density profile is continuous (there is no density jump), the boundary conditions between adjacent layers are simply the continuity of the vertical velocity and its derivative (see Eq. (\ref{eq:int1})):
\begin{align}
w_n &= w_{n+1},\label{eq:BC1_finite}\\
w'_n &= w'_{n+1}.\label{eq:BC2_finite}
\end{align}

Now, the function $\hat{W}$ reads
\begin{equation}
\hat{W}(z) = \left\{
\begin{array}{ll}
\mathcal{A}_0\,\text{e}^{k_{z,\text{a}}z} + \mathcal{B}_0\,\text{e}^{-k_{z,\text{a}}z} & \text{for} ~~ z>0,\\
\mathcal{C}_0\,\text{e}^{k_{z,\text{i}}z} + \mathcal{D}_0\,\text{e}^{-k_{z,\text{i}}z} & \text{for} ~ -l<z<0,\\[3mm]
\mathcal{A}_n\,\text{e}^{k_z(z+nL)} + \mathcal{B}_n\,\text{e}^{-k_z(z+nL)} &\\
\mathcal{C}_n\,\text{e}^{k_{z,\text{i}}(z+nL)} + \mathcal{D}_n\,\text{e}^{-k_{z,\text{i}}(z+nL)} & \text{for} ~ -(n-1)L-l<z\\
& <-nL, n=\{1,\dots,m\},\\[3mm]
\mathcal{A}_{m+1}\,\text{e}^{k_{z,\text{b}}(z+mL)} & \text{for} ~ z<-(mL+l).
\end{array}
\right.
\label{eq:what_finite}
\end{equation}

Using the boundary conditions given by (\ref{eq:BC1_finite})--(\ref{eq:BC2_finite}), we get recurrence relations between coefficients at each interface between a convective layer and a stably stratified interface that, after little algebra, can be put into matrix form as in section \ref{sec:FREE:maths}. We first get a relation between coefficients $(\mathcal{A}_n, \mathcal{B}_n)$ and $(\mathcal{C}_n, \mathcal{D}_n)$,
\begin{equation}
\left[
\begin{array}{c}
\mathcal{C}_n\\
\mathcal{D}_n
\end{array}
\right] =
{\mathbfsf{T}}_1 \left[
\begin{array}{c}
\mathcal{A}_n\\
\mathcal{B}_n
\end{array}
\right],
\label{eq:T1_relation}
\end{equation}
where
\begin{equation}
{\mathbfsf{T}}_1 = \frac{1}{2}\left[
\begin{array}{cc}
(1+K_{\text{i}}^{\text{c}}à)\Delta^{-1}_{\text{c},l}\Delta_{\text{i},l} & (1-K^{\text{c}}_{\text{i}})\Delta_{\text{c},l}\Delta_{\text{i},l}\\[3mm]
(1-K^{\text{c}}_{\text{i}})\Delta_{\text{c},l}^{-1}\Delta_{\text{i},l}^{-1} & (1+K^{\text{c}}_{\text{i}})\Delta_{\text{c},l}\Delta_{\text{i},l}^{-1}
\end{array}
\right].
\end{equation}
Here, we have defined
\begin{equation}
\Delta_{\upalpha,\upbeta} = \text{e}^{k_{z,\upalpha}\upbeta}.
\end{equation}
Here $\upbeta$ stands for either $l$ or $L$. Then, we get a second relation between coefficients $(\mathcal{A}_n, \mathcal{B}_n)$ and $(\mathcal{C}_{n+1}, \mathcal{D}_{n+1})$,
\begin{equation}
\left[
\begin{array}{c}
\mathcal{A}_n\\
\mathcal{B}_n
\end{array}
\right] =
\widetilde{\mathbfsf{T}}_2 \left[
\begin{array}{c}
\mathcal{C}_{n+1}\\
\mathcal{D}_{n+1}
\end{array}
\right],
\label{eq:T2_relation}
\end{equation}
where
\begin{equation}
\widetilde{\mathbfsf{T}}_2 = \frac{1}{2}\text{e}^{-\text{i}\tilde{\varphi}}\left[
\begin{array}{cc}
(1+K^{\text{i}}_{\text{c}})\Delta_{\text{c},L} & (1-K_{\text{c}}^{\text{i}})\Delta_{\text{c},L}\\[3mm]
(1-K_{\text{c}}^{\text{i}})\Delta_{\text{c},L}^{-1} & (1+K_{\text{c}}^{\text{i}})\Delta_{\text{c},L}^{-1}
\end{array}
\right].
\end{equation}
Equations (\ref{eq:T1_relation}) and (\ref{eq:T2_relation}) can be combined in order to get a recurrence relation between coefficients $(\mathcal{C}_{n}, \mathcal{D}_{n})$:
\begin{equation}
\left[
\begin{array}{c}
\mathcal{C}_n\\
\mathcal{D}_n
\end{array}
\right] =
{\mathbfsf{T}}_1
\widetilde{\mathbfsf{T}}_2 \left[
\begin{array}{c}
\mathcal{C}_{n+1}\\
\mathcal{D}_{n+1}
\end{array}
\right].
\end{equation}
Finally, as we did in section \ref{subsec:FREE:FSI_Tanalytic}, we can then express $\mathcal{A}_0$ in term of solely $\mathcal{A}_{m+1}$,
\begin{equation}
\mathcal{A}_0 = \left[\bm{\text{b}}_{\text{a}}^{\text{T}}\,\left(\mathbfsf{T}_1\widetilde{\mathbfsf{T}}_2\right)^{m} \bm{\text{b}}_{\text{b}} \right] \mathcal{A}_{m+1},
\end{equation}
where the left and right vectors are defined by
\begin{align}
\bm{\text{b}}_{\text{a}} &=\frac{1}{2}\left[
\begin{array}{c}
(1+K^{\text{i}}_{\text{a}})\\[2mm]
(1-K^{\text{i}}_{\text{a}})
\end{array}
\right],\\[2mm]
\bm{\text{b}}_{\text{b}} &=\frac{1}{2}\Delta^{-1}_{\text{b},l}\left[
\begin{array}{c}
(1+K_{\text{i}}^{\text{b}})\Delta_{\text{i},L}\Delta_{\text{i},l}\\[2mm]
(1-K_{\text{i}}^{\text{b}})\Delta_{\text{i},L}^{-1}\Delta_{\text{i},l}^{-1}
\end{array}
\right].
\end{align}
Therefore, the transmission coefficient is given by
\begin{equation}
T = \frac{k_{z,\text{b}}}{k_{z,\text{a}}} \left| \bm{\text{b}}_{\text{a}}^{\text{T}}\,\left(\mathbfsf{T}_1\mathbfsf{T}_2\right)^{m} \bm{\text{b}}_{\text{b}} \right|^{-2},
\label{eq:Tcoeff_finiteSize}
\end{equation}
where $\mathbfsf{T}_{2} \equiv \text{e}^{\text{i}\tilde{\varphi}}\widetilde{\mathbfsf{T}}_{2}$ (we have $|\text{e}^{\pm\text{i}\tilde{\varphi}}|=1$), and the ratio ${k_{z,\text{b}}}/{k_{z,\text{a}}}$ is given by Eq. (\ref{eq:kzb/kza}).

\subsubsection{Comparison with the reference model}\label{subsec:FREE:FSI_comp}

Now, we study the effect of having finite size stably stratified interfaces on the transmission coefficient. We maintain the angle $\Theta=\uppi/4$, and choose a rotation rate $\Omega=0.4\bar{N}$. The expression of the transmission coefficient is now given by Eq. (\ref{eq:Tcoeff_finiteSize}). The results for different aspect ratios $\varepsilon=0.01$, $0.1$ and $0.5$ for a single step embedded in a stably stratified medium are displayed on Fig. \ref{fig:T_finite}. The panels can directly be compared to Fig. \ref{fig:onestep_RRrotation}b, on which is displayed the single step case at the same colatitude and rotation rate, assuming infinitesimally thin interfaces.

As the aspect ratio is increased, a band of enhanced transmission around $\omega=f$ ($\approx 0.57\bar{N}$ here) appears, this being in agreement with \cite{GerkemaExarchou2008}. Note that this was already observed in the reference model when the transmission coefficient was plotted as a function of the vertical wave number, $k_z$. If we denote by
\begin{equation}
\omega_{\pm}^{(\text{i})} = \omega_{\pm}(N_{\text{i}})
\end{equation}
the frequency limits between which the incident wave is propagative in the interfaces, there is a frequency range over which the wave is propagative in the convective layers (this happens for $0<\omega<2\widetilde{\Omega}$) {and} the stably stratified interfaces (this happens for $\omega_{-}^{(\text{i})}<\omega<\omega_{+}^{(\text{i})}$). In our case, the intersection of those two frequency domains is such that this happens for
\begin{equation}
\omega_{-}^{(\text{i})} < \omega < 2\widetilde{\Omega}.
\end{equation}
\begin{figure}
\centering
\begin{tikzpicture}
\node[anchor=south west,inner sep=0] at (0,0)
    {\includegraphics[height=7.5cm]{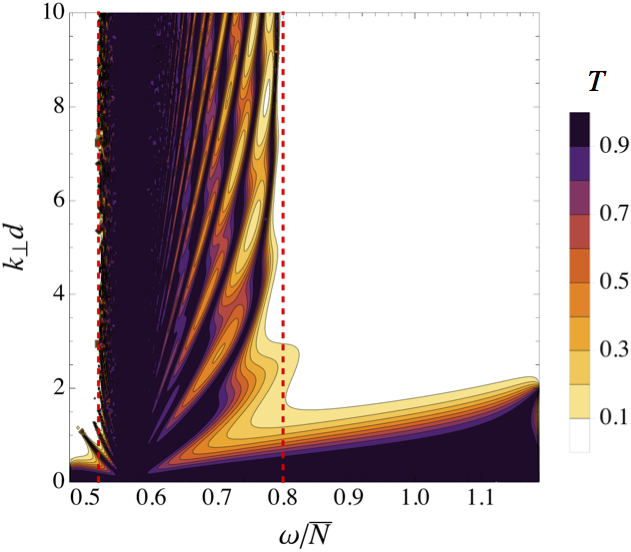}};
\draw[color=red, thick, ->] (0.35,5.25) node[thick, red, above]{$\omega=\omega_-^{(\text{i})}$} -- ++ (0.9,-0.25);
\draw[color=red, thick, ->] (4.85,5.35) node[thick, red, above]{$\omega=2\tilde{\Omega}$} -- ++ (-0.9,-0.25);
\end{tikzpicture}
\caption[Understanding the transmission coefficient for finite size interfaces]{Transmission coefficient for one step embedded in a stably stratified medium, with $\Omega=0.4\bar{N}$, $\Theta=\uppi/4$ and an aspect ratio $\varepsilon=0.5$. Transmission is enhanced within the range of frequency $\omega_-^{(\text{i})}<\omega<2\widetilde{\Omega}$, for which the waves are propagative both in the convective steps {and} in the stably stratified interfaces.}
\label{fig:T_finite_explanation}
\end{figure}
This is illustrated on Fig. \ref{fig:T_finite_explanation}, on which the frequency range above has been overplotted in dashed red lines.
Therefore, there is enhanced transmission for waves that are propagative in both convective layers and stably stratified interfaces, as one would expect. The transmission is inhibited for $\omega<\omega_-^{(\text{i})}$ for larger aspect ratios $\varepsilon$ -- leading to less symmetric transmission about $\omega=f$ than our reference model (which corresponds to $\epsilon\rightarrow 0$) -- because the waves are evanescent in the interfaces for $\omega<\omega_-^{(\text{i})}$. Thus, as we make $\varepsilon$ larger, the evanescent region becomes larger, which inhibits transmission.

Another difference with the reference model is that there is more transmission near $\omega_+$, and that the transmission peak is shifted to higher wave numbers when the aspect ratio is increased.

\section{Conclusions \& prospects}\label{SEC:conclusion}
\subsection{Summary of results}
We have studied the transmission of internal waves through a staircase-like density profile that could be produced by oscillatory double-diffusive convection in giant planet interiors. First, we analysed the free modes of oscillation of such a density staircase in a rotating planet by deriving the dispersion relation that describes them, generalising \cite{BelyaevQuataertFuller2015} to consider any latitude (they previously considered the effects of rotation at the pole and the equator). We then analysed the transmission of a wave through a density staircase consisting of one or multiple convective steps by extending \cite{Sutherland2016} to include the complete Coriolis acceleration {rather than neglecting its horizontal component, under the so-called `traditional approximation' which is not appropriate when studying convective layers of giant planets \citep{OgilvieLin2004}}.

We showed that the transmission of internal (inertial or gravito-inertial) waves is strongly affected by the presence of a density staircase in a frequency- and wavelength-dependent manner. {This is true even} if these waves are initially pure inertial waves. Large wavelength waves (with wavelengths $\lambda \gg D = md$, where $\lambda$ is either the vertical or horizontal wavelength and $D$ is the total size of the staircase) are unaffected by the staircase. Low-frequency (inertial or gravito-inertial) waves of any wavelength are perfectly transmitted near {the critical latitude $\theta_{\text{c}} = \sin^{-1}({\omega}/{2\Omega})$ (at which $\omega=f$)}, confirming a feature also obtained by \citet{GerkemaExarchou2008}. Otherwise, short-wavelength waves (with $\lambda \sim d \lesssim D$) are only efficiently transmitted if they are resonant with a free mode of the staircase (these are interfacial gravity or short-wavelength inertial modes, corresponding to the roots of the dispersion relation), and if not they are primarily reflected.

The frequency interval around $f=2\Omega\cos\Theta$ on which transmission remains close to unity widens when rotation is increased (see Fig. \ref{fig:onestep_CCrotation}). This means that for fast rotators, we expect a relatively broad frequency window for which layered semi-convection has no particular effect upon the propagation of internal waves near the critical latitude. Note that this perfect transmission at $\omega=f$ can only occur at any latitude if $\omega < 2\Omega$, i.e. only for inertial and sub-inertial gravito-inertial waves. We do not expect this to happen for higher frequency gravity waves, since there is no colatitude $\Theta$ for which $\omega=f$.

We also found that non-traditional effects could be of major importance. First, they modify the frequency spectrum of propagation of internal waves, allowing the propagation of sub-inertial waves inside the convective layers of the staircase. Thus, the transmission of pure inertial waves can be strongly affected by the presence of a density staircase. Second, they modify significantly the transmission near the inertial frequency $\omega=f$, with bands of perfect transmission departing from this frequency. 
These have been physically interpreted as being resonances between pure inertial waves propagating inside a convective step whose wavelength fits exactly inside a step, and the incident wave.

{We first modelled the staircase as having infinitesimally thin interfaces, which we also extended to consider interfaces of finite size. An analytical expression of the transmission coefficient has been derived in both set-ups}, and its behaviour analysed. In addition to what was already described with the first model {(that assumed infinitesimally thin interfaces)}, it has been found that transmission was significantly enhanced in the frequency range for which internal waves are propagative in both the convective steps and the stably stratified interfaces. As a result, perfect transmission was obtained near $\omega=f$ for any wavelength. Since this model is more realistic, this means that in reality almost any incident IWs or GIWs is perfectly transmitted at a special location, the critical latitude {(if its frequency is smaller than $2\Omega$)}. Otherwise it is strongly affected by the staircase and is only transmitted if it has a large wavelength.

\begin{figure}
\centering
\includegraphics[width=0.8\linewidth]{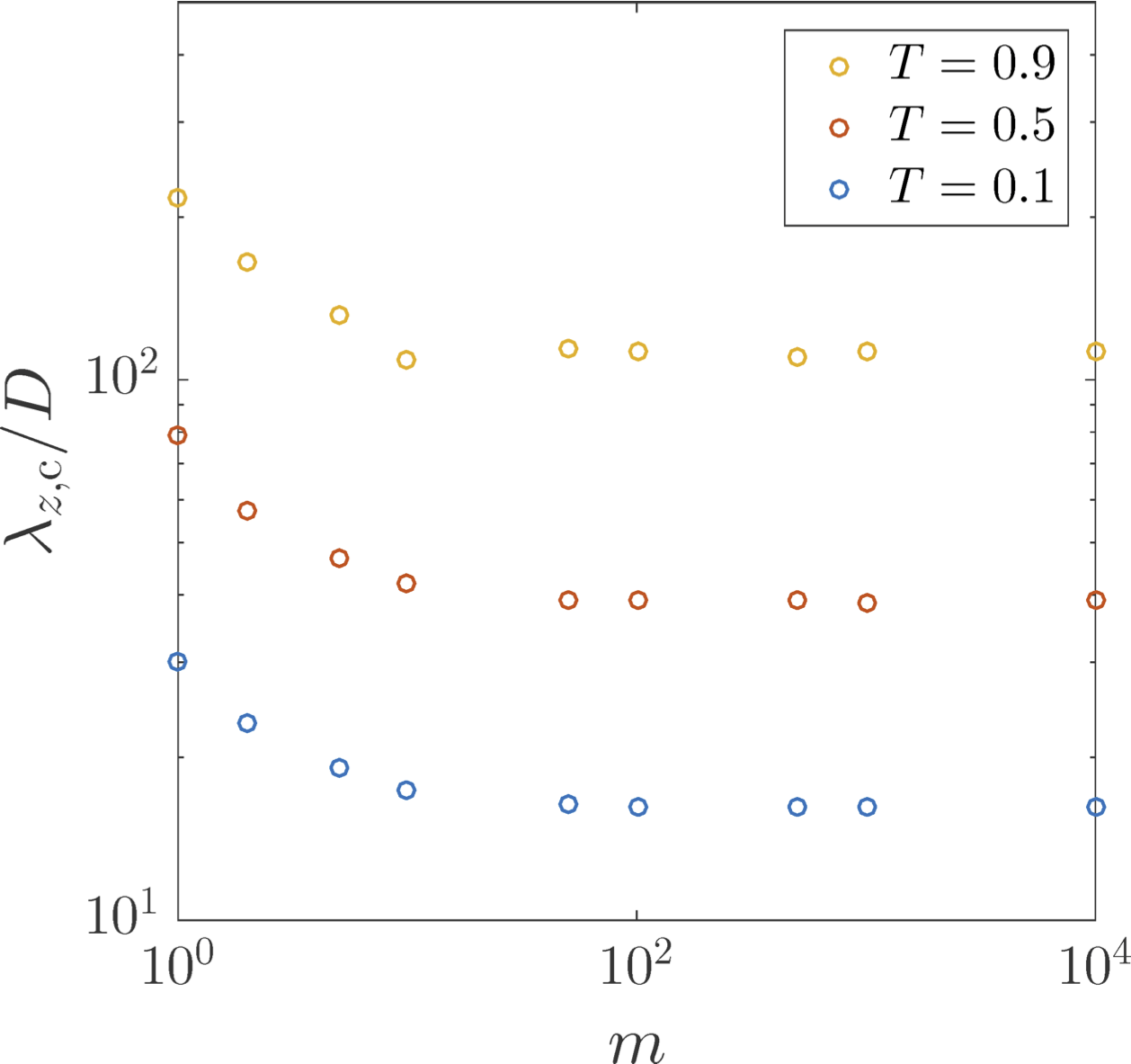}
\caption{Vertical cut-off wave-length $\lambda_{z,\text{c}}$ normalized by the size of the staircase $D = md$, as a function of the number of steps $m$, measured for three criteria on the transmission: $T=0.9$ (green), $T=0.5$ (red), $T=0.1$ (blue). These results were obtained in the case of a staircase embedded in a convective medium, with $\Omega=0.4\bar{N}$ and at the particular frequency $\omega=0.2\bar{N}$ {which was not a resonant mode}.}
\label{fig:lambda_cutoff}
\end{figure}

\subsection{Consequences for the seismology of planets}\label{sec:csqces}
A case from which we can draw relevant astrophysical conclusions is the case of multiple steps embedded in a convective medium, because gaseous giant planet envelopes are expected to be mostly convective. {It is also relevant for the study of} a portion of a more vertically extended staircase. {In that case when $\omega\neq f$, and the mode is not resonant with a free mode of the staircase, transmission was found to be inhibited for larger wavelength incident waves as we increased the number of steps, behaving somewhat linearly with $1/m$} (this qualitative behaviour being the same for the other top and bottom layer properties). 
We find that the cut-off (indexed by the subscript '$\text{c}$') wavelength below which transmission is less than a given threshold (when not resonant with a mode of the staircase) increases approximately linearly with the number of steps; or, alternatively, is approximately constant when normalized by the size of the staircase $D=md$. On Fig. \ref{fig:lambda_cutoff}, we show $\lambda_{z,\text{c}}/D$ as a function of the number of steps $m$, with a rotation rate $\Omega=0.4\bar{N}$, for different transmission thresholds $T=0.9, 0.5, 0.1$, at the particular frequency $\omega=0.2\bar{N}$. We see that, for a given criteria on the transmission coefficient, the cut-off vertical wavelength {only} depends on the total size of the staircase, $D$, for a large enough number of steps. For instance, {approximately} perfect transmission is obtained in that case for $\lambda_z > 10^2 D$. 

{Assuming that the entire gaseous envelope is semi-convective,} \cite{LeconteChabrier2012} estimated that the size of the convective steps should lie in a range
\begin{equation}
10^{-9} - 10^{-6} \lesssim \frac{d}{H_{\text{p}}} \lesssim 10^{-4} - 10^{-2},
\label{eq:LC12}
\end{equation}
where $H_{\text{p}}$ is the pressure scale-height, that is expected to be of the same order of magnitude as the planet's radius, $R$, in the deep interior of giant planets ($H_{\text{p}}\sim R$). {Note that the hypothesis of a fully semi-convective envelope gives us an upper limit on the number of steps. In particular, \cite{VazanEtal2016} found it to be unlikely because of the efficiency of upward mixing by overturning convection in the upper layers of the envelope. In any case, it provides us with some order of magnitude estimates. We also note that local numerical simulations by e.g. \cite{Radko2003} show merging of the double-diffusive layers over a finite time, until only one step remains in the local box. This differs from the hypothesis of assuming a large number of steps. However, we stress that those simulations are local and Boussinesq, so that the long-term evolution of double-diffusive convection in spherical geometry with a realistic density stratification is still unknown. Furthermore, the case of the Arctic Ocean on Earth seems to suggest that a staircase with a large number of steps can exist \citep{GhaemsaidiEtal2016}}.

{Following our reasoning from Eq. (\ref{eq:LC12}),} this means that only incident waves coming from above the staircase with a wavelength 
\begin{equation}
\lambda \gtrsim \lambda_{z,\text{c}},
\end{equation}
will be efficiently transmitted to deeper regions, where the cut-off frequency $\lambda_{z,\text{c}}$ is estimated to lie in a range
\begin{equation}
\left[10^{-7} - 10^{-4}\right]m \lesssim \frac{\lambda_{z,\text{c}}}{R} \lesssim \left[10^{-2} - 1\right]m.
\label{eq:crudeEstimate}
\end{equation}
Incident waves with $\lambda_z<\lambda_{z,\text{c}}$ are expected to be primarily reflected at the top of the staircase, unless $\omega \approx f$ or the incident wave is resonant with a free mode of the staircase. LC12 estimated that the number of steps in giant planet deep interiors lies in a range
\begin{equation}
\left[10^2 - 10^4\right] \lesssim m \lesssim \left[10^{6} - 10^{9}\right].
\end{equation}
For the bottom estimate $m=10^2$ steps the estimate given by Eq. (\ref{eq:crudeEstimate}) yields
\begin{equation}
\left[10^{-5} - 10^{-2}\right] \lesssim \frac{\lambda_{z,\text{c}}}{R} \lesssim \left[1 - 10^2\right].
\end{equation}
We note that the top range of the inequality above does not correspond to any wave that could propagate in giant planet interiors, since their wavelength is larger than the radius of the planet. More restrictive constraints on the layer sizes and their expected numbers would be needed to provide more precise estimates. However, the bottom range for $\lambda_{z,\text{c}}$ is of interest, since tidally excited inertial waves (in the presence of a core) are often thought to be of very short-wavelength \citep[e.g.][]{OgilvieLin2004}, so their propagation may be strongly affected by a staircase.

On the other hand, \citet{NettelmannEtal2015} estimate that in the helium rain region, convective layers should have a size $d \sim 0.1-1\,\text{km}$, corresponding to $10,000 - 20,000$ layers. Following the discussion above and taking a typical giant planet's radius $R \sim 10^5\,\text{km}$, this suggests that the cut-off vertical wavelength is approximately given by
\begin{equation}
\frac{\lambda_{z,\text{c}}}{R} \sim \left[1 - 10\right]
\label{eq:cutoff_HerainRegion}
\end{equation}
in the helium rain region. Because we expect tidal waves to have a vertical wavelength that is small compared to the planet's radius, Eq. (\ref{eq:cutoff_HerainRegion}) suggests that a region of layered semi-convection in the helium rain region \citep[as modelled by][]{NettelmannEtal2015} would act as a rigid wall for most internal waves propagating inside the planet's envelope, {unless $\omega \approx f$ or the incident wave is resonant with a free mode of the staircase}.
\\

Waves with shorter wavelengths than $\lambda_{z,\text{c}}$ will be strongly reflected and will not penetrate into deeper regions of the giant planet. This means that a tidally excited wave launched near the surface of a giant planet that has a region of layered semi-convection (e.g. just outside its core), could see the core's size artificially enlarged because it can be reflected at the top of the staircase region \citep[which could have a radial extent of the order of $10^3 - 10^4\,\text{km}$ according to][]{NettelmannEtal2015}, {rather than be reflected from the core itself} (which would happen in the absence of a density staircase). This could have an impact on the nature of tidal modes that develop inside giant planets containing regions of layered semi-convection, and of course on tidal dissipation.

{The (vertical and horizontal) wavelengths are related to the wave's frequency through the dispersion relation. Thus, an inertial wave that propagates in the convective envelope of a giant planet towards the center has a frequency that is set by Eq. (\ref{eq:disprelIW}). This means that the cut-off wavelengths $\lambda_{\perp,\text{c}}$ and $\lambda_{z,\text{c}}$ define a frequency cut-off through the dispersion relation of pure inertial waves,
\begin{equation}
\omega_{\text{c}} = \left[\frac{(\tilde{f}\lambda_{z,\text{c}} + f \lambda_{\perp,\text{c}})^2}{\lambda_{\perp,\text{c}}^2+\lambda_{z,\text{c}}^2}\right]^{1/2}.
\end{equation}
In the presence of layered semi-convection, potentially observable modes of different frequencies will see a different size of the core. This observation could be a signature of layered semi-convection just outside the rocky/icy core of giant planets.}

{On the other hand, if density staircases are also present in the helium rain region, the spherical shell contained between that and the core could act as a trap for pure inertial waves in a certain frequency range, modifying the nature of the modes that can be sustained in that particular region. However, this could not be observed by seismology of the planet’s surface.}

{Finally, we have seen that new modes can potentially be observed when invoking layered semi-convection compared to a fully convective envelope. These are the free modes of the staircase, first given by \citetalias{BelyaevQuataertFuller2015} and extended here to the case with rotation for any latitude through Eq. (\ref{eq:BQFdisprel_finite}).} {This includes the gravito-inertial modes with a sufficiently large radial wavelength that the staircase is seen as a stably stratified medium characterized by the mean buoyancy frequency $\bar{N}$.}

\subsection{Perspectives}

\begin{figure}
\centering
\includegraphics[width=\linewidth]{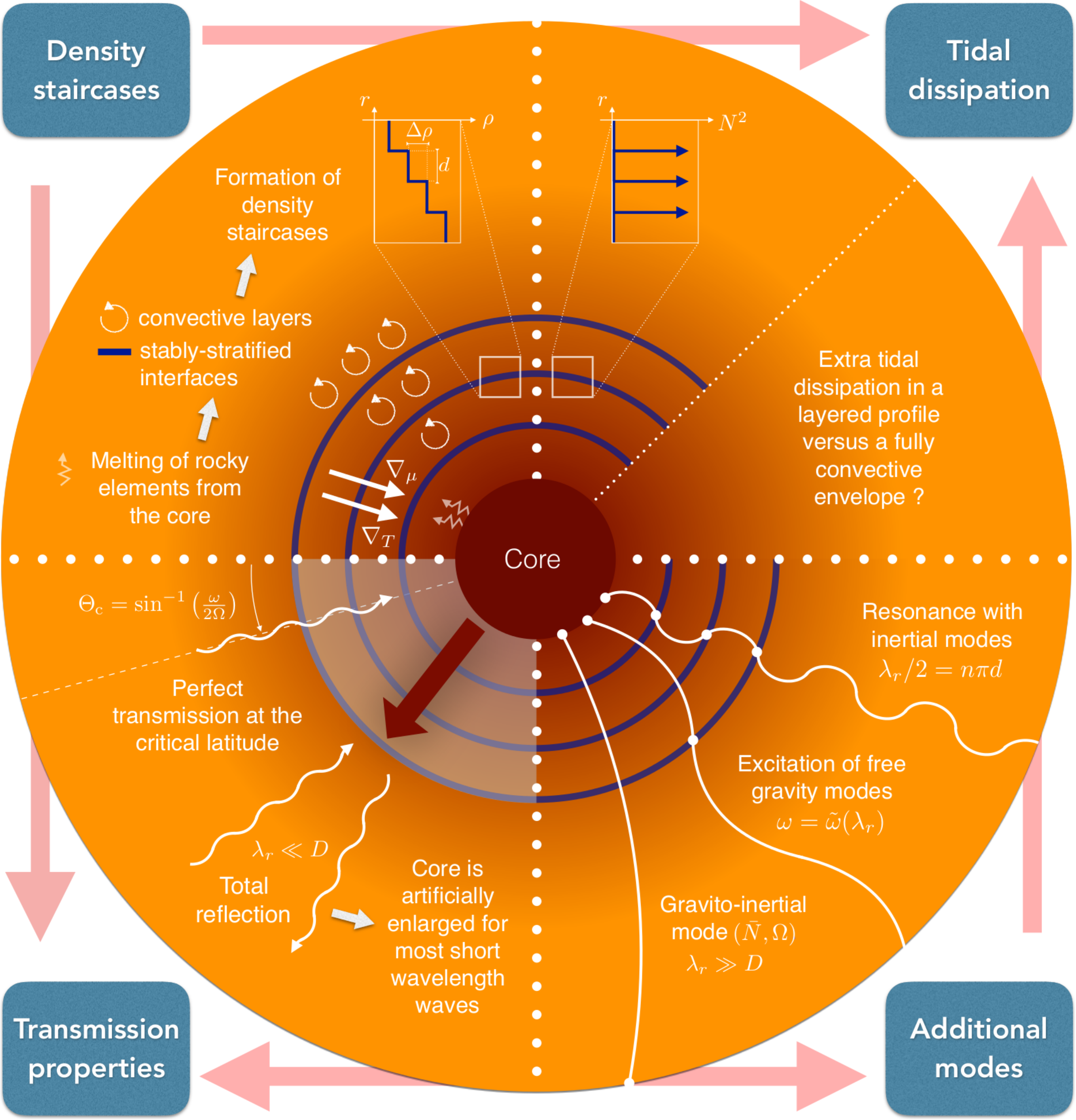}
\caption{Sketch presenting an illustrative summary of our motivations, results and future work. North-West quarter: how density staircases are possibly created (see Section \ref{SEC:intro}). South-West quarter: main properties of the transmission of internal waves (see Sections \ref{SEC:layeredCase} and \ref{sec:csqces}). South-East quarter: additional modes that could potentially be observed thanks to the seismology of planets (see Section \ref{sec:csqces}). North-East quarter: the question we address in our second paper.
}
\label{fig:finalSketch}
\end{figure}

{Our results help us to understand the impact of layered semi-convection on the propagation of internal waves at a qualitative level. In order to make quantitative predictions, one should extend this study to spherical geometry relevant for planets and stars.}
{In particular, it could be interesting to analyse probabilities of excitation of free modes of the staircase. This could allow us to determine e.g. what fraction of tidal waves are perfectly transmitted through the staircase region, but this is beyond the scope of our paper.}

{In addition, we have considered a density staircase established by the double-diffusive instability, as described by e.g \cite{LeconteChabrier2012}. Note, however, that it would be important in the near future to take into account the impact of rotation \citep{MollGaraud2017}, differential rotation \citep{Worthem1983} and magnetic fields on the formation of such a structure. Waves will also be affected by differential rotation \citep{Mathis2009,MirouhEtal2016} and magnetic fields \citep[e.g.][]{MathisDeBrye2011,Wei2016}. Moreover, non-linear effects such as the feedback of waves on the background should be taken into account.}
\\

{Finally, a synthetic sketch of our results is shown on Fig. \ref{fig:finalSketch}.} Based on this first study, we will explore in our second paper the tidal response and dissipation of a giant planet or star containing a staircase-like density profile resulting from the presence of layered semi-convection.

\begin{acknowledgements}
{We thank the referee for a constructive and helpful report.} QA was supported by ENS Paris-Saclay and CEA. QA and SM acknowledge funding by the European Research Council through ERC SPIRE grant 647383. AJB was supported by the Leverhulme Trust through the award of an Early Career Fellowship. QA was also partly supported by this Leverhulme Trust award.
\end{acknowledgements}

%
%

\bibliographystyle{aa} 
\bibliography{staircase} 

\begin{appendix}
\section[Forced Poincar\'e equation]{Derivation of the Poincar\'e equation}
\label{app:PoincareEq}
The linearised system we consider, when we adopt the Boussinesq approximation, is given by Eqs. (\ref{eq:momx})--(\ref{eq:energy}). The aim of this appendix is to demonstrate the Poincar\'e equation, used in Section \ref{subsec:GIW:propagation}.

First, taking the combination $\uppartial_y$(\ref{eq:momz}) $-~\uppartial_z$(\ref{eq:momy}) gives
\begin{equation}
\frac{\uppartial}{\uppartial t}\left(\frac{\uppartial w}{\uppartial y}-\frac{\uppartial v}{\uppartial z}\right) - (\bm{f}\cdot\bm{\nabla})u
= \frac{\uppartial b}{\uppartial y},
\label{eq:1stcomb}
\end{equation}
where $\bm{f} = (0, \tilde{f},f)$. 
Then, taking the combination $\uppartial_z$(\ref{eq:momx}) $-~\uppartial_x$(\ref{eq:momz}) and using equation (\ref{eq:cont}) gives
\begin{equation}
\frac{\uppartial}{\uppartial t}\left(\frac{\uppartial u}{\uppartial z}-\frac{\uppartial w}{\uppartial x}\right) - (\bm{f}\cdot\bm{\nabla})v
= - \frac{\uppartial b}{\uppartial x}.
\label{eq:2ndcomb}
\end{equation}
Then, taking the combination $\uppartial_x$(\ref{eq:momy}) $-~\uppartial_y$(\ref{eq:momx}) and using equation (\ref{eq:cont}) gives
\begin{equation}
\frac{\uppartial}{\uppartial t}\left(\frac{\uppartial v}{\uppartial x}-\frac{\uppartial u}{\uppartial y}\right) - (\bm{f}\cdot\bm{\nabla})w
= 0.
\label{eq:3rdcomb}
\end{equation}
Then, by taking the combination $\uppartial_t\left(\uppartial_y\text{(\ref{eq:1stcomb})} - \uppartial_x\text{(\ref{eq:2ndcomb})}\right)$ and using equations (\ref{eq:cont}), (\ref{eq:energy}) and (\ref{eq:3rdcomb}), we finally obtain an equation on the vertical component of the velocity $w$
\begin{equation}
\dfrac{\uppartial^2}{\uppartial t^2} \nabla^2 w + (\bm{f} \cdot \bm{\nabla})^2 w + \left[N^2 \nabla_{\perp}^2\right] w
= 0,
\label{eq:forced_PoincareSimplified}
\end{equation}
where $\nabla_{\perp}^2 \equiv \dfrac{\uppartial^2}{\uppartial x^2} + \dfrac{\uppartial^2}{\uppartial y^2}$.

\end{appendix}

\end{document}